\documentclass[aps,prd,a4paper,superscriptaddress,nofootinbib,preprint,11pt]{revtex4-1}
\pdfoutput=1
\usepackage{color}
\usepackage[table]{xcolor}
\usepackage[colorlinks=true,citecolor=blue,linkcolor=blue,
breaklinks=true]{hyperref}
\usepackage{amsmath,amssymb}
\usepackage{graphicx}
\usepackage{slashed}
\usepackage[caption=false,labelformat=simple]{subfig}
\usepackage{wrapfig}
\usepackage{placeins}
\usepackage{multirow, makecell}
\usepackage{booktabs}

\newcommand{\gsim}{\gtrsim}

\usepackage{url}
\usepackage{enumitem}
\usepackage{multirow}
\usepackage{array}
\usepackage{tabulary}
\usepackage{float}

\usepackage{ulem}
\renewcommand{\emph}{\textit}

\newcolumntype{?}{!{\vrule width 1pt}}
\newcolumntype{M}[1]{>{\centering\arraybackslash}m{#1}}
\newcommand{\lag}{\mathcal L}
\newcommand{\cred}[1]{{\color{red}#1}}

\begin{document}
\preprint{HRI-RECAPP-2023-12}
%%%%%%%%%%%%%%%%%%%%%%%%%  Title  %%%%%%%%%%%%%%%%%%%%%%%%%%%%%%%%%%%%%%
\title{%The Dark Side of a Neutrinophilic $U(1)$ Model \\
 Heavy Neutrino as Dark Matter in a Neutrinophilic $U(1)$ Model 
%{\color{red} (Exploration of) Dark Matter Aspects of a Neutrinophilic $U(1)$ Model} \\
%{%\color{blue} Neutrinophilic Model: Unveiling its Dark Side/Dark Matter Aspects
}
%{\color{magenta} Addressing Dark Matter Puzzle in a Neutrinophilic $U(1)$ Model}}

%%%%%%%%%%%%%%%%%%%%%%%%%  Authors  %%%%%%%%%%%%%%%%%%%%%%%%%%%%%%%%%%%%
\author{Waleed Abdallah}
\email{awaleed@sci.cu.edu.eg} 
\affiliation{Department of Mathematics, Faculty of Science, Cairo
University, Giza 12613, Egypt.}

\author{Anjan Kumar Barik}
\email{anjanbarik@hri.res.in}

\author{Santosh Kumar Rai}
\email{skrai@hri.res.in}
\affiliation{Regional Centre for Accelerator-based Particle Physics,
Harish-Chandra Research Institute,\\ A CI of Homi Bhabha National
Institute, Chhatnag Road, Jhunsi, Prayagraj 211019, India.}

\author{Tousik Samui}
\email{tousik.pdf@iiserkol.ac.in}
\affiliation{Department of Physical Sciences, Indian Institute of
Science Education and Research Kolkata,\\ Mohanpur 741\,246, India.}

%%%%%%%%%%%%%%%%%%%%%%%%%  Abstract  %%%%%%%%%%%%%%%%%%%%%%%%%%%%%%%%%%%
\vspace{1.5cm}
\begin{abstract}
	
We study the prospect of heavy singlet neutrinos as a dark matter (DM) candidate within a neutrinophilic
$U(1)$ model, where the Standard Model (SM) is extended with a $U(1)$ gauge symmetry, and neutrino mass and oscillation parameters are explained through an inverse see-saw mechanism. The lightest of the heavy neutrinos plays the role of the DM while the newly introduced scalars and the extra gauge boson $Z'$ act as
mediators between the dark sector and the SM sector. 
%These mediators play an important role in obtaining the $Z'$-funnel and the scalar-funnel region and obtaining the correct DM relic density (DMRD).
We show the range of model parameters where this DM candidate can be accommodated in the Weakly Interacting Massive Particle (WIMP) or Feebly
Interacting Massive Particle (FIMP) scenario. The observed DM relic density is achieved via the new gauge boson and singlet scalar
portals in the WIMP scenario whereas within the FIMP scenario, these two particles assume a distinct yet pivotal role in generating the observed relic density of dark matter.
\end{abstract}

\maketitle

%%%%%%%%%%%%%%%%%%%%%%%%%  Introduction  %%%%%%%%%%%%%%%%%%%%%%%%%%%%%%%
\section{Introduction}
Astrophysical observations from Coma cluster\,\cite{Zwicky:1933gu},
galaxy rotation curves\,\cite{1939LicOB..19...41B,1980ApJ...238..471R},
gravitational lensing\,\cite{Massey:2010hh} and bullet
cluster\,\cite{Clowe:2006eq} provide strong evidence that there is a large amount of invisible matter in our universe. This matter 
is believed to interact very weakly with the visible sector particles described successfully by the Standard Model (SM) and is called the dark matter (DM). 
%the non-luminous nature with respect to the SM interactions suggests a non-SM origin for this matter, and it is called dark matter (DM).
Although very little information is available for the DM to us, %about this non-luminous matter,
it reveals some broad features; that it has gravitational interactions and it should be 
non-relativistic or "cold" so that it can
form structures in the universe. 
Additionally, WMAP\,\cite{WMAP:2012fli} data indicates that DM contributes roughly 27\% of the universe's energy budget, compared to only 5\% for visible matter. The energy density of the DM is expressed in terms of its relic
density (RD), which is measured to be
$0.1198\pm0.0012$\,\cite{Planck:2018vyg}. %with $68 \%$ C.L..

Despite its key importance in modern cosmology, the nature of the DM remains elusive to us. While the DM is known to have gravitational interactions, other interactions, {\it viz.} strong
and electroweak interactions of the DM sector are expected to be very
feeble to accommodate its non-observation in the direct detection
experiments\,\cite{XENON:2018voc,XENON:2019rxp,PandaX-4T:2021bab}. Two popular hypothesized scenarios containing DM are (a)
Weakly Interacting Massive Particle
(WIMP)\,\cite{Kolb:1990vq,Jungman:1995df,Bertone:2004pz}, and (b) Feebly
Interacting Massive Particle (FIMP)\,\cite{Hall:2009bx}. The WIMP
scenarios hypothesize a freeze-out mechanism, wherein the DM that was abundant
in the early phase of the universe froze out from the thermal bath due to the expansion of the
universe and remained dark to the visible sector except for its gravitational interaction. On the other hand, in the feebly interacting
scenario, the DM is considered to be absent in the early universe and produced later from the SM particles. The interaction strength of the DM candidate with the visible sector particles should be feeble enough so as not to
over-shoot the currently observed DM relic density. 
A crucial difference between these two scenarios is that the WIMPs are thought to be in the thermal
bath before they freeze out while the FIMPs were never in thermal equilibrium. There are many attempts to accommodate DM within BSM frameworks (see for
example~\cite{PhysRevD.56.1879,Dutra:2015vca,Mizukoshi:2010ky,Fayet:2006xd,An:2014twa,Pospelov:2007mp,Chakraborty:2021tdo,Dey:2022whc}) primarily via the WIMP scenario. On the other hand, attempts to explain the DM in the FIMP
scenario are relatively new (see for
example~\cite{Hall:2009bx,Gondolo:1990dk,Yaguna:2011qn,Chu:2011be,Shakya:2015xnx,Pandey:2017quk,Elahi:2014fsa,Bernal:2017kxu,Merle:2013wta,Ghosh:2017vhe,Cosme:2021baj,Cosme:2020mck,Bhattacharyya:2018evo,Abdallah:2019svm}).
In this work, we study a neutrinophilic model~\cite{Abdallah:2021npg},
which can accommodate WIMP and FIMP in different parameter regions. This
model considers an extra $U(1)$ gauge symmetry, the simplest
gauge extension of the SM. Models with an extra $U(1)$ symmetry can have their roots in grand unified theories \cite{Robinett:1981yz,Robinett:1982tq,Langacker:1984dc,Hewett:1988xc}, where one starts with
groups of higher ranks that leave an additional $U(1)$ after the spontaneous symmetry breaking. 
On the other hand, in the more simplistic bottom-up approach, one adds the $U(1)$ to the SM gauge group. The typical feature of such $U(1)$ models is the appearance of an extra massive gauge
boson~$Z'$ and additional scalars responsible for the spontaneous breaking of the additional symmetry. 
%The role of such $Z'$ boson as a mediator is sometimes crucial.

The model in consideration is primarily motivated to provide a solution to the neutrino mass and mixing puzzle. The light
neutrinos get their mass via the inverse see-saw mechanism in the presence of heavy sterile neutrinos that provide the Majorana term to felicitate the tiny masses. 
%which  after the symmetry breaking of an
%extra $U(1)_X$ symmetry with the introduction of an extra Higgs doublet and real scalar singlet. 
The lightest heavy sterile neutrino can act as a cold DM (CDM) in this scenario. The production of such DM in the early universe and their 
interaction with the SM particles then depends on its interactions via the $Z'$ and the new scalars in the model. The signature of this model at the Large Hadron Collider (LHC) was studied in Ref.~\cite{Abdallah:2021npg, Abdallah:2021dul} where the role of gauge-kinetic mixing was considered.  
%with a gaugekinetic mixing (GKM) of $\mathcal{O} (10^{-2})$. 
%In that study, the new gauge boson $Z'$ was considered to be produced from the quarks owing to the large value of GKM. When the new gauge boson is lighter than the SM gauge boson, a smaller value of GKM is needed to evade all the constraints on the new gauge boson. Hence, the production from quarks is negligible. Therefore, an alternative approach was suggested in Ref.~\cite{Abdallah:2021dul}. In that, we have used scalar mixing to produce it from SM Higgs. 
%We also found that lepton flavor violating
%decay of $Z'$ is also possible at the one-loop level in this model,
%which can have a potential impact on the experimental searches for~$Z'$.
%Another implication of a similar model in the context of ultralight
%mediator has also been studied in
%Refs.~\cite{He:2020zns,Berbig:2020wve}. However, 
In the latter work, we also showed the viability of the heavy neutrinos as DM candidates in the model. In this work, we discuss in more detail the prospects of the heavy sterile neutrinos as DM and their role as a WIMP or FIMP.

The rest of the article is organized as follows. In
section~\ref{sec:model}, we describe the model briefly with a focus on
the DM aspect of the model. Section~\ref{sec:wimp} contains the
discussion of the WIMP scenario and the detection of such a model in the
current and upcoming experiments. In section~\ref{sec:fimp}, we explore
the possibility of the FIMP scenario in a certain parameter space of the
model. We then summarize and conclude in section~\ref{sec:summ}.

%%%%%%%%%%%%%%%%%%%%%%%%%    The model   %%%%%%%%%%%%%%%%%%%%%%%%%%%%%%%
\section{The model} \label{sec:model}
We consider a neutrinophilic
model~\cite{Abdallah:2021npg,Abdallah:2021dul}, which is an extension
of the SM gauge symmetry with a new $U(1)_X$ gauge symmetry. The SM particles
do not transform under the new gauge symmetry. The new particles in this model
are a second Higgs doublet, a scalar singlet, and three vector-like
fermions. The charge assignment of the particles is shown in
Table~\ref{tab:particle}.
%=======================================================================
\begin{table}[!h]
\begin{center}
\begin{tabular}{|c|c|c|c|c|c|}
\hline 
Fields  & $SU(3)_C$ & $SU(2)_L$ & $U(1)_Y$ & $U(1)_X$ & Spin \\
\hline 
$N_L^i$ & 1 & 1 & 0 & $q_x$  & 1/2 \\ [1mm]
\hline 
$N_R^i$ & 1 & 1 & 0 & $q_x$ & 1/2 \\ [1mm]
\hline
$H_1$  & 1 & 2 & $-1/2$ & 0 & 0 \\ [1mm]
\hline 
$H_2$ & 1 & 2 & $-1/2$ & $-\,q_x$ &  0 \\ [1mm]
\hline 
$S$ & 1 & 1 & 0 & $2q_x$ &  0 \\ [1mm]
\hline 
$S_2$ & 1 & 1 & 0 & $ q_x$ &  0 \\ [1mm]
\hline 
\end{tabular}
\end{center}
\caption{New fields and their charge assignments under the SM gauge
group and $U(1)_X$.}
\label{tab:particle}
\end{table}
%=======================================================================

The gauge invariant Lagrangian containing the new
fields and the SM like Higgs doublet (denoted by $H_1$) is given as
\begin{eqnarray}
\lag &\supset& \left(D_\mu H_1 \right)^\dagger D_\mu H_1 + \left(D_\mu H_2 \right)^\dagger D_\mu H_2 + \left(D_\mu S \right)^\dagger D_\mu S - \mu_1 H_1^\dagger H_1\! -\! \mu_2 H_2^\dagger H_2 \!-\! \mu_s S^\dagger S  
   \nonumber \\ && +\{\mu_{12} H_1^\dagger H_2 + {\rm\, h.c.} \} -  \lambda_1 (H_1^\dagger H_1)^2 - \lambda_2 (H_2^\dagger H_2)^2  
             - \lambda_s (S^\dagger S)^2- \lambda'_{12} \left|H_1^\dagger H_2\right|^2
   -   \lambda_{12} H_1^\dagger H_1 H_2^\dagger H_2 
   \nonumber \\
 && - \lambda_{1s} H_1^\dagger H_1 S^\dagger S - \lambda_{2s} H_2^\dagger H_2 S^\dagger S - \{ Y_\nu\,\overline l_L H_2 N_R \!+\! Y_R S \overline N_R N_R^C \!+\! Y_L S \overline N_L N_L^C \!+\! {\rm\, h.c.}\}.\label{eqn:lag}
\end{eqnarray}
%An explicit but soft $U(1)_X$ breaking term, via 
%$\mu_{12}$ term, is present in the Lagrangian to provide mass to the pseudoscalar after the spontaneous breaking of electroweak and $U(1)_X$ symmetry. 
Note that the $\mu_{12}$ term which {\it apriori} breaks the $U(1)_X$ symmetry in the Lagrangian is dynamically generated through an additional scalar $S_2$. $S_2$ is singlet under the SM gauge symmetry and has $U(1)_{X}$ charge opposite to the $H_{2}$ so that a gauge invariant term $\mu_{3}H_1^\dagger H_2S_2$ can be added in the Lagrangian and therefore the softly broken term  $\mu_{12} = \mu_{3} \langle S_2 \rangle $ can be obtained when $S_2$ acquires vacuum expectation value. As the new scalar would simply introduce extra terms in the Lagrangian, it is assumed to be heavy to keep the model minimal and has been excluded from the Lagrangian shown in Eq.~(\ref{eqn:lag}).
%The additional particle content in the model can affect our analysis hence the scalar $S_2$ can be made heavy such that it does not affect our study. 
The masses 
for the scalars are obtained after the spontaneous symmetry
breaking of the scalar potential term of Eq.~(\ref{eqn:lag}).
Considering the vacuum expectation values (vev) for the scalars $H_1$,
$H_2$, and $S$ to be $v_1$, $v_2$, and $v_s$, respectively, the mass
matrix for the CP-even scalar becomes
\begin{eqnarray}
M_H^2 = \begin{pmatrix}
2\lambda_1 v_1^2 + \mu_{12}\frac{v_2}{v_1}& (\lambda_{12}+\lambda'_{12})v_1 v_2 -\mu_{12} &~& \lambda_{1s}\,v_1 v_s \\
(\lambda_{12}+\lambda'_{12})v_1 v_2 -\mu_{12} & 2\lambda_2 v_2^2 + \mu_{12}\frac{v_1}{v_2} &~& \lambda_{2s}\,v_2 v_s \\
\lambda_{1s}\,v_1 v_s & \lambda_{2s}\,v_2 v_s &~& 2\lambda_s v_s^2
\end{pmatrix} \, , 
\end{eqnarray}
After diagonalization via $3\times 3$ unitary mixing matrix $Z^h$, the
physical eigenstates are represented by $h_1$, $h_2$, and $h_s$ to
represent the states dominated by the neutral CP-even components of
$H_1$, $H_2$ and $S$, respectively. In our scenario, we identify $h_1$ to
be the observed 125-GeV scalar~\cite{ATLAS:2012yve,CMS:2012qbp} at the LHC. The other two are kept heavier than 125~GeV. The masses for pseudo-scalar
$A$ and the charged scalar $H^\pm$ can be expressed as
\begin{eqnarray}\label{csmass}
M_A^2 &=&\dfrac{\mu_{12}}{v_1 v_2}v^2 = \frac{2\mu_{12}}{\sin2\beta},\\
M_{H^\pm}^2&=&\left(\frac{\mu_{12}}{v_1 v_2}-\frac{\lambda_{12}'}{2}\right)v^2 = m_A^2 - \frac{\lambda_{12}'}{2}v^2,
\end{eqnarray}
where $v=\sqrt{v_1^2+v_2^2} \simeq 246$~GeV. In the general two Higgs
Doublet model type setups, it is conventional to parametrize the vevs
$v_1$ and $v_2$ in terms of $v$ and $\tan\beta\equiv v_2/v_1$ \footnote{In the most common two Higgs Doublet Model convention prefers to align the SM Higgs doublet with
$H_2$. Hence, a higher $\tan\beta$ value is preferable in those setups.
Our convention is opposite to the most common convention; therefore, a
low $\tan\beta$ region is preferable.}. 

A gauge-kinetic mixing (GKM) term of the form $\frac{1}{2}\tilde{g} B^{\mu\nu}
C_{\mu\nu}$ between $U(1)_Y$ gauge boson $B_\mu$ and the new gauge boson
can be introduced in the Lagrangian as it is gauge invariant~\cite{delAguila:1995rb, Chankowski:2006jk}. The kinetic term for the gauge boson with the inclusion of the GKM term can be written as 
\begin{eqnarray}
	\lag \supset - \frac{1}{4} G^{a,\mu\nu} G_{\mu\nu}^a - \frac{1}{4} W^{b,\mu\nu} W_{\mu\nu}^b -\frac{1}{4} B^{\mu\nu} B_{\mu\nu} - \frac{1}{4} C^{\mu\nu} C_{\mu\nu} + \frac{1}{2} \tilde{g} B^{\mu\nu} C_{\mu\nu} \, , 
\end{eqnarray}
The spontaneous breaking of gauge symmetry leads to massive gauge bosons in the particle spectrum. The $SU(2)\times U(1)$ gauge symmetry is spontaneously broken to $U(1)_{\text{em}}$ when $H_1$ and $H_2$ acquired vev, whereas the additional $U(1)_{X}$ gauge symmetry is spontaneously broken when $S$ and $H_2$ obtain vev. The new $U(1)_{X}$ gauge boson mixes with the SM neutral gauge bosons. %diagonalizing the neutral gauge boson mass matrix we find one physical state remains
%massless. This massless state is identified as the photon. The two other states are massive. 
The masses of the physical gauge bosons $Z$ and
$Z'$ can be expressed as
\begin{eqnarray}
M_{Z,Z'}^2 &=& \frac{1}{8}\Big[g_z^2 v^2 + {g'_x}^2 v^2 + 4 g_x g'_x v_2^2 + 4 g_x^2(v_2^2+ 4 v_s^2)\Big] \\
& &  \mp  \frac{1}{8}\sqrt{\Big({g'_x}^2 v^2 + 4 g_x g'_x v_2^2 + 4 g_x^2(v_2^2+ 4 v_s^2) - g_z^2 v^2\Big)^2 + 4 g_z^2 \Big(g'_x v^2 + 2 g_x v_2^2\Big)^2}  \nonumber \, , 
\end{eqnarray}
where $g_z=\sqrt{g_1^2+g_2^2}$, and
$g'_x=\frac{g_1\tilde{g}}{\sqrt{1-\tilde{g}^2}}$. The coupling constants
for  $U(1)_Y$, $SU(2)$ and $U(1)_X$ gauge groups are $g_1$, $g_2$ and
$g_x\sqrt{1-\tilde g}$, respectively\footnote{The redefinition of the
original coupling $g_x$ was needed to receive simplified expressions;
for more details, see Ref.~\cite{Abdallah:2021npg}.}. The mixing angle
between $Z$ and $Z'$ is given by
\begin{equation}\label{tanthetap}
\tan 2 \theta' = \dfrac{2g_z \left(g'_x v^2 + 2g_x v_2^2\right)}{{g'_x}^2 v^2 + 4 g_x g'_x v_2^2 + 4 g_x^2(v_2^2+ 4 v_s^2) 
- g_z^2 v^2}\simeq \frac{(2g_x\tan^2\beta+g'_x)M_Z^2}{M_{Z'}^2-M_Z^2},
\end{equation}
The electroweak precision observables, as well as precise measurement of
$Z$ boson decay width puts an upper limit of $10^{-3}$ on the mixing
angle $\theta'$ \cite{ParticleDataGroup:2020ssz}. 

Recall the terms for the neutrinos in the Lagrangian density in Eq.~(\ref{eqn:lag}):
\begin{eqnarray}
\lag &\supset& -\ Y_\nu\,\overline{l_L} H_2 N_R
 - Y_R S \overline N_R N_R^C - Y_L S \overline N_L N_L^C - {M}_N\overline{N}_L N_R + \text{h.c.}
\label{eqn:lag-neutrino}
\end{eqnarray}
where $N_R$ and $N_L$ are 3-generations of sterile neutrinos and $Y_\nu,
Y_L,$ and $Y_R$ are Yukawa couplings of the form $3\times 3$ matrices.
After symmetry breaking, the mass term for the neutrino sector
becomes
\begin{eqnarray}
\lag_\nu^\text{mass} = -\frac{v_2}{\sqrt 2}Y_\nu \overline{\nu}_L N_R - \frac{v_s}{\sqrt 2}Y_R\overline{N_R^C} N_R - {M}_N \overline{N}_L N_R - \frac{v_s}{\sqrt 2}Y_L\overline{N_L^C} N_L + {\rm h.c.}
\end{eqnarray}
The neutrino mass matrix in $\left(\nu_L\ N_R^C\ N_L \right)^T$ basis,
therefore, becomes
\begin{equation} 
{\cal M}_{\nu} = \left( 
\begin{array}{ccc}
0 &m_D^T  &0\\ 
m_D  &m_R  &{M}_N\\ 
0 &{M}_{N}^{T} &m_L \end{array}
\right), 
\end{equation}
with $m_D= v_2 Y_\nu/\sqrt{2}$, $m_R=\sqrt{2} v_s Y_R$ and $m_L=\sqrt{2}
v_s Y_L$. For $m_L, m_R \ll m_D, {M}_N$, the expressions for
masses of the neutrinos become
\begin{eqnarray}\label{nuH_mass}
m_{\nu_\ell} & \simeq & \frac{m^2_D\, m_L}{{M}_N^2+m_D^2},\label{mnul}\\
m_{\nu_{H,H'}} & \simeq & \frac{1}{2}\left(\frac{{M}_N^2\,m_L}{{M}_N^2+m_D^2}+ m_R\right) \mp \sqrt{{M}_N^2+m_D^2}\,,
\end{eqnarray}
where we represent $\nu_\ell$ to be light neutrinos and $\nu_{H,H'}$ to be
the heavier ones. For simplicity we have chosen $Y_R=0$. The $\mathcal{O}(0.1)$~eV neutrino mass can be
achieved with the following choices \[Y_\nu \sim \mathcal{O}(0.1),
\quad{M}_N \sim 1~\text{TeV},\quad \text{and}\quad m_L \sim
\mathcal{O}(10^{-6})~\text{GeV}.\]
We find that the model can accommodate a DM candidate from a pair of
heavy neutrinos $\nu_{4(5)}$. As there is no symmetry which prevents them from decay we have the decaying DM scenario in this model. These heavy neutrinos can decay to $l^{\pm} W^{\mp}$, $\nu Z$ due to their mixing with the SM neutrinos. As these decay channels of DM can yield high energy neutrino flux therefore the strongest bound on the life time of the DM comes from the IceCube collaboration which is $\mathcal{O}(10^{28})\,\text{s}$ \cite{Higaki:2014dwa,BhupalDev:2016gna,ReFiorentin:2016rzn,DiBari:2016guw}.  Therefore, the lightest of them ($\nu_{4}$) becomes
stable or long-lived if the corresponding Yukawa $Y_{\nu}$ couplings are  very small. For simplicity we consider $Y_{\nu}$ and $M_N$ to be diagonal. The total decay width of the lightest heavy neutrino when $M_4 > M_{Z}$ is approximately given by \cite{Atre:2009rg}
\begin{equation}
    \Gamma_{\nu_{4}} \approx \left|V_{14}\right|^2 \frac{3G_F M_{4}^3}{8\pi \sqrt{2}}
\end{equation}
 with $\left|V_{14}\right| = \mathcal{O}(\frac{m_{D}}{M_{N}})$ \cite{Abdallah:2021npg}. The IceCube bound on $\Gamma_{\nu_{4}}$ then translates into an upper bound for $Y_{\nu} < 4.7 \times 10^{-26}\times\frac{1}{\sin{\beta}}\sqrt{\frac{\text{GeV}}{M_{4}}}$. 
%$\lesssim 3\times 10^{-26}({\rm GeV}/M_N)^{1/2}$~\cite{BhupalDev:2016gna,ReFiorentin:2016rzn,DiBari:2016guw}. 
A viable choice that leads to a stable $\nu_4$ can be $Y_{\nu_{11}} \simeq 10^{-27}$ and
$Y_{\nu_{1j}}=0$ has been considered through out the subsequent analysis. The $\nu_{5}$ also has a similar
set of interactions as it is connected to $\nu_{4}$ via $M_{N_{11}}$.
A degenerate $\nu_5$ will therefore be also stable like $\nu_{4}$. In this work, we
concentrate on one component DM scenario. Therefore, we choose $Y_{{L}_{11}}$ in such a way that a mass splitting of a few MeV to a few GeV between $\nu_{4}$ and
$\nu_{5}$ is generated, that allows $\nu_{5}$ to decay to $\nu_{4}$ and a light SM 
fermion anti-fermion pair via off-shell $Z$ or $Z'$. Since there is a mixing between
the SM Higgs boson and the other two BSM scalars, the above can also be
mediated by these heavy scalars. As the interaction between DM and visible sector is mediated by the the singlet scalar and the $Z'$
%The lightest of them ($H_2$) is primarily dominated by the singlet $h_s$. 
, the DM production and annihilation are dominated by the $s$-channel diagrams with propagator $Z'$ and $h_i$'s. 
%In both the WIMP and FIMP scenarios, the most
%interesting DM phenomenology happens when the mass of the lightest sterile neutrino 
%is approximately half of $M_{Z'}$ or $M_{h_s}$. In the
%upcoming section, these two specific conditions will be referred to as
%`$Z'$-funnel' and `scalar-funnel'.

%%%%%%%%%%%%%%%%%%%%%%%%%       WIMP     %%%%%%%%%%%%%%%%%%%%%%%%%%%%%%%
\section{WIMP scenario}\label{sec:wimp}
Weakly interacting massive particles have been proposed to address the
DM puzzle, and in this scenario the WIMPs are usually massive compared to
the bath particles. Their scattering cross-sections are in the range of weak
interaction cross-sections suggesting that the WIMP was in thermal
equilibrium with SM particles in the early universe. As the universe
cools down because of the Hubble expansion, the equilibrium number
density of a particle species changes with time. When the temperature of
the universe becomes comparable to WIMP mass, the rate of the creation of DM from the annihilation of the bath particles is suppressed. This happens because the
fraction of the number density of equilibrium particles that possess
such high momentum to produce DM pair is Boltzmann suppressed. The co-moving number 
density of DM continues to change until the rate of all the DM number changing processes 
become much smaller compared to the Hubble expansion rate. 

The evolution of the number density $n_k$ of a particle species $k$ due
to the expansion of the universe and its interaction with other
particles is described by Boltzmann equation~\cite{Gondolo:1990dk}       
\begin{equation}
\frac{d n_k}{dt} + 3 H n_k = R_{k}(t), \label{eqn:boltzmann}
\end{equation}
where $H$ is the Hubble expansion rate and $R_{k}(t)$ is the collision term containing decay and (co)annihilation processes.
When the dark sector contains additional particles with masses close to
the mass of the DM, the Boltzmann equation for the number density of these additional particles
need to be taken into consideration, as the co-annihilation processes
can change the total dark sector density. We can obtain the present DM
number density by solving the coupled Boltzmann equations for the dark
sector particles. The coupled form of the equations comes via the
collision term, which incorporates decay and (co)annihilation processes \cite{Griest:1990kh}.
For a particular particle species, one can then calculate, via the
evolution equations, its present relic density (RD). This is defined as the
ratio of the present energy density to the critical energy density \cite{Belanger:2018ccd},
\begin{equation*}
\Omega_{X} = \frac{\rho_0}{\rho_c} =\frac{m_{X} n^{0}_{X}}{\rho_c} = \frac{m_{X} Y^{0}_{X} s_{0}}{\rho_c},\, \rho_c \equiv \frac{3H_0^2}{8\pi G},
\end{equation*}
where $\rho_c$ is the critical energy density, $s_{0} = 2.8912 \times 10^9\,$m$^{-3}$ gives the 
entropy density, and $H_0 = 100\, h \, {\rm km}\, {\rm s}^{-1}\, {\rm Mpc}^{-1}$ with
$h = 0.678(9)$, is the Hubble constant in present time.
%is the energy density for the universe considering it to be flat. 
The present yield $Y^0_{X}$ for particle species $X$ is $Y^0_{X} \equiv \frac{n_{X}^0}{s_0}$.

In this work, we consider the lightest ($\nu_{4}$) of the heavy sterile neutrinos to be the DM
candidate. While the next lightest state ($\nu_{5}$) mass is kept close to the $\nu_{4}$ mass, 
the remaining sterile neutrinos are kept much heavier than these two. 
So we ignore the heavier states as only $\nu_4$ and $\nu_5$ will play the most
important role in our analysis for DM. The $\nu_{4}$ and $\nu_{5}$ interact with the 
SM particles only through the $Z'$ and the singlet dominated scalar $h_s$ and they can (co)annihilate to the visible sector through the
$s$-channel via $Z'$ and $h_s$. There is significant enhancement in the (co)annihilation 
cross-section when the heavy neutrino mass becomes half of the $s$-channel mediator's mass. 
This gives a funnel-like shape in the DM relic density distribution 
near $M_{Z'}/2$ or $M_{h_s}/2$  when plotted as a function of DM mass. 
Therefore, we have two situations which we refer to as $Z'$-funnel and scalar-funnel 
scenarios depending on the mediator responsible for the funnel shape. 
In the following analysis, we study the annihilation channels of both the aforementioned scenarios. 

We used {\tt SARAH}~\cite{Staub:2013tta}
to implement the Lagrangian and {\tt SPheno}~\cite{Porod:2011nf} for the generation of a spectrum of parameters and masses. We have then used
{\tt micrOMEGAs}~\cite{Belanger:2010pz} to calculate the DM abundance.
%it is also possible to get $Z' Z'$, $h_s Z'$ and $h_s h_s$
%final state through the $t$-channel exchange of $\nu_{4(5)}$ from dark
%sector annihilation. This scenario has been discussed along with the
%scalar-funnel subsection because of the same final state in the two
%scenarios.
%%%%%%%%%%%%%%%%%%%%%%%%%  Z'-funnel  %%%%%%%%%%%%%%%%%%%%%%%%%%%%%%%%%%%
\subsection{$Z'$-funnel}
%As we discussed in the previous section, the WIMP goes out of equilibrium
%as the temperature of the universe falls below its mass, and then they
%continue to annihilate among themselves to the visible sector. Due to the
%expansion of the universe, after a certain time, this process becomes
%insignificant to change their comoving number density, which we observe
%as the present DMRD. Hence, (co)annihilation processes are
%crucial for maintaining observed relics. If the cross-sections of these
%processes are too small, then the observed DMRD would be larger than the
%present DMRD. On the other hand, a larger annihilation cross
%section will lead to an under-abundance of the DM. 
In this section, we study the DM freeze-out when they annihilate to the visible sector,
primarily through the $s$-channel exchange of $Z'$. 
%In order to have the $Z'$-mediated DM annihilation, the $Z$-$Z'$ mixing angle $\theta'$ has to be
%nonzero. This is because 
%The $Z'$ couples with the SM fermions one term is proportional to the GKM parameter $g'_x$, and the
%other is proportional to $\theta'$. 
The expression for the coupling of $Z'$ with SM fermions and scalars are given in Appendix~\ref{app:A}. Note that 
the SM fermion coupling to the $Z'$ depends on both the GKM parameter $g'_x$ and the $Z$-$Z'$ mixing angle $\theta'$. The precise measurement of $Z$ boson decay width at LEP put a strong constraint on the value of $\theta'$ to be below $<10^{-3}$~\cite{ParticleDataGroup:2020ssz}. This bound suppresses the coupling strength of $Z'$ with SM fermions containing $\theta'$. However the GKM parameter $g'_x$ can still have large values and dictate the couplings of $Z'$ 
to the SM fermions, thus enabling the DM to primarily annihilate to the SM fermions via the $Z'$ propagator. The $Z'$ coupling to SM gauge bosons also depend on $\theta'$ and its coupling to $W^+W^-$ and $h_iZ$ are given in the appendix.
% A natural choice of the $Z$-$Z'$ mixing angle to be small (), such that it does not modify the $Z$ boson couplings significantly with the SM fields.
%This can be ensured by keeping 
% GKM to be $\mathcal{O}(0.1)$ and small $\tan\beta \lesssim 10^{-2}$ (see Eq.~\ref{tanthetap}). %(has been explained already, right?)
% One can see from Eq.~(\ref{tanthetap}) that there are mainly two sources of nonzero mixing angle $\theta'$; one is pseudo-scalar mixing angle $\beta$ and the other is the GKM $g'_x$.
% One may choose to Different DM phenomenology 
In the vanishing GKM scenario, the possibility of DM annihilating to the SM fermions via $Z'$ is less. However, the presence of non-zero coupling of $Z'$ to the scalars of the model allows the DM to annihilate into these scalars. 
We explore these two scenarios {\it viz.} (a) nonzero GKM and (b) vanishing GKM in detail.

%========================================================================
\begin{figure}[t!]
\begin{center}
\includegraphics[width=9cm,height=4cm]{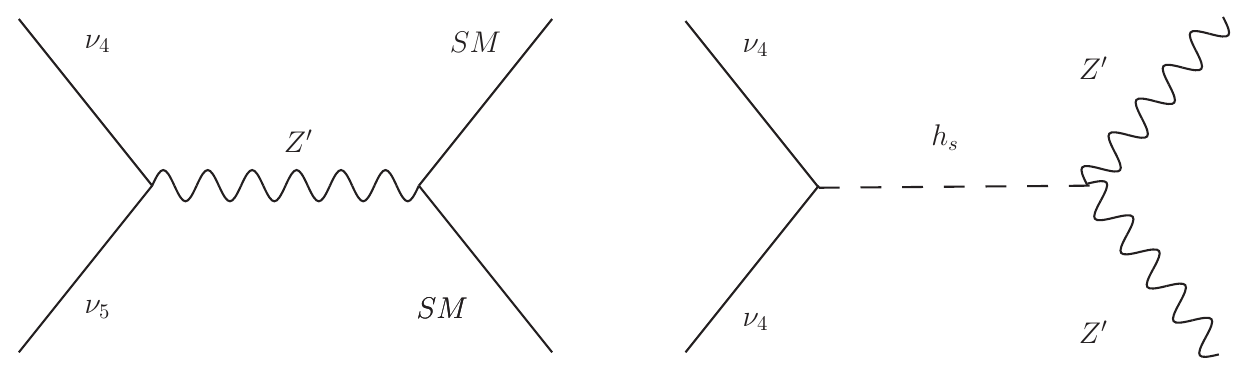}
\end{center}
\vspace{-0.5cm}
\caption{DM (co)annihilation process through $Z'$ and $h_{s}$ resonances with SM particles and $Z'$s in the final state respectively.}
\label{Fig-DM-vPhM-1}
\end{figure}
%========================================================================

%========================================================================
\subsubsection{Non-zero GKM} \label{Nonzero GKM}
\begin{figure}[h!]
\begin{center}
\includegraphics[width=8cm,height=6cm]{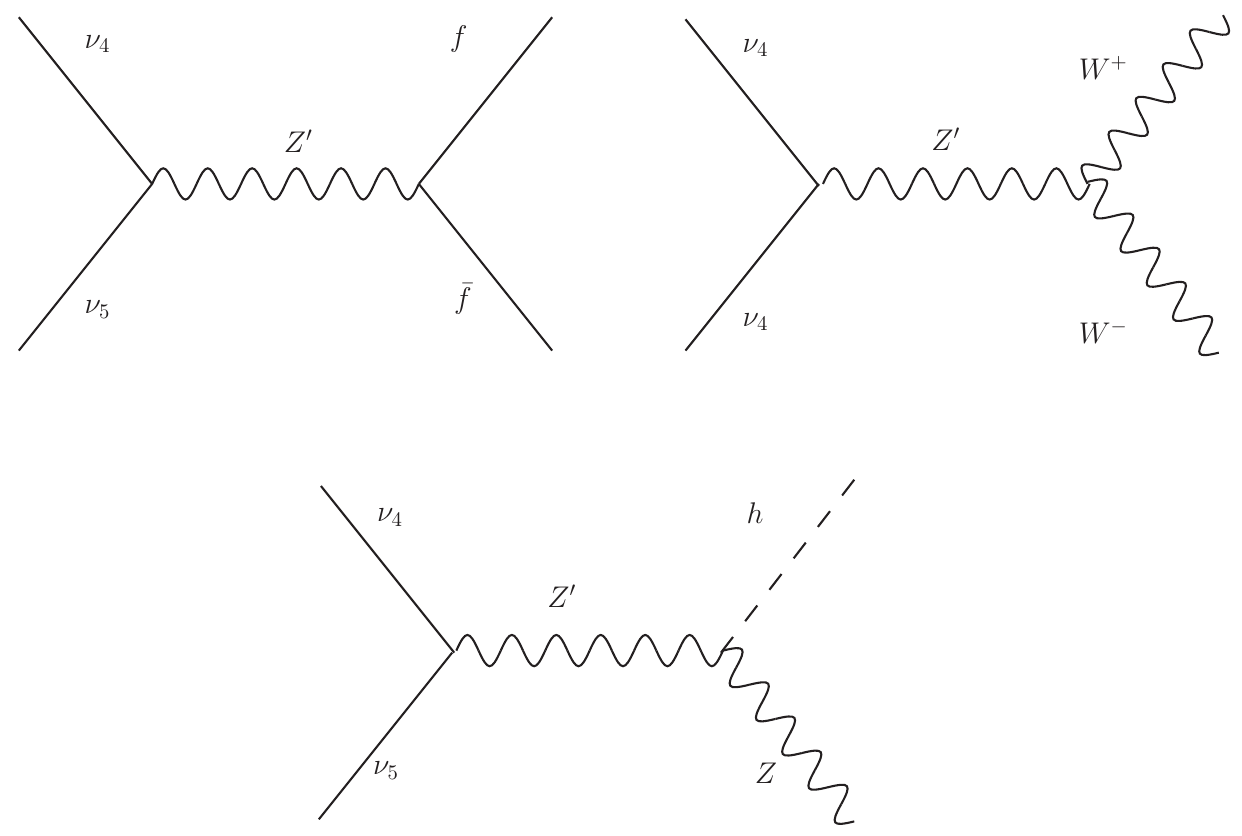}
\end{center}
\vspace{-0.5cm}
\caption{Feynman diagrams for $\nu_4,\nu_5$ co-annihilation to SM particles via $Z'$.}
\label{Fig-DM-vPhM-Zp-funnel}
\end{figure}

In this scenario we keep a non-vanishing value for the gauge kinetic mixing.
%We note that the value of $\theta'$ remains below $10^{-3}$ for $g'_x$ up to $\mathcal{O}(0.1)$ while
%a smaller value for $\tan\beta \lesssim 10^{-2}$ is preferable. 
In our WIMP scenario the masses for both the DM and the $Z'$ are chosen to be in the range of a few hundred GeVs. 
We have further checked that $\nu_{5}$ is not stable but is very long-lived. 
%with a total decay width of $\mathcal{O}(10^{-30})$~GeV. 
%In this parameter region, 
The relevant annihilation channels for our study are
\begin{eqnarray}
\nu_4\nu_5\to W^+W^-, \qquad \nu_4\nu_5\to h Z,\qquad \text{and}
\quad \nu_4\nu_5\to f\bar{f},
\end{eqnarray}
and they contribute via non-zero $Z$-$Z'$ mixing and GKM. The
Feynman diagrams for the above annihilation processes are shown in
Fig.~\ref{Fig-DM-vPhM-Zp-funnel}.
%========================================================================
\begin{figure}[h!]
\begin{center}
\subfloat[]{\includegraphics[width=8cm,height=6cm]{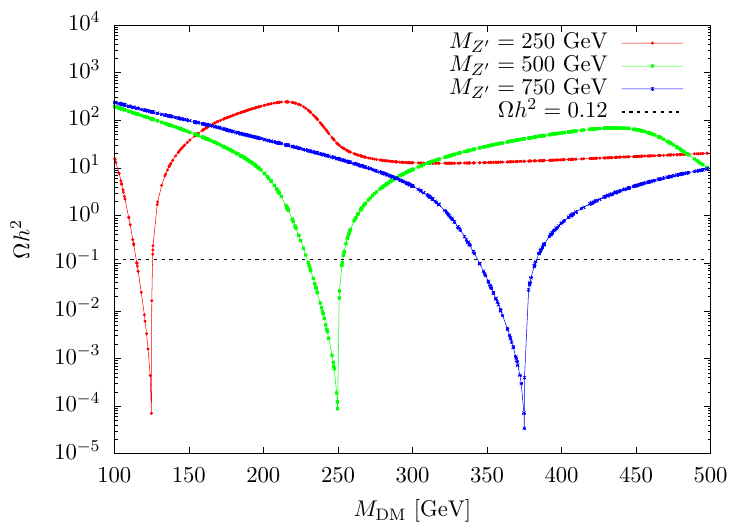}\label{Fig-DM-vPhM-2a}}
\subfloat[]{\includegraphics[width=8cm,height=6cm]{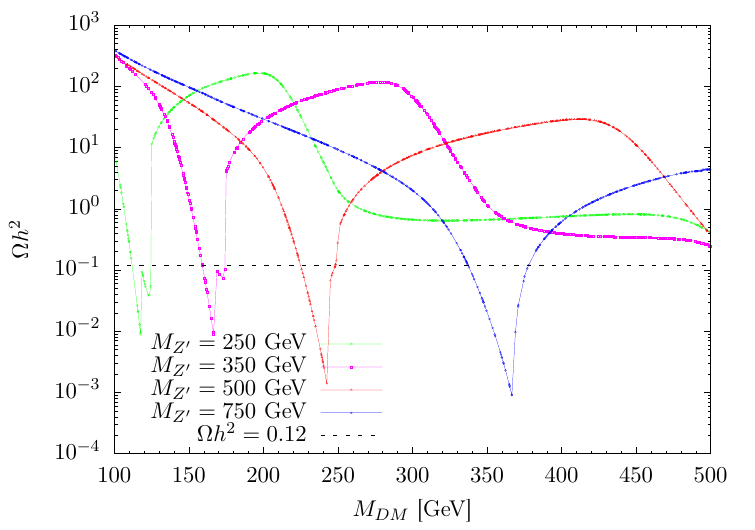}\label{Fig-DM-vPhM-2b}}
\end{center}
\vspace{-0.5cm}
\caption{The Dark Matter Relic Density (DMRD), denoted as $\Omega h^2$, is plotted against the Dark Matter (DM) mass for three different values of $M_{Z'}$. The mass gap between $\nu_4$ and $\nu_5$ is approximately of the order of a few MeV on the left plot and around 10 GeV on the right plot.}
\label{Fig-DM-vPhM-2}
\end{figure}
%========================================================================
To demonstrate the $Z'$-funnel, we first show, in
Fig.~\ref{Fig-DM-vPhM-2a},  the variation of the DM relic density $\Omega h^2$ as
a function of the mass of the DM. We choose three different values of $M_{Z'} = \{250,500,750\} \,\text{GeV}$
while the GKM parameter $g'_{x} = \{4,8,12\}\times 10^{-3} $ to show the relic density dependence on the propagator mass and coupling strengths. The mass gap between $\nu_4$ and $\nu_5$ is of $\mathcal{O}(\text{MeV})$, with $v_S = 1$ TeV,
$Y_{L_{11}} = 10^{-5}$. The $\nu_4 \nu_4 Z' $, $\nu_4 \nu_5 Z' $ and $\nu_5 \nu_5 Z' $ couplings strength depend on the mass splitting between $\nu_4$ and $\nu_5$ 
\begin{figure}[b]
\begin{center}		
\includegraphics[width=7cm, height=5cm]{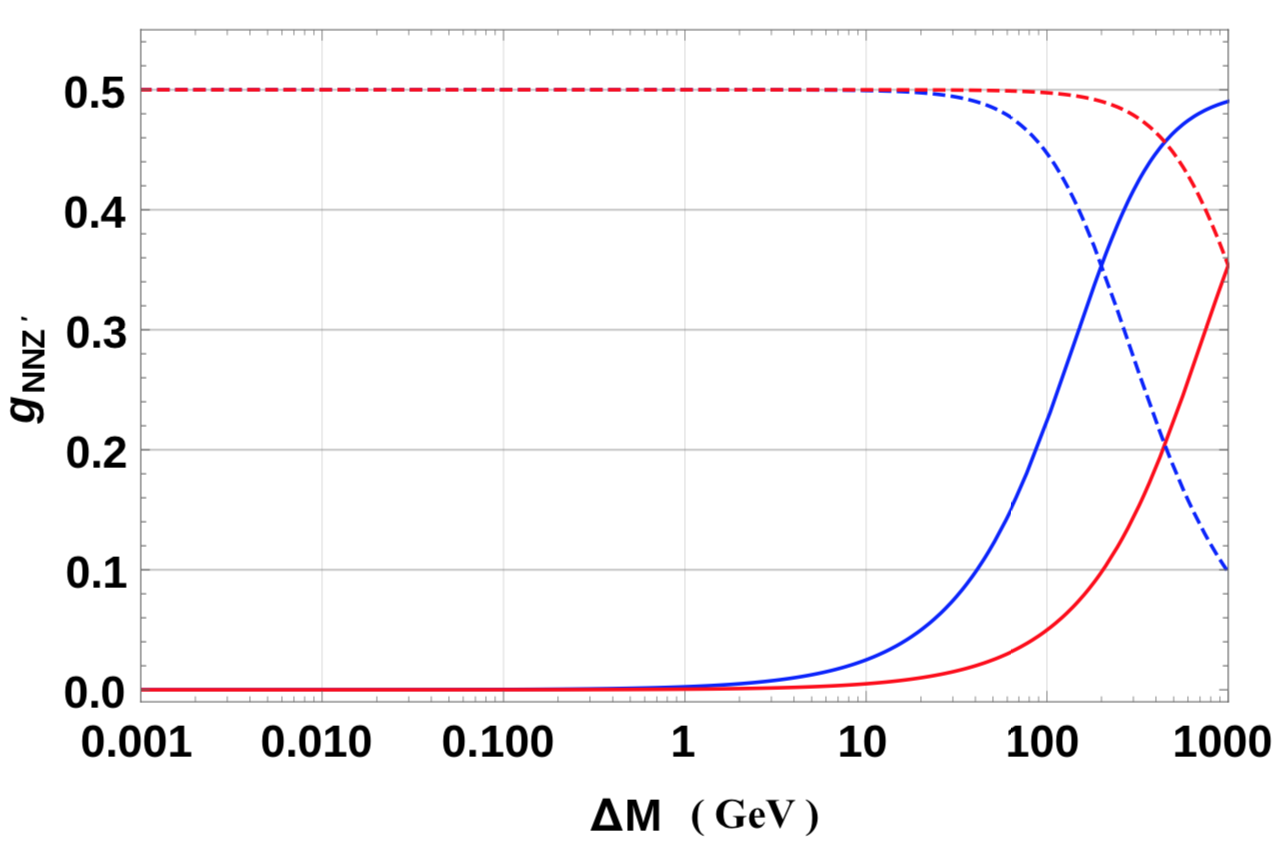}
\caption{ The variation of $\nu_4\nu_4Z'$ (solid line) and $\nu_4\nu_5Z'$ (dashed line) couplings as a function of $\Delta M = M_{\nu_5} - M_{\nu_4}$, with $g_x = 1$. The results are presented for two values of $M_{N_{11}}$: 100 GeV (blue) and 500 GeV (red).  }	
\label{NNZp_coupling}
\end{center}
\end{figure}
as shown in Fig.\ref{NNZp_coupling}. The dominant channel is therefore $\nu_{4} \nu_{5} \to f \bar{f}$ 
as they co-annihilate. A clear dip is visible in the DM relic at $M_\text{DM}\simeq M_{Z'}/2$. This is due 
to the $s$-channel resonance enhancement of the annihilation cross-section. An interesting
feature is that the funnel becomes wider as we go to a higher mass of
$Z'$ because of the increase in decay width of the $Z'$ boson. The DM relic density curve is not symmetric about the lowest point of $Z'$ funnel. In the region $M_\text{DM} < M_{Z'}/2$, as DM mass approaches the funnel region
the $\Omega h^2$ decreases since the DM annihilation can produce on-shell  $Z'$ as it possesses some kinetic
energy to reach up to the resonance.  As soon as we go
to the region $M_\text{DM}>M_{Z'}/2$, two DM particles cannot produce a single
$Z'$ on-shell as it is kinematically forbidden. As a result, it is
sharply increasing on the right of the resonance. We notice a fall in DM relic density near $M_\text{DM} \simeq M_{Z'}$. This is because of the
 t-channel DM annihilation processes $\nu_4\nu_4\to Z^{'}Z^{'}$ and $\nu_5\nu_5\to Z^{'}Z^{'}$ become kinematically allowed. 

We now consider the scenario where we have a $\mathcal{O}$(10~GeV) mass difference between $\nu_4$ and $\nu_5$. 
To calculate the relic density we scan over the following ranges of the relevant parameters:
\begin{eqnarray}
&M_{N_{11}}\in[100,500]~\text{GeV},\qquad g_x\in[0.2,1.0],\qquad
g'_x \in[-0.05,0.05], \nonumber \\
&\tan\beta = 10^{-2}, \quad \text{and} \qquad v_s = 502.5~\text{GeV}. \label{eqn:scanrange1}
\end{eqnarray}
Note that the smaller value for the vev $v_s$ allows a larger coupling strength $g_x$, compared to the previous scenario where the vev was 1 TeV. This has a direct effect on the total decay width of the mediating gauge boson $Z'$. Fig.~\ref{Fig-DM-vPhM-2b} shows the DM relic density as a function of the DM mass for four different values of $M_{Z'}$: 250, 350, 500, and 750~GeV with $g'_{x} = \{4,4,8,12\}\times 10^{-3} $ respectively. For the mass gap of $\mathcal{O}$(10~GeV) between $\nu_5$ and $\nu_4$, with $Y_{L_{11}} =  2\times 10^{-2}$, the distribution of the relic density is very similar to the in Fig.~\ref{Fig-DM-vPhM-2a}. As expected, the DM relic density has a dip near $M_{\rm DM}\simeq M_{Z'} - M_{\nu_5}$ due to the co-annihilation of $\nu_4$ and $\nu_5$ via $Z'$ channel. Additionally, a second dip appears near $M_{\rm DM}\simeq M_{Z'}/2$. This second dip is much more prominent for  $M_{Z'} = 250$~GeV and $M_{Z'} = 350$~GeV. 
Note that with a larger mass splitting between $\nu_4$ and $\nu_5$, there is a non-zero and sizeable $\nu_4\,\nu_{4} Z'$ interaction. However, as the DM mass increases, the coupling strength diminishes for a given mass difference between $\nu_4$ and $\nu_5$ (see Fig. \ref{NNZp_coupling}). As we go to higher values of the DM mass, the $\nu_4$ and $\nu_4$ annihilation process is not efficient enough to obtain the correct DM relic density. Therefore the depth of the second funnel decreases with an increase in DM mass. For $M_{Z'} = 500$~GeV and $M_{Z'} = 750$~GeV, the annihilation effect is less prominent.

\begin{figure}[h!]
\begin{center}
%\subfloat[]{\includegraphics[width=8cm,height=6cm]{MZpm2MDM-Oh2-thetap.png}\label{Fig-DM-vPhM-3a}}~~\subfloat[]{\includegraphics[width=8cm,height=6cm]{mdm_oh2_gxp.pdf}\label{Fig-DM-vPhM-3b}}\\
\subfloat[]{\includegraphics[width=8cm,height=6cm]{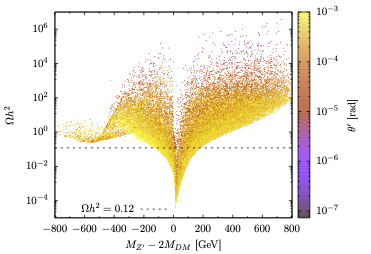}\label{Fig-DM-vPhM-3c}}~~\subfloat[]{\includegraphics[width=8cm,height=6cm]{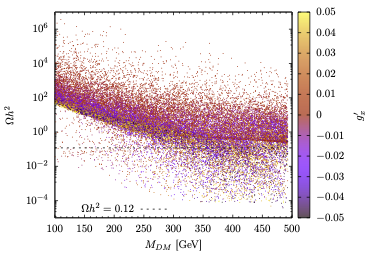}\label{Fig-DM-vPhM-3d}}
\end{center}
\vspace{-0.5cm}
\caption{(a) Dip in the DMRD occur at $M_{Z'} \simeq 2 M_\text{DM}$. (b) The DMRD $\Omega h^2$ as a function of DM mass and GKM $g'_{x}$.}
\label{Fig-DM-vPhM-3}
\end{figure}

We have performed a scan over the parameter regions mentioned in Eq.~(\ref{eqn:scanrange1}). We show the DM relic density in the $\Omega h^2$-($M_{Z'}-2M_\text{DM}$)
plane in Fig.~\ref{Fig-DM-vPhM-3c} for $|m_{\nu_5}-m_{\nu_4}|$
of $\mathcal{O}$(10~GeV). The plot shows that the DM relic density can be satisfied only when
$M_{\rm DM}\simeq M_{Z'}/2$, which we call $Z'$-funnel region.
Furthermore, in general, the DMRD is less in the funnel region compared
to the regions left and right to it. This is expected since the
annihilation of DM occurs though on-shell $Z'$ boson with a larger rate
and, therefore, the DMRD becomes less after the
freeze-out. 
\iffalse
On the left side of the plot, the DM annihilation can not
occur through  on-shell $Z'$ since the DM candidate is heavier than
$M_{Z'}/2$ in this region. On the right side of the plot, on the
other hand, the DM annihilation is possible through on-shell $Z'$, but
this requires a velocity larger than the mean thermal velocity of the DM.
This calls for a large Boltzmann suppression factor. Therefore, below and
above the funnel region, the DMRD after its
freeze-out remains high due to less annihilation in the early phase of
the universe. 
\fi
Clearly, most of the points are outside the funnel region have overabundant 
DMRD value. However, many points in the funnel region satisfy the
observed DMRD.

%the correct relic density for DM can be satisfied for various values of $M_{\rm DM}$.
%========================================================================
%========================================================================

% In the scan range, the dominant annihilation channels are as follows:
% \begin{eqnarray}
% \nu_4\nu_5\to W^+W^-, \qquad \nu_4\nu_5\to h_i Z^{(')},\qquad \text{and}
% \quad \nu_4\nu_5\to f\bar{f}.
% \end{eqnarray}
% These annihilation processes depend on the  $Z$-$Z'$ mixing angle and
% GKM parameter.
The same data points have been plotted in $\Omega h^2$-$M_\text{DM}$ plane in
Fig.~%~\ref{Fig-DM-vPhM-3c} 
\ref{Fig-DM-vPhM-3d} for $|m_{\nu_5}-m_{\nu_4}| \sim 10$ GeV, and $g'_x$ on the colour bar.
% From the colour bar, it is evident that negative values of $g'_x$ are preferable.
The red points,
which are close to $g'_x = 0$, are overabundant as, in that region, the
interaction of SM fields with $Z'$, which only depend on $\theta'$, is
very small. Hence, the DM annihilation via this channel is not efficient
enough to give the correct DMRD. We also notice that, there are
a large number of points satisfying DMRD in the heavier DM
region compared to the lighter DM region. A comparison
between the Figs.~\ref{Fig-DM-vPhM-3c} and~\ref{Fig-DM-vPhM-3d} reveals
that heavier $Z'$ prefers the correct- and under-relic points. This is
because, in the light $M_{Z'}$ region, the value of $g'_x$ needs to be
small in order to keep the $Z$-$Z'$ mixing angle $\theta'$ below
$10^{-3}$. As a result, DM annihilation is suppressed, leading to higher
RD. On the other hand, a heavy $Z'$ can accommodate higher
values of GKM while staying within the $\theta'$ upper limit. Hence, the
DMRD is less due to its high annihilation with high GKM. 
\subsubsection{Vanishing GKM}
% The previous subsection shows the role of $Z'$ in DM annihilation when GKM is nonzero.
We now move to study the DM phenomenology when $g'_{x} \simeq 0$. For
this scenario, we have varied the following parameters in the given
range:
\begin{equation}
M_{N_{11}}\in[100,500]~\text{GeV}, \qquad g_x\in[0.1,1.0],\qquad \text{and}\quad \tan\beta\in [0.01,3],
\end{equation}
while fixing $Y_{L_{11}} = 2 \times 10 ^{-2}$ and other scalar sector parameters.
As $\mu_{12}$ has been kept fixed, the masses of physical scalars from
$H_2$ can be varied by the parameter $\tan\beta$. The scalar quartic
couplings $\lambda_{12}$ and $\lambda'_{12}$ are kept small
($5\times10^{-3}$), so that the physical scalars from $H_2\, (h_2,A,H^{\pm})$ will have nearly degenerate
mass. Since GKM is vanishingly small, the $Z'$ will interact with the SM
fermions only through $\theta'$ (see Appendix~\ref{app:A} for Feynman
rules). Therefore, this mixing angle depends on the VEV of $H_2$, i.e. on
$\tan\beta$ and $M_{Z'}$, as can be seen from Eq.~(\ref{tanthetap}). One
can then check that, for a fixed value of $M_{Z'}$, the value of
$\theta'$ increases as $\tan\beta$ increases. Hence, $Z'$ coupling with
the SM fermions become stronger as we increase $\tan\beta$. Therefore,
to satisfy the observed DMRD, one needs an increased $\tan\beta$,
which enhances DM annihilation via $Z'$ to the SM modes. However, such a
large value of $\tan\beta$ is ruled out for lighter $Z'$ by the
requirement of $\theta' < 10^{-3}$. In addition large $\tan\beta$ can also alter the SM Higgs signal
strength.
% as before mixing SM fermions were not interacting with scalars from second Higgs doublet.
In the region scanned for $\tan\beta\in [0.01,3]$, we find that the allowed range of $\tan\beta\in [0.01,0.14]$ that satisfies the
constraints from various scalar searches at collider experiments and
Higgs signal strength. We have checked these experimental constraints 
using {\tt HiggsBounds}~\cite{Bechtle:2008jh,Bechtle:2011sb} and {\tt HiggsSignals}~\cite{Bechtle:2013xfa}. 

%========================================================================
\begin{figure}[h!]
\begin{center}
\subfloat[]{\includegraphics[width=5cm,height=4cm]{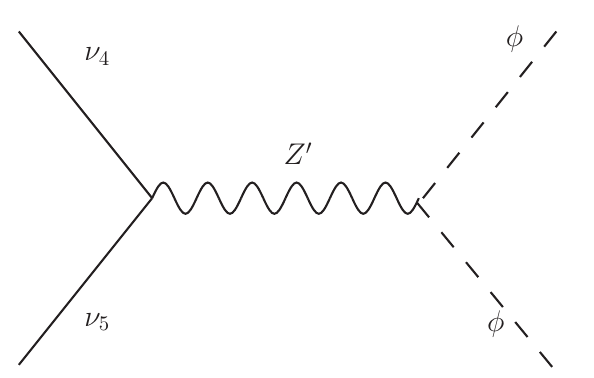}
\includegraphics[width=5cm,height=4cm]{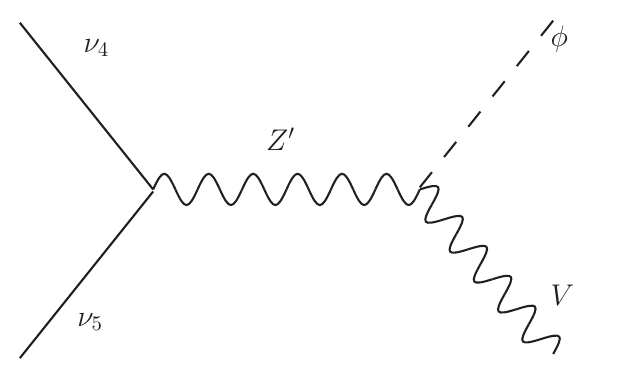}\label{Fig-DM-vPhM-4a}} \\ %\label{Fig-DM-vPhM-4b}}\\
\subfloat[]{\includegraphics[width=8cm]{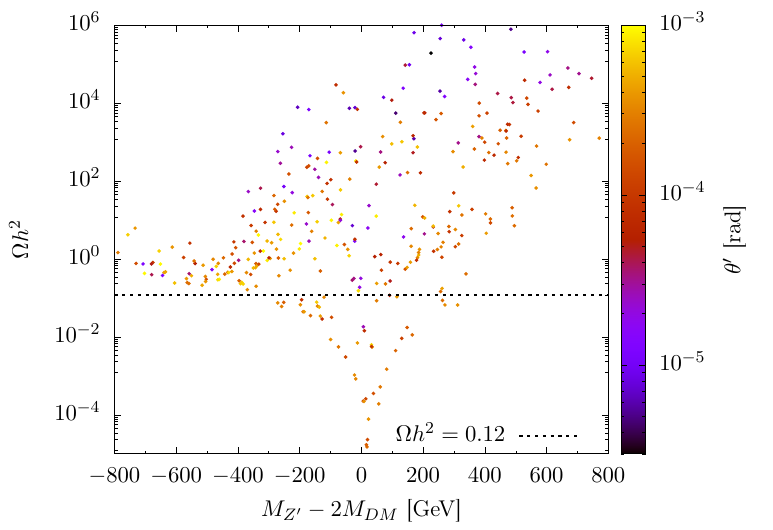}\label{Fig-DM-vPhM-4e}}
\subfloat[]{\includegraphics[width=8cm]{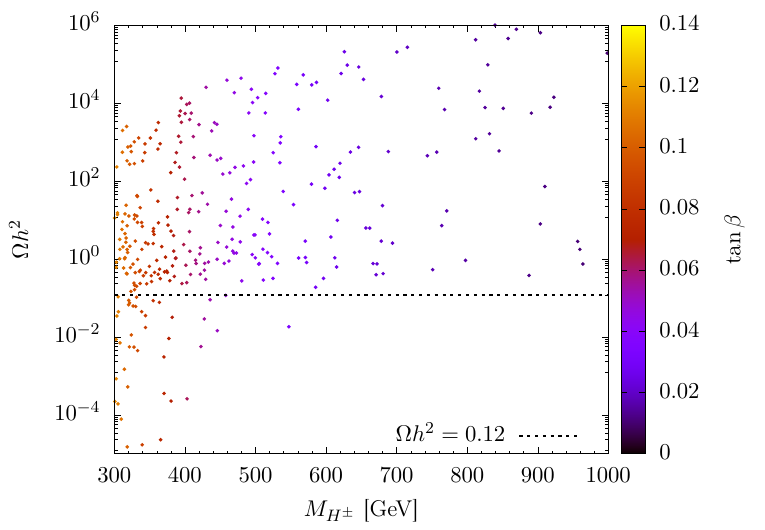}\label{Fig-DM-vPhM-4f}}\\
\subfloat[]{\includegraphics[width=8cm]{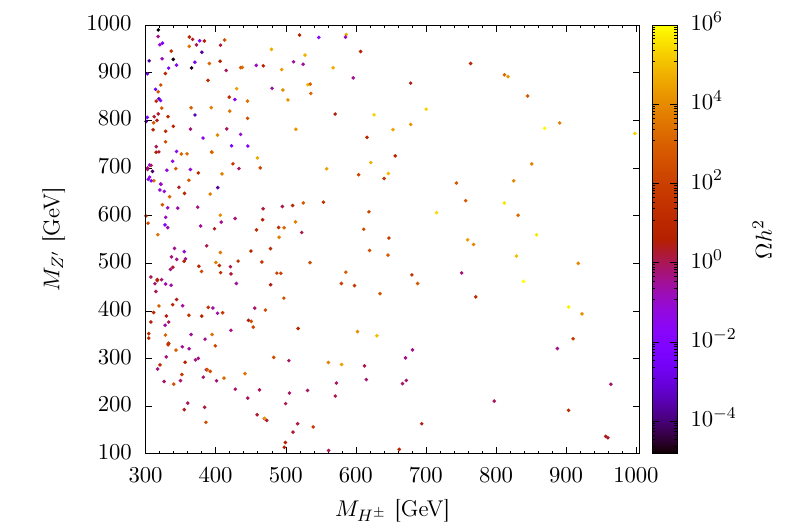}\label{Fig-DM-vPhM-4g}}
\end{center}
\vspace{-0.5cm}
\caption{(a)~Feynman diagrams for DM annihilation via $Z'$ resonance for vanishing GKM, where $\phi =h_1,h_2,A,H^\pm\ $ and $ V =Z,W^\pm $. (b)~Dip in the DMRD $\Omega h^2$ due to resonant production of $Z'$. (c)~Higher values of $\tan\beta$ give lighter $h_{2}, H^\pm, A$, hence DM can annihilate to them via $Z'$ resonance. (d) Scanned points in $M_H^{\pm}-M_{Z'}$ plane with DMRD in colour bar.}
\label{Fig-DM-vPhM-4}
\end{figure} 
%========================================================================

The small value of $\theta'$ would prevent the DM from annihilating to the SM
fields, which could have led to an overabundant scenario. However, coupling of the $Z'$ to the scalars 
comes to our rescue. Therefore, in the vanishing GKM scenario, the most dominant DM (co)annihilation processes, 
through $Z'$ propagator are
\begin{eqnarray}
\nu_{4} \nu_{5} \to H^+ H^-/ h_2 A/h_2 Z/H^{\pm} W^{\mp}/h_1 A.
\end{eqnarray}
Some representative Feynman diagrams of these processes are shown
in Fig.~\ref{Fig-DM-vPhM-4a}. To show the funnel region, we have plotted
the parameter points in the $\Omega h^2$-$(M_{Z'}-2M_\text{DM})$ plane in Fig.~\ref{Fig-DM-vPhM-4e}.
All these points satisfy the scalar sector constraints.
%, which have been checked by using {\tt HiggsBounds}~\cite{Bechtle:2008jh,Bechtle:2011sb}
The first two processes $(\nu_4\nu_5 \to H^+ H^-/ h_2 A)$ are dominant when
$M_{Z'} > 2 M_{H^\pm}$ and $M_{Z'} > M_{h_2}+M_A$. 
This is illustrated in Fig.~\ref{Fig-DM-vPhM-4f}, where the DMRD is lower in the lighter $M_{H^\pm}$ region. Since $\mu_{12}$ is fixed, therefore, a smaller $M_{H\pm}$ corresponds to larger $\tan\beta$, which is evident from the figure and the equation \ref{csmass} .
 On the other hand, the vertex containing two gauge
bosons and one scalar involves two gauge couplings and a vev, making
the third and fourth processes, $(\nu_4\nu_5 \to h_2 Z/ H^\pm W^\mp)$ subdominant due to
the small vev of $H_{2}$ doublet. Therefore, in this region, the
DM annihilation rate through these processes is higher for a smaller mass of charged Higgs.  The last process,
{i.e.}~$\nu_4\nu_5 \to h_1 A$, depends on  the mixing angle between the neutral CP-even components
of $H_1$, $H_2$. Thus, for larger $\tan\beta$, this process has significant contribution in achieving the correct DMRD. 

In the $M_{Z'} < 2 M_{H^\pm}$ region, we rely on the last three processes to
achieve the correct DMRD, if they are kinematically allowed. We need a larger
$\tan\beta$ so that coupling strength involving two gauge bosons and a second doublet scalar
 get enhanced due to a larger $H_2$ vev. From Fig. \ref{Fig-DM-vPhM-4g}, we find that some points with $M_{Z'} < 2 M_{H^\pm}$ in the lighter region of $M_{H^\pm}$ satisfy the relic density. 
 In high $M_{H^{\pm}}$ region ($>500$~GeV), DM annihilation through these processes is ineffective due to small $\tan\beta$ . Given the scan range of $M_{Z'}\in [100,1000]$ GeV, in the region  $M_{H^{\pm}} > 500 $ GeV, $Z'$ can not decay to $ H^+ H^-/ h_2 A$, leading to an overabundant scenario.

%========================================================================
\begin{figure}[h!]
\includegraphics[width=8cm,height=6cm]{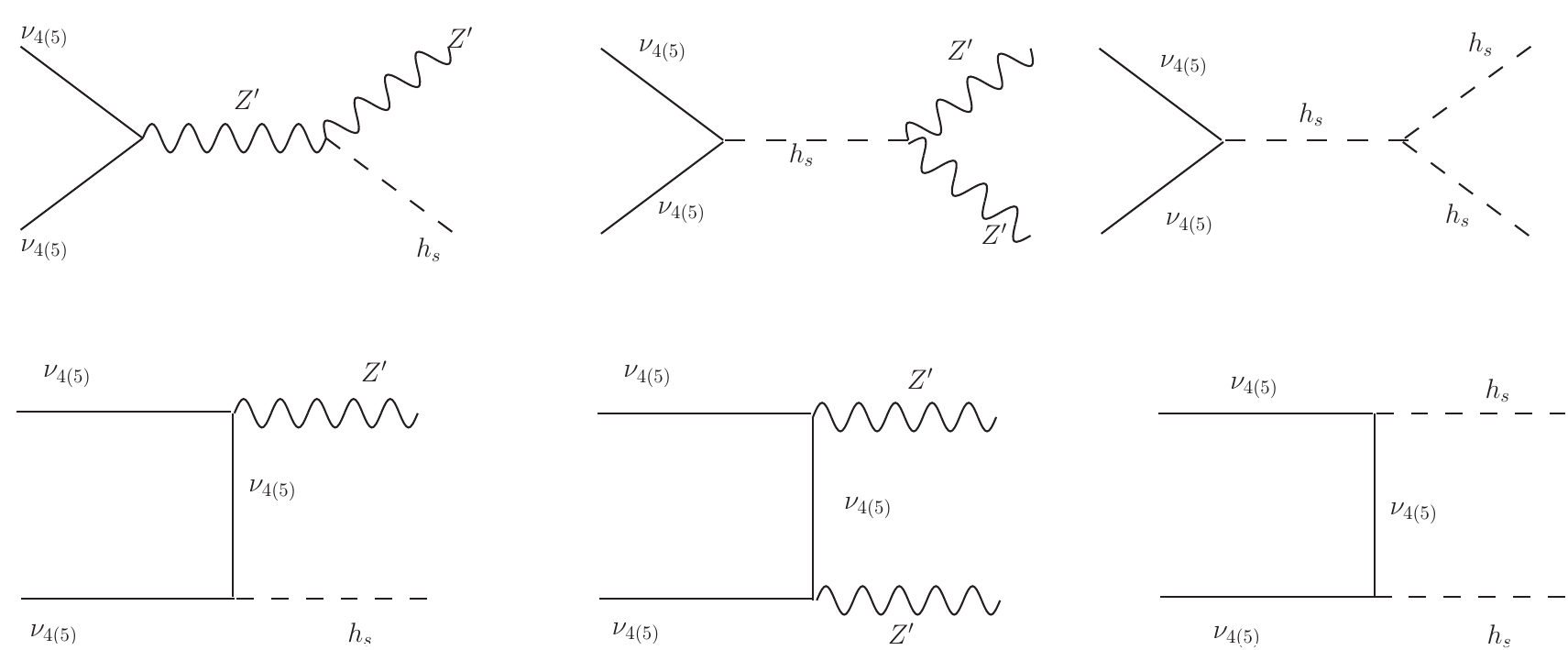}
%\reflectbox{\includegraphics[width=6cm,height=8cm,angle=90,origin=c]{vphm.pdf}}
\caption{The dominant annihilation channels: $\nu_4\nu_5\to h_{s} Z^\prime$ ($t$-channel via $\nu_4$ and $\nu_5$), $\nu_4\nu_4\to Z^\prime Z^\prime, \nu_5\nu_5\to Z^\prime Z^\prime$. Note that it is possible for  DM annihilation through  $t$-channels: $\nu_4\nu_5\to h_{s}h_{s}$, $\nu_5\nu_5\to h_{s}h_{s}$ and $\nu_4\nu_4\to h_{s}h_{s}$.}
\label{Fig-DM-vPhM-5}
\end{figure}
%========================================================================

%========================================================================
\begin{figure}[h!]
\subfloat[]{\includegraphics[width=8.5cm]{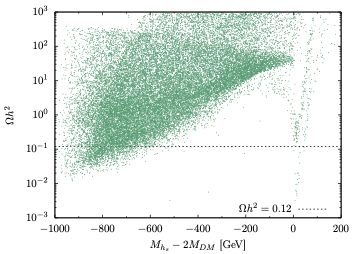}\label{Fig-DM-vPhM-6a}}
\subfloat[]{\includegraphics[width=8cm]{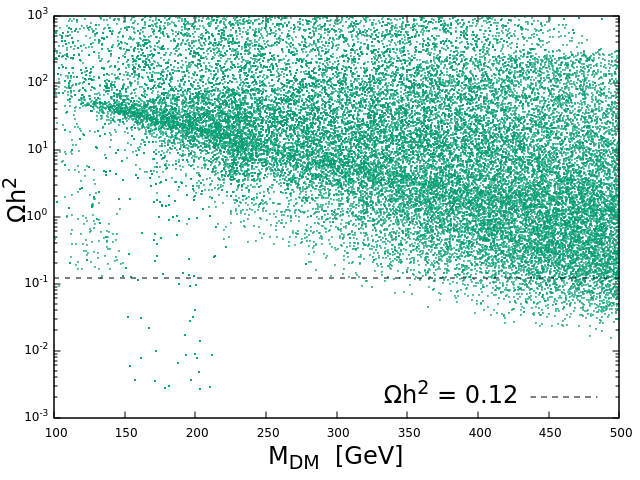}\label{Fig-DM-vPhM-6b}}\\ %[0.1cm]
\subfloat[]{\includegraphics[width=8.5cm]{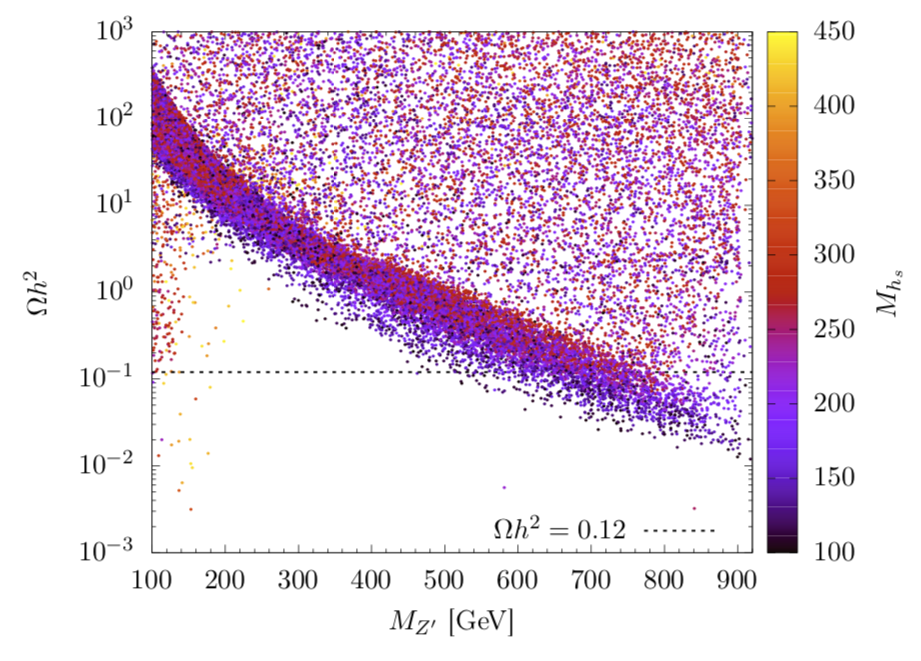}\label{Fig-DM-vPhM-6c}}
%\subfloat[]{\includegraphics[width=8.5cm]{Mzp_oh2_yl.pdf}\label{Fig-DM-vPhM-6d}}
\caption{(a) Dip at $M_{h_{s}} = 2 M_\text{DM}$ implies, $h_{s}$ is resonantly produced from DM annihilation. In region $M_{h_{s}} - 2 M_\text{DM} < 400$~GeV, a DM pair dominantly annihilates to $h_{s} h_{s}$, $h_{s}Z'$. (b)~The DMRD $\Omega h^2$ as function of $M_\text{DM}$. (c)~The DMRD $\Omega h^2$ as a function of $M_{Z'}$ and $M_{h_{s}}$. For lighter $Z'$, a DM pair annihilates to $Z' Z'$ via $h_{s}$ resonance while for heavy $Z'$, it annihilates to $h_{s}Z'$.} %(d)~The $\Omega h^2$ as a function of $M_{Z'}$ and $Y_{L_{11}}$.
\label{Fig-DM-vPhM-6}
\end{figure}
%========================================================================
%%%%%%%%%%%%%%%%%%%%%%%%%  scalar-funnel  %%%%%%%%%%%%%%%%%%%%%%%%%%%%%%%
\subsection{Scalar funnel}
In the scenarios where the $Z'$ has weaker interaction with the SM particles or the
annihilation processes  discussed earlier are not
kinematically feasible, the DM annihilation via $Z'$ is not
efficient enough to provide the observed DMRD. In such cases, we
rely on the scalar sector parameters to obtain the correct DMRD. The $\nu_{4}$ and $\nu_{5}$ primarily  interact with the singlet
scalar (neglecting the coupling with $h_2$, as it is of
$\mathcal{O}(10^{-27})$) through the Yukawa coupling $Y_{L_{11}}$. Additionally, they can couple to other scalars through scalar mixing.
As a result, $s$-channel processes are possible where $\nu_4$ and $\nu_5$
annihilate via $h_s$  to the SM and other visible BSM
particles. Moreover, $\nu_4$ and $\nu_5$ can annihilate to $h_s Z',~ Z'Z',\text{~and~} h_s h_s$ through t-channels when these processes are kinematically allowed.

The Feynman diagrams contributing to these processes are shown in
Fig.~\ref{Fig-DM-vPhM-5}. The top row of the figure illustrates all the
$s$-channel processes, with the top middle diagram potentially responsible for scalar-funnel DM
annihilation. On the other hand, the bottom row shows the diagrams with
the $\nu_{4(5)}$ in the $t$-channel. Note that the $Z'$
propagator in the diagram corresponding to the annihilation channel
$h_s Z'$, which is shown in the top-left panel, cannot be on-shell. The
same is true for the diagrams contributing to
$\nu_{4(5)}\nu_{4(5)}\to h_s h_s$ annihilation channel. These channels,
therefore, do not contribute to the funnel region.

 For the study of DM phenomenology, we vary the relevant parameters in the following ranges:
\begin{eqnarray}
Y_{L_{11}} \in [0.005,0.5], \qquad\qquad M_{N_{11}} \in [100,500]~\text{GeV}, \nonumber \\
\qquad M_{Z'} \in [100,1000]~\text{GeV}, \quad\text{and}\quad M_{h_{s}} \in [100,450]~\text{GeV}.
\end{eqnarray}
In this study, we consider $g'_{x} = 0$ so that the $Z'$ coupling to
the SM fields is negligible and choose $\tan\beta=10^{-4}$ to make masses of $h_2, A, H^\pm$  around 10~TeV when $\mu_{12} = 10^4~{\rm GeV}^2$. This setup minimises the contribution of $Z'$ resonance to the DMRD.
The
 process ($\nu_{4(5)} \nu_{4(5)} \to Z' Z'$) becomes  significant 
in  the DM sector particles annihilation, when $h_s$ is resonantly produced and decays to two $Z'$ bosons, thereby creating
a funnel in DMRD. When the $s$-channel scalar
resonance does not provide substantial annihilation, the other DM annihilation processes with
$t$-channel and $s$-channel  are effective involving $h_s Z'$, $Z' Z'$
and $h_s h_s$ in the final state.
The  annihilation process ($\nu_4 \nu_5 \to h_s Z'$) happens without
 scalar mixing as the singlet scalar was charged under $U(1)_X$ . This is the most
dominant annihilation channel in this parameter region of our scan, occurring via both $t$-channel and $s$-channel. The $s$-channel diagram, contains $Z' Z' h_s$ vertex, which is proportional to $v_s g_x^2$, making this process the most significant contributor.   
\iffalse
In the  ($\nu_4 \nu_5 \to Z' Z'$) and
 ($\nu_4 \nu_5 \to h_s h_s$) processes, the strength of Yukawa
coupling $Y_{L_{11}}$ determines the dark sector and singlet scalar
coupling. This coupling also controls the mass splitting between
$\nu_{4}$ and $\nu_{5}$. Therefore, taking higher values of the Yukawa
coupling enhances the annihilation process. However, in that case, the
co-annihilation channel would not be effective due to the large mass
splitting between $\nu_{4}$ and $\nu_{5}$. 
\fi

Fig.~\ref{Fig-DM-vPhM-6} shows the variation of the DMRD with
$M_\text{DM}$, $M_{h_s}$ and $M_{Z'}$. In order to show the funnel region,
the parameter points are plotted in
$\Omega h^2$-$(M_{h_s}-2M_\text{DM})$ plane in Fig.~\ref{Fig-DM-vPhM-6a}
where the funnel-like shape appears near $M_{h_s}-2M_\text{DM}\simeq 0$
region. In this region,
the dominant
channel for these points is $\nu_4 \nu_4 \to h_{s} \to Z' Z'$ through the
singlet scalar $s$-channel process. The points to the left of the funnel region correspond to the
$\nu_4 \nu_5 \to h_s Z'$ process where the DM pair is annihilating to $h_s$
and $Z'$. In Fig.~\ref{Fig-DM-vPhM-6b}, we show the same parameter
points but in the $\Omega h^2$-$M_\text{DM}$ plane. In this plot, one can
see that the points corresponding to the funnel region are spread near
$M_\text{DM}\simeq M_{Z'} \in [100,200]$~GeV. However, the region with 
DM mass above 500~GeV has correct- or under abundant relic density because of their
annihilation to $h_s Z'$ final state. These points correspond to
$M_{Z'} > 500$~GeV as can be seen from Figs.~\ref{Fig-DM-vPhM-6c} . From these two panels,
we find some points are under-abundant when $M_{Z'} \in [100,200]$~GeV with $M_{h_s} \in [250,450]$~GeV. 
These points belong to the scalar funnel region. The scalar funnel region can be extended to heavier $M_{Z'}$ by increasing $M_{h_s}$.
%In this region,
%the dominant
%channel for these points is $\nu_4 \nu_4 \to h_{s} \to Z' Z'$ through the
%singlet scalar $s$-channel process.
As the singlet scalar has a mixing of
$\mathcal{O}(10^{-8})$ and smaller coupling with the other two CP
even scalars, DM annihilating to SM particles or two light scalars through
singlet scalar gives a negligible contribution to the the DMRD.   

Fig.~\ref{Fig-DM-vPhM-6c} shows that there are some under-abundant points
with $M_{Z'} \in [500,900]$~GeV and 
$M_{h_s} \in [100,300]$~GeV. The dominant channel for these points is
$\nu_4 \nu_5 \to h_{s} Z'$ with $Z'$ in $s$-channel. For $\frac{M_{h_{s}}}{2}< M_{Z'} < 500$~GeV,
$g_x$ is not large enough for considerable DM annihilation. Therefore, overabundance of the DMRD occurs since the annihilation cross-section is proportional to $g^4_x$. 

\subsection{DM detection}
%========================================================================
\begin{figure}[h!]
\begin{center}
\subfloat[]{\includegraphics[width=8.5cm,height=6.5cm]{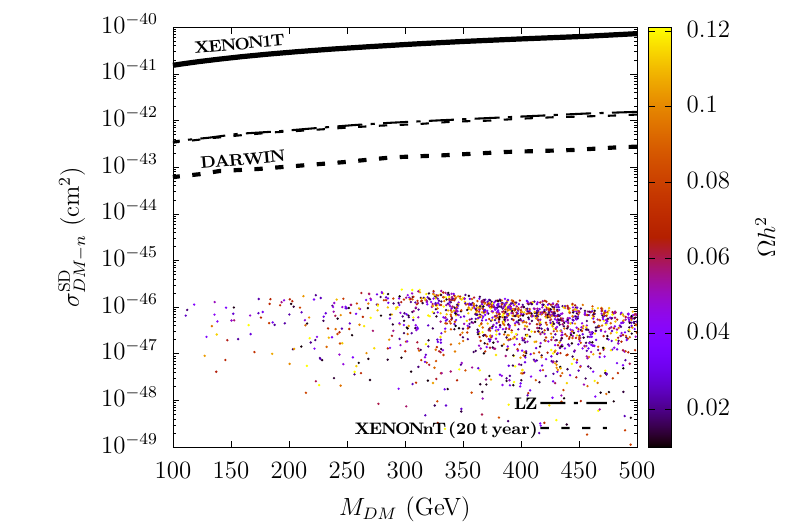}\label{Fig-DM-vPhM-7a}}
~\subfloat[]{\includegraphics[width=8cm,height=6.5cm]{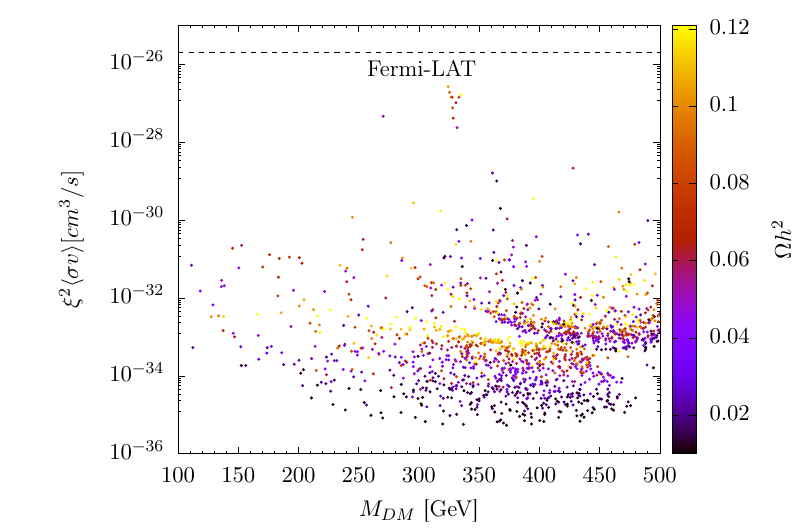}\label{Fig-DM-vPhM-7b}}
\end{center}
\vspace{-0.5cm}
\caption{(a) Scatter plot of spin-dependent cross-section of DM and nucleons. (b) Scatter plot of thermally averaged annihilation cross-section. For both plots, the scan range is set according to Eq.~(\ref{eqn:scanrange1}).}
\label{Fig-DM-vPhM-7}
\end{figure}
%========================================================================
In the previous sections, we have discussed how to achieve the observed DMRD in the right ballpark via different annihilation channels. Now, we move to the
observational aspects of the DM candidate in our model and 
the constraints coming from the DM direct and indirect detection experiments. 

In the scenario, where the DM  $(\nu_{4})$ and $\nu_{5}$ have nearly degenerate masses with a mass splitting of $\mathcal{O} (\text{MeV})$, giving them the
pseudo-Dirac nature, the $\nu_{4}\nu_{4}$ coupling with the
$Z'$ becomes very small. While the aforementioned scenarios result in the observed DMRD for $M_\text{DM} \in [100,500]$~GeV, it is also important to check for their signal in the direct detection~(DD) experiments. In our scenario we find that the spin-dependent 
DM-nucleon scattering cross-section measured at the DD experiments is very small. This is because both the $Z$ and $Z'$ coupling to the DM pair is small as mentioned above. Since the mixing angle between the CP-even component of $S$ and $H_1$ is considered to be less than $10^{-3}$, the spin-independent cross-section is also very small due to this minimal scalar mixing. The DM candidate interacts with the neutral gauge bosons $Z$ and $Z'$  through axial vector coupling, resulting in no contribution to the spin-independent DM-nucleon scattering from the gauge boson-mediated channels.
%generated by the axial vector coupling, becomes very small.
%Therefore, its value is close to the {\it neutrino floor}~\cite{Freedman:1973yd,Billard:2013qya,OHare:2016pjy,OHare:2021utq}.The inelastic scattering cross-section between the DM and nucleons where $\nu_5$ and a nucleon
%in the final state becomes negligible because the cold DM has kinetic energy of
%$\mathcal{O}(10-100)$~keV.

 %As we have consider mixing angle between CP even component of $S$ and $H_1$ to be less than $10^{-3}$, the spin-independent cross-section is also
%very small due to this small scalar mixing. The  DM candidate couples to the neutral gauge bosons $Z$ and $Z'$ only via axial vector coupling, resulting in no contribution to the spin-independent DM nucleon scattering due to the gauge boson-mediated channels. 

%Therefore, to enhance the $\nu_4\nu_4Z'$ coupling, we need to deviate from the pseudodirac nature of the DM. Thus, we introduce
Note that a mass splitting of $\mathcal{O}(10)$ GeV between $\nu_{4}$ and $\nu_{5}$ (by setting $Y_{L} \sim \mathcal{O}(10^{-2})$), increases the spin-dependent cross-section of DM-nucleon interactions. This enhances the detection potential of the DM in DD experiments. The significant mass splitting between $\nu_{4}$ and $\nu_{5}$ also suppresses the co-annihilation channel, leading to DM overabundance. For fixed values of $M_{Z'}$, $M_\text{DM}$, and $M_{\nu_5}$, the DD cross-section can be enhanced by increasing $g_x$, though this would alter the dark matter relic density.  In Fig.~\ref{Fig-DM-vPhM-7a}, we have shown the spin-dependent DM-nucleon cross-section via scatter plot. The points shown in the figure satisfy the DMRD, $\theta'$ constraint,
DD and indirect detection cross-section bound. From the DD experiments, the bounds on the DM-nucleon cross-section comes from {\tt XENON1T} experiment~\cite{XENON:2018voc,XENON:2019rxp}. This upper limit on the cross-section has been shown in Fig.~\ref{Fig-DM-vPhM-7a}, and the parameter points are clearly well below the upper limit. The plot also shows the projected sensitivity for several upcoming DM direct detection experiments, such as {\tt DARWIN} \cite{DARWIN:2016hyl}, {\tt LZ} \cite{LZ:2018qzl}, and {\tt XENONnT}\cite{XENON:2020kmp}.

 Indirect detection of DM involves observing visible particles resulting from DM annihilation. The {\tt Fermi-LAT} and {\tt MAGIC}~\cite{MAGIC:2016xys, Daylan:2014rsa} collaborations consider $\text{DM}\, \text{DM} \to b\,\bar{b}$ and  $\text{DM}\, \text{DM} \to \tau\, \bar{\tau}$ channels to constrain the thermal average annihilation cross section times relative velocity between two DM particles denoted by $\langle \sigma v \rangle$. The DM annihilation scales with the square of local DM density, $\xi=\frac{\Omega_\text{DM}}{0.12}$~\cite{Barman:2020vzm}. Thus, we have shown $\xi^2 \langle \sigma v \rangle$ as a function of DM mass and its RD, along with the {\tt Fermi-LAT} bound at  $\langle \sigma v \rangle = 2 \times 10^{-26}\, {\rm cm}^3/{\rm s}$  in Fig.~\ref{Fig-DM-vPhM-7b}. In the scenario with sizeable GKM, the co-annihilation of $\nu_4$ and $\nu_5$ to the SM particles via $Z'$ resonance is the most dominant channel in obtaining DMRD,  resulting in a small number of visible particles from DM-DM annihilation. In the singlet scalar resonance region, DM primarily annihilates to $Z' Z'$. To increase DM annihilation to $\tau\bar{\tau}$ and $b\bar{b}$, we need to enhance the singlet scalar's mixing with the SM Higgs.
%%%%%%%%%%%%%%%%%%%%%%%%%%%%%Section IV  FIMP SCENARIO  %%%%%%%%%%%%%%%%%%%%%%%%%%%%%%%%%%%%
\section{FIMP scenario} \label{sec:fimp}
Searches for thermal DM have
been going on for more than a decade, and so far, we do not have any
observations confirming the nature of DM. Consequently, alternative scenarios that can explain the observed DMRD are being explored. A scenario thus arises, when a DM candidate has a non-thermal nature that leads to non-thermal production
of the observed DMRD. These DM candidates have very tiny coupling strengths with visible sector and can easily escape the direct
and indirect detection bounds. Particles that show this kind of properties are the so-called FIMPs. 

In the non-thermal production mechanism, it is assumed that the particles, which
are out of thermal equilibrium, have a very negligible initial abundance
and have a feeble interaction with the thermal bath particle. The DM
candidates, in this case, are produced from the bath particles via
$2\to 2$ scattering processes and from the decay of any bath particle to
the dark sector.  Additionally, BSM particles which are not in equilibrium may decay into dark sector particles. As the DM candidate has negligible initial abundance, the reverse process remains insignificant.
Therefore, the DMRD is generated from zero to its current
value. This scenario is thus called the freeze-in mechanism. 

This is in
contrast to the WIMP scenario, where the DM remains in the thermal bath
until its freezes out. In the WIMP scenario, increased coupling to thermal bath particles keeps DM in equilibrium longer and increases annihilation to bath particles, thereby reducing the total yield in the relic density. On the contrary, in the FIMP scenario, as the DM candidate
is not in thermal equilibrium with the bath particles, increasing its
coupling with bath particles leads to more DM production.
Hence, it will increase the total DMRD in the
FIMP scenario. One can then easily infer that the coupling should always
be below a certain threshold value above which it can enter into the
thermal bath. For a particle to be out of equilibrium throughout the
evolution of the universe, the rate of conversion of bath particles to
the DM particle must remain negligible compared to the Hubble expansion
rate.

In a $B_1 B_2 \to Y \to X X'$ process, when the $s$-channel resonance through particle $Y$ occurs, the cross-section of this process can be written as \cite{Belanger:2018ccd} 
\begin{equation}\label{crossx}
	\sigma(s) = \frac{g_{_Y}}{g_{B_1} g_{B_2}}\frac{1}{C_{B_1B_2}}\frac{4\pi^2 m_{_Y}}{(p^{\rm CM}_{B_1,B_2})^2}\frac{\Gamma_{Y\to B_1B_2}\times\Gamma_{Y\to XX'}}{\Gamma_{\rm tot.}}\,\delta(s-m^2_{_Y}),
\end{equation} 
where $g_{_i}$ is the number of degrees of freedom of the $i^{\rm th}$ initial particle, $p^{\rm CM}_{1,2}$ are the momenta of initial particles at the 
centre of mass frame and $s = (p^{\rm CM}_{B_1}+p^{\rm CM}_{B_2})^2$. The $\Gamma$'s are the partial decay widths of $Y$ in the respective channels and $\Gamma_{\rm tot.}$ is its total decay width. The factor $C_{B_1B_2}$ is the combinatorial factor, equals to $\frac{1}{2}$ if two particles are identical otherwise it is equal to 1. 
Ignoring the back process, the collision term of the Boltzmann equation (see Eq.~(\ref{eqn:boltzmann})) at temperature $T$ can be written as \cite{Belanger:2018ccd} 
\begin{equation}\label{sch-reso}
R(B_1 B_2\to Y \to X X')=\frac{T g_{_Y}}{2\pi^2}m^2_{_Y}\frac{\Gamma_{Y\to B_1\,B_2}\times\Gamma_{Y\to X\,X'}}{\Gamma_{\rm tot.}}\tilde{K_1}(x_Y,x_1,x_2,0,\eta_1,\eta_2),
\end{equation}
where, for $i^{\rm th}$ particle, $x_i = m_i/T$ and $\eta_i = \pm e^{\mu_i/T}$ for bosons/fermions ($+/-$) with chemical potential $\mu_i$. The function $\tilde{K}_1$ is given by \cite{Belanger:2018ccd}
\begin{equation}
\tilde{K}_1 (x_1, x_2, x_3, \eta_1, \eta_2, \eta_3) = \frac{1}{16\pi^2 p^{\rm CM}_{1,2} T} \int \prod_{i=1}^3 \left(\frac{d^3 p_i}{E_i}\frac{1}{e^{E_i/T} - \eta_i}\right) e^{E_1/T} \delta^4(p_1-p_2-p_3).
\end{equation}
In the standard FIMP scenario, the evolution of the DM density begins after the reheating phase. During this phase the inflaton field decays, creating a thermal bath, from which the non thermal DM is generated. Throughout the analysis, we have kept the reheating temperature $T_R$ to be $10^{10}$~GeV. We work in the regime where the mediator that produces DM from its decay has a mass much smaller than the reheating temperature. The DMRD is very much unaffected by variation of reheating temperature unless it is close to the mediator mass~\cite{Blennow:2013jba,Abdallah:2019svm}.

As mentioned previously, in this work, we consider the lightest heavy right-handed neutrino~($\nu_{4}$) as a possible DM
candidate. In the WIMP scenario, we have already seen that $\nu_4$ can be
stable because of its small Yukawa couplings $(Y_{\nu_{11}} = 10^{-27}$
and $Y_{\nu_{12}} = 0 = Y_{\nu_{13}})$. In the FIMP scenario however, the couplings,
%\footnote{In the absence of the $Z$-$Z'$ mixing
%and the singlet scalar mixing with any other scalars, $\nu_4$ has only
%%%term $M_{N_{11}}$. }
$g_x$ and
$Y_{L_{11}}$ must be small enough to keep $\nu_4$ out of
thermal bath. Additionally the couplings should not be very small such that the observed DMRD cannot be achieved.  
%The value of  $Y_{L_{11}}$ can be larger if the singlet scalar stays out of thermal equilibrium. 

In the WIMP scenario, we have seen that there is a
small mass difference between $\nu_4$ and $\nu_5$. Being the lightest,
the $\nu_4$ is stable, and $\nu_5$ decays to $\nu_4$ and a pair of SM
fermions via the exchange of off-shell $Z$, $Z'$ or $h_s$.  In the FIMP scenario, due
to small values of $g_x$ and $Y_{L_{11}}$, $\nu_5$ can be
a long-lived particle.
If both $Z'$ and $h_s$ are out of thermal equilibrium, the lifetime of
$\nu_5$ can be larger than the age of the universe as the GKM and the
singlet scalar mixing angles with other scalars are very small. In
order to avoid this additional DM component, we have chosen the value of
$v_s\,Y_{L_{11}}$ (see Eq.~(\ref{nuH_mass})) such that it creates
a mass difference between $\nu_4$ and $\nu_5$ large enough to make the decay $\nu_5\to\nu_4 Z'/\nu_4 h_s$  kinematically possible.

Some points to note about the other particles, {\it viz.} the heavy neutrinos $(\nu_6,\nu_7,\nu_8,\nu_9)$ will be in the thermal bath because of their sizeable coupling with the second Higgs doublet and SM neutrinos.  Similarly the scalars $A,h_2$ and $H^{\pm}$ are in thermal equilibrium as they couple to the SM gauge bosons with electroweak strength. 
Since the DM ($\nu_4$) interacts only with $Z'$ and $h_s$ it can be produced from their exchange as well as decay. The production of the DM can happen when the $Z'$ and $h_s$ are either in thermal equilibrium or out of thermal equilibrium.
We shall consider both these scenarios for DM production in the following subsections.

%========================================================================
\subsection{The DM production via $Z'$}

In this section, we consider DM production through the decay or exchange
of $Z'$ only while keeping $h_s$ out of thermal equilibrium with negligible coupling to the DM. The $Z'$ interacts with the SM particles via GKM ($g'_x$) and its mixing
with the SM $Z$-boson ($\theta'$). As the DM is out of the thermal bath, its couplings including the $U(1)_X$ gauge coupling $g_x$ must be very small. As a result, the interaction of $Z'$ with heavy neutrinos becomes weak. So $g'_x$ and $\theta'$ will determine whether the $Z'$ will remain in the thermal bath. Thus, two distinct cases emerge where (a) the $Z'$ is out of thermal equilibrium, and, (b) the $Z'$ is in thermal equilibrium.

We find that the DM candidate ($\nu_4$) becomes non-thermal for $g_x <  10^{-9}$ and $Y_{L_{11}} <  10^{-9}$. This range of $g_x$ makes $Z'$ very light unless the $U(1)_X$ symmetry breaking scale, $v_s$ is chosen to be very large. We set $v_s \sim \mathcal{O}(10^{13})$ GeV to achieve a $Z'$ mass within the range of a few GeV to a few TeV. As the couplings and mixings involved here are very small, the $Z'$ mass is easily allowed by the constraints obtained from 
$Z'$ searches (Drell-Yan) at the LHC~\cite{ATLAS:2017fih,CMS:2019buh,ACCOMANDO:2013zz}. 
 We vary
$g'_x \in [10^{-14}, 10^{-11}]$ that keeps $Z'$ out of thermal equilibrium.
In addition, the scalar mixing matrix elements $Z^h_{13}$, $Z^h_{23}$ are
chosen in the range of $\mathcal{O}(10^{-15}-10^{-10})$ and $Y_{L_{22}} \simeq Y_{L_{33}}\simeq\mathcal{O}(10^{-11})$ to ensure that the singlet
scalar $h_s$ never attains thermal equilibrium. 
 In this scenario, we fix all the remaining heavy neutrino masses beyond $\nu_5$  to be $m_{\nu_{H}} \simeq 4~\text{TeV}$. The scalar sector values are chosen as $M_{h_{2}} = M_A=M_{H^\pm} = 1.5~\text{TeV},\, \tan\beta = 10^{-4}$ and $v_{s} = 10^{13}~\text{GeV}.$

\begin{figure}[t!]
\begin{center}
\subfloat[]{\includegraphics[width=8cm, height=6cm]{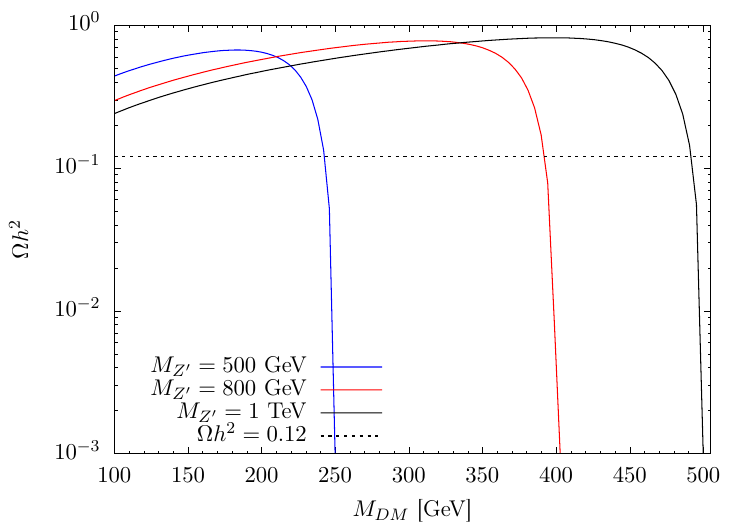}\label{FIMPDM-vPhM-1a}}
\subfloat[]{\includegraphics[width=8cm, height=6cm]{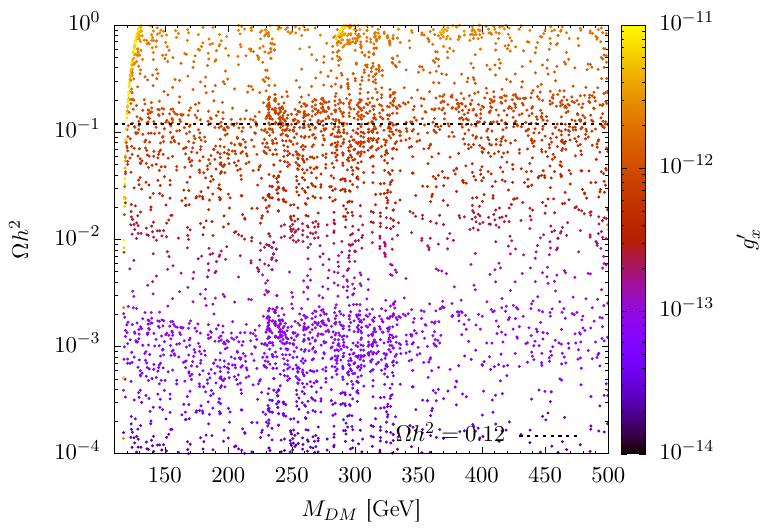}\label{FIMPDM-vPhM-1b}}\\
%\subfloat[]{\includegraphics[width=8cm, height=6cm]{"oh2_mdm_2_Lmzp_new"}}
%\includegraphics[width=8cm, height=6cm]{"mzp_oh_gpx"}
\caption{(a)~The DMRD $\Omega h^2$ versus $M_\text{DM}$ for $M_{Z'}= 500,800,1000$~GeV at $g'_x = 2\times 10^{-12}$. (b)~Scatter plot presenting the DMRD $\Omega h^2$ as a function of  $M_\text{DM}$ and $g'_{x}$.}
\label{FIMPDM-vPhM-1}
\end{center}
\end{figure}

In Fig.~\ref{FIMPDM-vPhM-1a}, we show the DMRD ($\Omega h^2$) as a function of the DM mass $M_\text{DM}$ for three different values of $M_{Z'}$ with a fixed value of $g'_x = 2\times 10^{-12}$. Here, we see that the observed DMRD can be achieved near $M_{Z'}\simeq 2 M_\text{DM}$. In the overabundant region the correct relic can be achieved by varying the GKM.  The figure clearly shows a sharp drop in DMRD when $M_\text{DM}$ exceeds $M_{Z'}/2$, indicating that for $M_\text{DM} > M_{Z'}/2$, the off-shell production of $Z'$ contributes negligibly to the DMRD. In the present scenario, since the GKM is non-zero, reversing the channels that contribute to the DMRD in the WIMP section under a non-zero GKM condition assists in achieving the correct relic density for the FIMP.  The dominant DM production channels are $f \bar{f} \to Z' \to \nu_4 \nu_4$,
$W^+ W^- \to Z' \to \nu_4 \nu_4$, and $h_1 Z \to Z' \to \nu_4 \nu_4$. The DM production due to the heavy neutrinos annihilation via an offshell $Z'$ is
negligible since $m_{\nu_{H}} \simeq 4~\text{TeV} \gg M_{Z'}$. To suppress the DM production from the singlet-dominated CP-even scalar in the $Z'$ driven scenario, we keep $h_s$ lighter than the heavy neutrinos and the corresponding scalar sector mixing angles very small. 

%========================================================================
%========================================================================

In the region where $2M_{\text{DM}} < M_{\text{mediator}} < T_{R}$, the dominant contribution to the DM production occurs when the centre-of-mass energy $\sqrt{s}$ is approximately equal to $M_{\text{mediator}}$ (see Eq.~(\ref{crossx})). In this region the final DMRD is relatively insensitive to  $T_{R}$~\cite{Blennow:2013jba}. In Fig.~\ref{FIMPDM-vPhM-1a}, we have considered three benchmark points (BPs) with three values of $M_{Z'}$. We find when there is enough phase space available for $Z'$ to decay into a pair of DMs, the BP with lightest $Z'$ yield in largest DMRD. Which can be explained as follows, referring to equation (\ref{sch-reso}) the magnitude of $\Gamma_{Z'\to {\rm SM~SM}} \times {\rm BR}_{Z'\to {\rm DM~DM}}$ is higher for larger masses of $Z'$ for given value for DM mass and the temperature-dependent function $\tilde{K}_{1}$ is more significant for lighter masses of $Z'$. Consequently, the DM production is maximised when $M_{Z'}$ is heaviest until the temperature of the universe matches the $Z'$ mass. However, as the temperature $T$ drops below the mediator mass, the production of DM from bath particles is considerably more suppressed for a heavy mediator compared to a lighter one. Thus, for $T< 500$~GeV, the BP with $M_{Z'} = 500$~GeV yields more DM compared to the other two BPs, resulting in the highest RD for a given~$M_{\text{DM}}$, if there much phase space available for $Z'$ decay to a DM pair. 

In Fig.~\ref{FIMPDM-vPhM-1b}, we show the variation of $\Omega h^2$ as a function of the DM mass and the GKM. In this scatter plot, we vary $M_\text{DM} \in [100, 500]$~GeV,
$M_{Z'} \in [100, 1000]$~GeV and $g'_x \in [10^{-14}, 10^{-11}]$. We find that the observed DMRD can be
obtained for $g'_x \simeq \mathcal{O} (10^{-12})$ . For the higher values of
the GKM ($>10^{-12} $), the coupling of $Z'$ to the SM particles becomes stronger,
which leads to an overabundance of the DMRD. Similarly, the smaller values
of the GKM ($<10^{-12} $) would give under-abundant DMRD. % If we decrease $v_s$ then for the same range of $M_Z'$ we need larger $g_X$ hence smaller GKM is needed to achieve the observed relic density of DM.

Now, we explore the case where  $Z'$ is in thermal equilibrium, achieved by making the GKM
$g'_x$ large while keeping $\theta'< 10^{-3}$. In
this region, we achieve the correct DMRD mainly through two
processes: (a) via  $Z'$ decay and (b) via $h_s$ resonance through $Z'Z'$
and $Z Z$ annihilation processes. We will discuss the $h_s$ resonance process in the next subsection.

%========================================================================
%%\subsection{$Z'$ in thermal bath}
\begin{figure}[h!]
\begin{center}		
{\includegraphics[width=8cm, height=6cm]{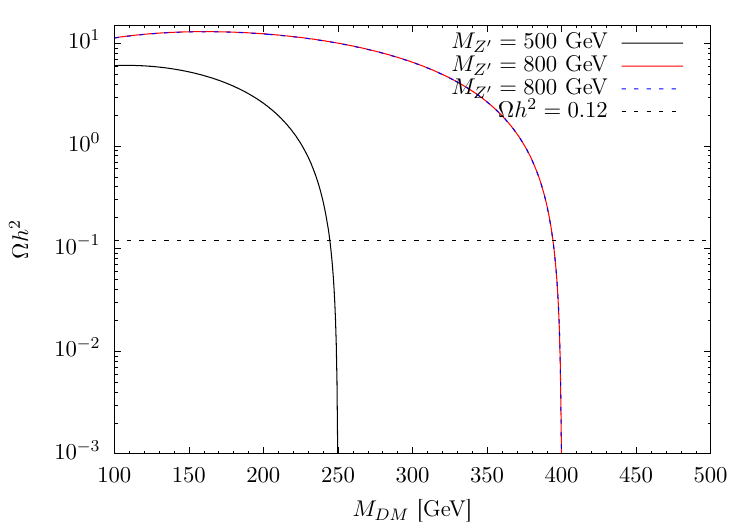}}
%\subfloat[]{\includegraphics[width=8cm, height=6cm]{"mdm_oh2_h3reso"}\label{FIMPDM-vPhM-4b}}
\caption{Variation of the DMRD as a function of the DM mass when $Z'$ is in thermal equilibrium.}	
\label{FIMPDM-vPhM-4a}
\end{center}
\end{figure}
%========================================================================
In Fig.~\ref{FIMPDM-vPhM-4a}, we show the DMRD for two different masses of $Z'$.  The observed DMRD can be obtained when a pair of DM is produced from decay of $Z'$. The black and red solid curves, respectively, are for $M_{Z'} = 500$~GeV and $M_{Z'} = 800$~GeV with $g'_x = 5 \times 10^{-4}$. The blue dashed curve is for $M_{Z'} = 800$~GeV with $g'_x = 5 \times 10^{-5}$. The red solid and blue dashed curves are overlapping although the red curve has one order of magnitude larger GKM as compared to the blue dashed curve. This demonstrates that varying the GKM does not significantly alter the DMRD as long as $Z'$ is in the thermal bath. This is because, the branching ratio~(BR) of $Z'$ to SM particles is close to one. 
%Changing GKM by one order of magnitude does not change the BR to the SM modes significantly, and the partial decay width of $Z'$ to a DM pair is also unaffected by this change. 
The production of non-thermal particles  from the decay of a thermal bath particle depends on its partial decay width to that mode~\cite{Belanger:2018ccd}. Therefore, the distribution of DMRD remains similar for these two BPs. Finally, the black curve with $M_{Z'} = 500$ GeV shows a smaller $\Omega h^2$ compared to the blue and red curves with $M_{Z'} = 800$ GeV. This is because, for a given DM mass, the partial decay width of a heavier $Z'$ to a DM pair is larger, as it is proportional to $g_x^3$.
 
%%%%%%%%%%%%%%%%%%%%%%%%%%%%%%%%%%%%%%%%%%%%%%%%%%%%%%%%%%%%%%%%%%%%%%%%%
\subsection{The DM production via $h_s$}
As the DM candidate couples to the singlet scalar $h_s$ via the Yukawa coupling
$Y_{L_{11}}$, it can potentially be produced from $h_s$ decay. In such scenarios, we have two
cases: (a) $h_s$ is out of thermal equilibrium, and (b) $h_s$ is in thermal equilibrium. Note that in this subsection the $Z'$ contribution to the DM production is negligible due to our assumption that $M_{Z'}<2 M_{\rm DM}$. The mixing angles, governed by the quartic couplings $\lambda_{1s}$
and $\lambda_{2s}$ are responsible for keeping
the singlet scalar $h_s$ in the thermal bath.
\iffalse
In the scenario, where the singlet scalar is in
the thermal bath, it mostly decays to a pair of SM particles, making its BR to DM pairs so small that the process goes out of equilibrium.In the second case, where the scalar is out of the thermal bath, its BR to DM is larger, and its coupling to bath particles is so small that its production from the thermal bath goes out of equilibrium. Hence, the correct DMRD can be achieved in widely separated regions. Hence, the correct DMRD can be achieved in widely separated regions.
\fi
In this subsection, we fix $m_{\nu_{H}} \simeq 4~\text{TeV},\, M_{h_{2}} = M_A=M_{H^\pm} = 1.5~\text{TeV},\,M_{Z'}=100~{\rm GeV},\, \tan\beta = 10^{-4}$ and $v_{s} = 10^{13}~\text{GeV}.$

%========================================================================
\begin{figure}[t!]
\begin{center}		
\subfloat[]{\includegraphics[width=8cm, height=6cm]{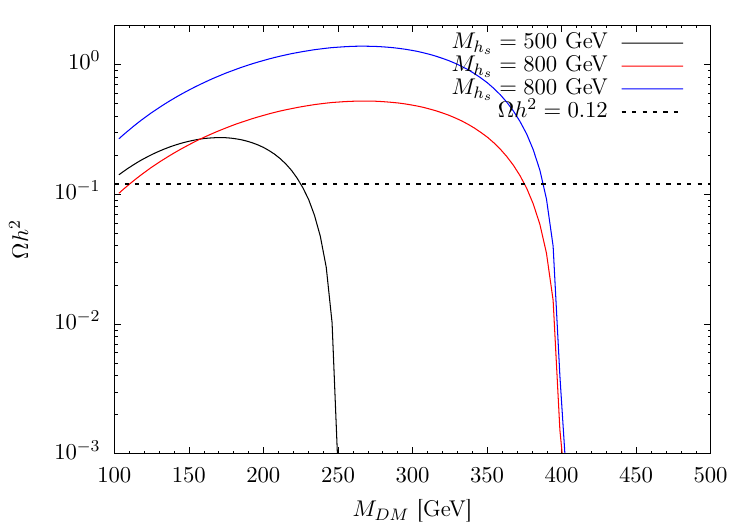}\label{FIMPDM-vPhM-2a}}
\subfloat[]{\includegraphics[width=8cm, height=6cm]{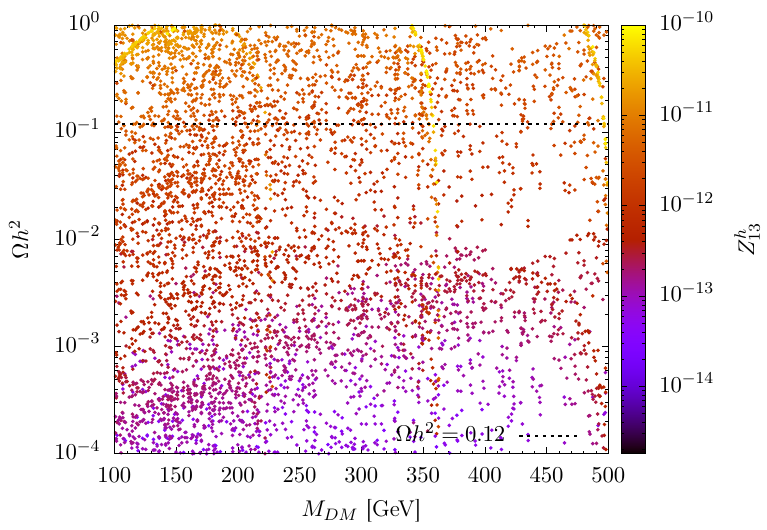}\label{FIMPDM-vPhM-2b}}
\caption{(a)~The DMRD $\Omega h^2$ versus $M_\text{DM}$ for $M_{h_s}= 500,800$~GeV. (b)~Scatter plot presenting the DMRD $\Omega h^2$ as a function of  $M_\text{DM}$ and $Z_{13}^h$. In both panels, $M_{Z'} = 100$~GeV.}
\label{FIMPDM-vPhM-2}
\end{center}
\end{figure}
%========================================================================
We first consider the non-thermal $h_s$ case. In this case, $h_s$ can be produced from the thermal bath via two types of processes: (i) from
the SM particles due to the mixing of singlet scalar with other CP-even scalars,
and (ii) from the non-SM particles such as the heavy neutrinos $\nu_H$ and the thermal $Z'$. In the first scenario, we set
$\lambda_{1s} \in [10^{-24},10^{-19}] , \lambda_{2s}= 0$ and $Y_{L_{11}} = 8.5\times10^{-11}$, which generate non-zero small mixing
between $h_s$ and the other scalars.\footnote{In this study, $\lambda_{2s}$ is fixed at zero, as by considering a non-zero $\lambda_{2s}$ this does not add any new processes for the $h_s$ production form the thermal bath. Moreover, the coupling strength of $h_2$ to the SM particles is $\tan\beta$ suppressed.} The mixing angle of the singlet
scalar with the SM-like Higgs $h_1$ is very small ($Z^h_{13} =\mathcal{O}(10^{-15}-10^{-10})$),
while the mixing angle with $h_2$ is even more suppressed than  $Z^h_{13}$ due to
$\lambda_{2s} = 0$. The singlet scalar  $h_s$ is kept lighter than the heavy neutrinos, preventing its single on-shell production from them.  The dominant channels for the DM
production are $W^+ W^-/h_{1}h_{1}/ZZ/t \bar{t} \to h_s \to {\rm DM\ DM}$. In Fig.~\ref{FIMPDM-vPhM-2a}, we show the variation of $\Omega h^2$ as a function of the DM mass. In this
figure, the black curve represents $M_{h_s} = 500$~GeV, and the blue
and red curves represent $M_{h_s} = 800$~GeV. For the black and red
curves $Z^h_{13} = 5.0\times10^{-12}$ while for the
blue curve $Z^h_{13} \simeq 10^{-11}$. The distribution of relic density shows that the DM production is significant when $h_s$ is
produced on-shell and then decays to a DM pair. When $M_\text{DM} > M_{h_s}/2$ the DM pair production is now away from the $h_s$ resonance and therefore the contribution from the off-shell $h_s$ becomes negligible. Initially, the DMRD grows with increasing DM mass because the energy density is a monotonically increasing function of the DM mass. However, due to phase space suppression (as the DM mass approaches $M_{h_s}/2$) the production rate drops, resulting in the decrease of DMRD. As in the case of $Z'$, here too we observe that the DMRD can be satisfied when the $h_s$ is produced on-shell from the thermal bath and decays to a pair of DM. In the overabundant region, the observed DMRD can be satisfied by varying $Z^h_{13}$. Although both the blue and the red
curves correspond to the same value of $M_{h_s}$, they differ in the DMRD because of the difference in mixing angle of $h_s$ with the SM-like Higgs. The blue curve, having a larger mixing angle $Z^h_{13}$, results in more $h_s$ production from the thermal bath, thereby yielding a larger DMRD.  In Fig.~\ref{FIMPDM-vPhM-2b}, a
scatter plot shows the DMRD as a function of  $M_\text{DM}$ and $Z_{13}^h$. In this scatter plot, we scan over the relevant parameters in the following range
\begin{equation}
M_\text{DM} \in [100,500]~{\rm GeV}, \qquad M_{h_s} \in [100,3500]~{\rm GeV}, \qquad {\rm and}\quad Z^h_{13} \in [10^{-14},10^{-10}]. \nonumber
\end{equation}
We find that the correct DMRD can be achieved for the whole range of $M_\text{DM}$ in this scan with $Z^h_{13} \in[10^{-12},10^{-11}]$, as shown in the figure palette.

\iffalse\cred{The Majorana fermion,  $\nu_{4}$, primarily originates from the right-handed neutral fermion. However, for a constant value of $Y_{L_{1}}$, an increase in $M_{N}$ results in a greater mixing angle with the newly added left-handed neutral fermion. Consequently, as the value of $M_{\nu_4}$ rises, the coupling of $\nu_{4} \nu_{4} h_{s}$ increases, as $Y_{R}=0$.

The decay width of $Z'$ to dark matter (DM) pairs is proportional to $\sqrt{1-\frac{4 M_{DM}^2}{M_{Z'}^2}}\left(1+\frac{2 M_{DM}^2}{M_{Z'}^2}\right)$, while the decay of $h_{s}$ to DM pairs is proportional to $\left(1-\frac{4 M_{DM}^2}{M_{h_{s}}^2}\right)^{\frac{3}{2}}$. Consequently, when $M_{Z'}$ and $M_{h_s}$ are equal, the Dark Matter Relic Density (DMRD) maximizes at higher values of $M_{DM}$ when produced from $M_{Z'}$ compared to $M_{h_s}$.}\fi
%========================================================================
\begin{figure}[t!]
\begin{center}
\includegraphics[width=8cm, height=6cm]{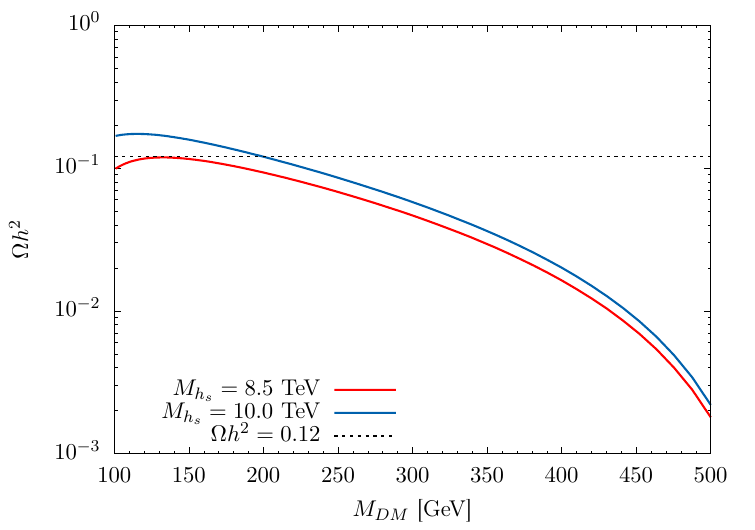}~~\includegraphics[width=8cm, height=6cm]{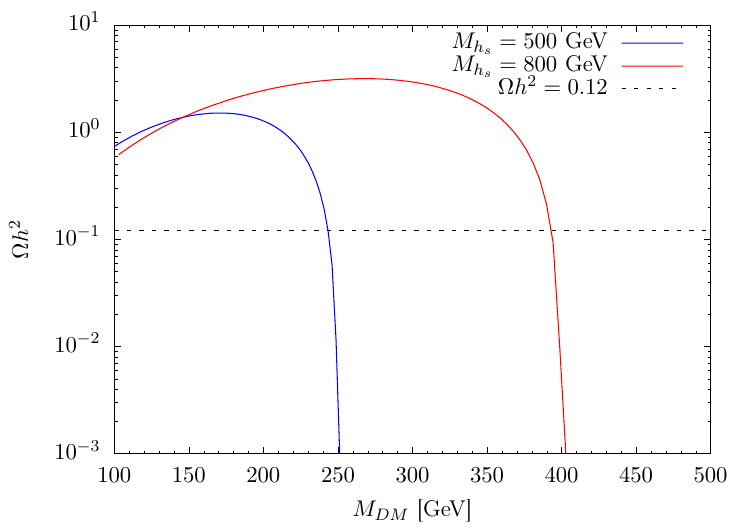}
\caption{Variation of the DMRD versus $M_{\rm DM}$ for a non-thermal $h_{s}$ production from heavy neutrinos annihilation (left panel) and $Z'Z'$ annihilation (right panel). Here, $M_{Z'}=100$~GeV and the masses of four heavy neutrinos are
$3.4$~TeV, $3.9$~TeV, $4.1$~TeV, and $4.7$~TeV. }	
\label{FIMPDM-vPhM-6}
\end{center}
\end{figure}
%========================================================================
In the scenario with vanishing scalar mixing, where the singlet scalar cannot be produced from SM particles, its production relies on pair of heavy neutrino annihilation and $Z'$ annihilation.
For the processes $\nu_{i} \nu_{j} \to h_s \to \nu_{4} \nu_{4}$\footnote{The contribution of $\nu_{\ell}\, \nu_{\ell} \to h_s \to \nu_{4} \nu_{4}$ to the DMRD is very small because the heavy neutrinos $\nu_H$ have small mixings with the active ones $(\nu_\ell)$.} 
where $i,j=6-9$, the $h_s$ mass is chosen such that it would be heavier than 
the sum of the masses of any two heavy neutrinos, so the scalar can be produced from their annihilation. In Fig.~\ref{FIMPDM-vPhM-6} (left panel), we show the variation of
$\Omega h^2$ as a function of $M_\text{DM}$ for two different values of
$M_{h_s}= 8.5,\,10$~TeV. In this figure, we vary
$Y_{L_{11}} \in [8\times 10^{-12},\,2\times 10^{-10}]$ to
keep $M_\text{DM} \in [100,\,500]$~GeV and fix
 $Y_{L_{22}} = 1.6\times 10^{-11}$,
$Y_{L_{33}} = 9.4\times 10^{-11}$. For $M_{h_{s}} = 8.5$~TeV, the production of $h_s$ from the heaviest heavy neutrinos is kinematically forbidden, resulting in a slightly smaller DMRD compared to $M_{h_{s}} = 10$~TeV. 

For the process $Z' Z' \to h_s \to \nu_{4} \nu_{4}$,  where $Z'$ is in thermal equilibrium but cannot produce a DM pair (i.e., $M_{Z'}<2M_{\rm DM}$), there is a possibility of achieving the correct DMRD via
singlet scalar resonance if it is feasible kinematically. 
In Fig.~\ref{FIMPDM-vPhM-6} (right panel), we plot $\Omega h^2$ as a function
of $M_\text{DM}$ for two different values of $M_{h_s}$, fixing $g'_x = 10^{-4}$ and
$Y_{L_{11}} = 8.5 \times 10^{-11}$. 
Here, again we see that the correct DMRD can be satisfied when
$M_\text{DM} < M_{h_s}/2$. 
\begin{figure}[h!]
\begin{center}		
\includegraphics[width=8cm, height=6cm]{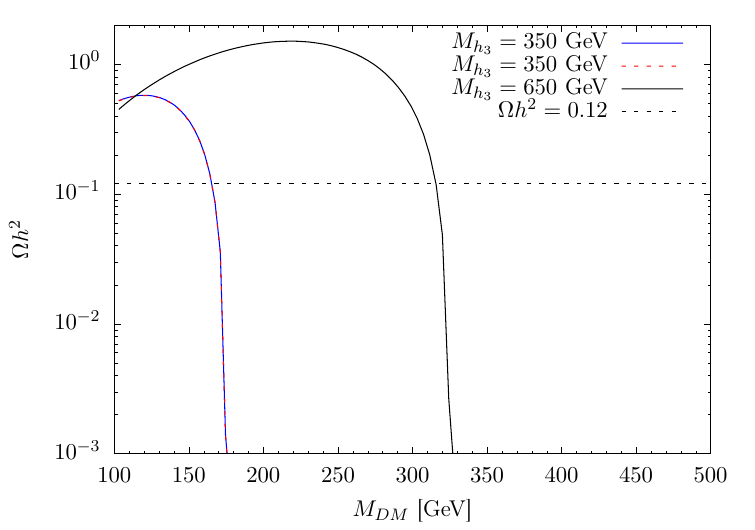}
\caption{Variation of the DMRD as a function of the DM mass when $h_s$ is in thermal equilibrium.}	
\label{FIMPDM-vPhM-3}
\end{center}
\end{figure}
%========================================================================

Now, we consider the second case where $h_s$ is in thermal equilibrium. The singlet scalar $h_s$ will be in thermal equilibrium if it has large enough mixing with the other CP 
even scalar components. For example, a mixing of $Z^h_{13}\simeq\mathcal{O}(10^{-4} )$ is large enough for the $h_s$ to remain in thermal equilibrium with the bath particles. Note that 
very large values of $Z^h_{13} \gsim 10^{-2}$ is likely to get constrained from the SM 
Higgs measurements at the LHC~\cite{ATLAS:2016neq}. In
Fig.~\ref{FIMPDM-vPhM-3}, we show the DMRD as a function
of $M_{\rm DM}$. The blue and black curves correspond to
$M_{h_s} = 350$~GeV and $M_{h_s} = 650$~GeV, respectively, with $Z^h_{13}\simeq 9\times 10^{-4}$. In this figure,
we find that the observed DMRD is achievable in the region
$M_\text{DM}<M_{h_s}/2$. The rapid fall in the DMRD for
$M_\text{DM} > M_{h_s}/2$ shows its production is via off-shell $h_s$. The
red dashed curve is also with $M_{h_s} = 350$~GeV but with a different
value of $Z^h_{13}\simeq 9\times 10^{-5}$. The blue
curve has one order larger coupling strength with the SM particles as
compared to the red curve. However, both of them give the same amount of
the DMRD. The reason is similar to the case discussed in the previous section for the thermal $Z'$-boson scenario. 

%%%%%%%%%%%%%%%%%%%%%%%%%%%%%%%%%%%%%%%%%%%%%%%%%%%%%%%%%%%%%%%%%%%%%%%%%
\section{Summary and Conclusion}\label{sec:summ}
In this article, we consider a $U(1)$ extension of the SM gauge symmetry, where the SM particles are singlet under this newly added abelian gauge symmetry. This model includes an additional doublet scalar, three vector-like SM gauge singlet fermions, and a singlet scalar, all charged under the new $U(1)$ gauge group. The new singlet and doublet
scalars are responsible for the spontaneous breaking of the new gauge
symmetry. The second Higgs doublet and the SM singlet fermions couple to SM leptons with a Yukawa coupling. Due to this interaction term, the SM neutrinos become massive at 
tree level when the second Higgs doublet acquires a vev.  In this work we assume that 
the lightest of the heavy neutrinos acts as the DM candidate. To make this happen, we 
require that the coupling of the lightest of the heavy neutrinos ($Y_\nu$) must be 
smaller than $\mathcal{O}(10^{-27})$ so that its lifetime becomes larger than the age 
of the universe. We have studied the viability of both the WIMP and FIMP mechanisms in this model to achieve the observed DMRD.

When the DM is thermal, its RD can be achieved via $Z'$ mediated (co)annihilation channels if
GKM and $Z$-$Z'$ mixing angle are non-negligible. If these parameters
are vanishingly small, the correct DMRD can be obtained through $s$-channel singlet scalar
 resonance. The scenario where the singlet scalar and $Z'$ are lighter than the DM, the observed DMRD is achieved when the DM annihilates to $h_s$ and $Z'$ in the final state. 
 The non-thermal DM can be obtained by making all its interactions with particles in the thermal bath feeble. We consider both thermal and non-thermal nature of the new gauge boson $Z'$ and the singlet scalar $h_{s}$ to show how these affect the DM production, when they mediate the interaction between DM and visible sector. 

The study in this work sheds light on the distinctive feature of having DM candidates within the framework of the neutrinophilic $U(1)$ model and thereby contributes to the ongoing discourse of DM candidates. The study aims to broaden the area of phenomenological aspects 
of the neutrinophilic model, which has already been studied in a wide range of aspects including collider signals, neutrino mass and mixings, and its role in lepton flavour 
violating signals. Future directions include an exploration of the phenomenological study of light DM between mass range $\mathcal{O}(\text{keV}-\text{GeV})$. Notably, the future study would also focus on how thermal correction to masses and couplings affect the feasible DM parameter space as well as the possible collider studies of the DM in this model.

\begin{acknowledgements}
AKB and SKR would like to acknowledge the support from the Department of Atomic Energy (DAE), India, for the Regional Centre for Accelerator-based Particle Physics (RECAPP), Harish Chandra Research Institute. TS thanks ICTS, Bengaluru and IISc, Bengaluru, for their hospitality when part of this work was being carried out. 

\end{acknowledgements}
\appendix
% \section{Appendix}
%\begin{appendix}
\section{Feynman rules }\label{app:A}

Below, we outline the relevant couplings of the model. We define $s_W \equiv \sin\theta_W$ and $c_W \equiv \cos\theta_W$, where $\theta_W$ is the Weinberg angle. Similarly, define $s_{\theta'} \equiv \sin\theta'$ and $c_{\theta'} \equiv \cos\theta'$, where $\theta'$ is the $Z$-$Z'$ mixing angle. Additionally, $T_3$ denotes the isospin, and $Q_f$ represents the electric charge of the fermions. The projection operators are given by $P_{L/R} = \frac{1 \mp \gamma_5}{2}$.

\begin{wrapfigure}[6]{l}{0.15\textwidth}
	\includegraphics[width=0.15\textwidth]{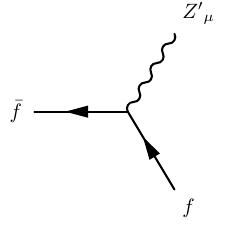}
\end{wrapfigure}
 %\vspace{5mm}
{\small\begin{eqnarray*} 
	i\left(\frac{e \, s_{\theta'}}{s_{W} c_{W}} \left(T^3 - Q_f s^2_{W}\right) + g_x' c_{\theta'}\left(T^3 - Q_f\right)\right)\!\gamma^\mu P_L - i\left(\frac{e \, s_{\theta'}}{s_{W} c_{W}} Q_f s^2_{W} + g_x' c_{\theta'}Q_f\right)\!\gamma^\mu P_R 
	\end{eqnarray*}}

\begin{wrapfigure}[5]{l}{0.16\textwidth}
	\includegraphics[width=0.15\textwidth]{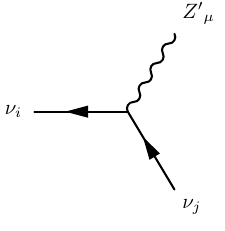}
\end{wrapfigure}~\\[-32pt]
{\small\begin{eqnarray*} 
	\frac{i}{2}\left(\left(\frac{e \, s_{\theta'}}{2 \, s_{W} c_{W}} + \frac{g_x'}{2} \, c_{\theta'}\right)\sum_{k=1}^{3} {\cal N}_{ik}{\cal N}^*_{jk} -  g_x \, c_{\theta'} \left(-\sum_{k=7}^{9} {\cal N}_{ik}{\cal N}^*_{jk}+\sum_{k=4}^{6} {\cal N}_{ik}{\cal N}^*_{jk}\right)\right)\!\gamma^\mu P_L \\
	-  \frac{i}{2}\left(\left(\frac{e \, s_{\theta'}}{2 \, s_{W} c_{W}} + \frac{g_x'}{2} \, c_{\theta'}\right)\sum_{k=1}^{3} {\cal N}_{ik}^*{\cal N}_{jk} -  g_x \, c_{\theta'} \left(-\sum_{k=7}^{9} {\cal N}^*_{ik}{\cal N}_{jk}+\sum_{k=4}^{6} {\cal N}^*_{ik}{\cal N}_{jk}\right)\right)\!\gamma^\mu P_R,
\end{eqnarray*}}

\vspace{8pt}
\noindent
where ${\cal N}$ is the neutrino mixing matrix. We note that $\nu_i$ for $i=1,2,3$ are identified as the light neutrinos and rest are heavy neutrinos. These neutrinos are Majorana fermions written in 4-component notation.
\begin{wrapfigure}[4]{l}{0.16\textwidth}
	\includegraphics[width=0.18\textwidth]{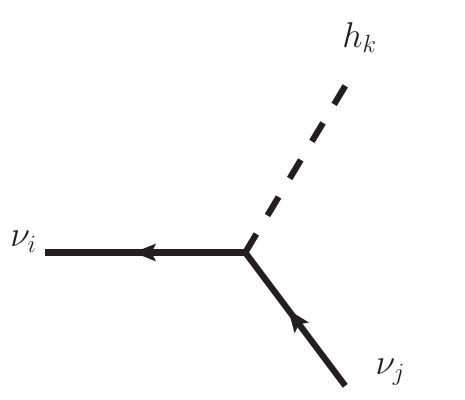}
\end{wrapfigure}~\\
{\small\begin{eqnarray*}
-\frac{i}{\sqrt{2}} \left(\sum_{a=1}^{3} \sum_{b=1}^{3} {\cal N}^*_{i6+a}{\cal N}^*_{j6+b} {Y_{L}}_{ab}\right) Z_{h_{k3}}\! P_L -\frac{i}{\sqrt{2}}\left(\sum_{a=1}^{3} \sum_{b=1}^{3} {\cal N}_{i6+a}{\cal N}_{j6+b} {Y_{L}}^*_{ab}\right)Z_{h_{k3}}\! P_R    
\end{eqnarray*}}

\vspace{12pt}
\begin{wrapfigure}[4]{l}{0.18\textwidth}
	\includegraphics[width=0.18\textwidth]{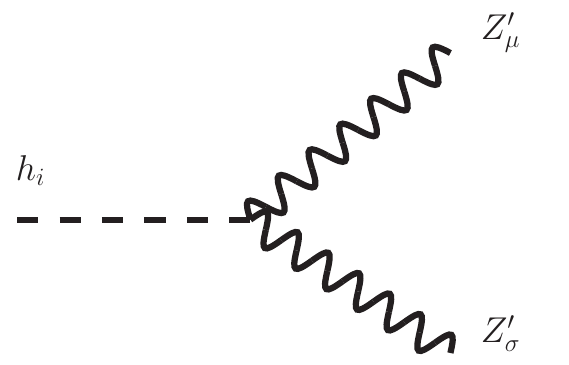}
\end{wrapfigure}~\\[-24pt]
\begin{flalign*}
    &\frac{i}{2}\left(v_1\left(\left(g_1 s_{W} + g_2 c_{W}\right)s_{\theta'}+g_x'c_{\theta'}\right)^2 Z_{h_{i1}} + v_2\left(\left(2g_x+g_x'\right)c_{\theta'} \right.\right. &&\\ & \left.\left. +\left(g_1s_{W} + g_2 c_{W}\right)s_{\theta'}\right)^2 Z_{h_{i2}}+16 v_s \left(g_x c_{\theta'}\right)^2 Z_{h_{i3}}\right) g_{\sigma \mu} &&
\end{flalign*}

\begin{wrapfigure}[4]{l}{0.18\textwidth}
	\includegraphics[width=0.18\textwidth]{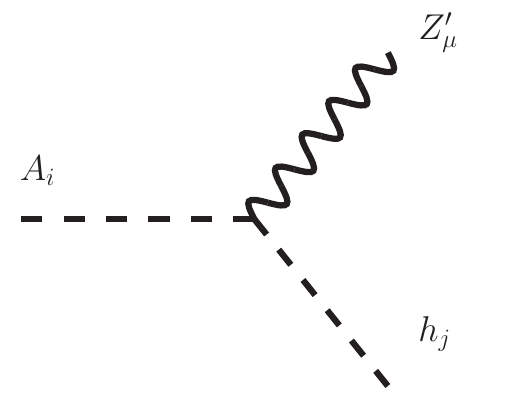}
\end{wrapfigure}~\\[-20pt]
\begin{flalign*}
    &-\frac{1}{2} \left(\left(\left(g_1 s_{W} + g_2 c_{W}\right)s_{\theta'}+g_x'c_{\theta'}\right)Z_{A_{i1}}Z_{h_{j1}}+ \left(\left(2g_x+g_x'\right)c_{\theta'}\right.\right. &&\\ & \left.\left. +\left(g_1s_{W} + g_2 c_{W}\right)s_{\theta'}\right)Z_{A_{i2}}Z_{h_{j2}}-4g_x c_{\theta'}Z_{A_{i3}}Z_{h_{j3}}\right) \left(p^{A_{i}}_{\mu}-p^{h_{j}}_{\mu}\right)&&
\end{flalign*}

\begin{wrapfigure}[4]{l}{0.2\textwidth}
	\includegraphics[width=0.18\textwidth]{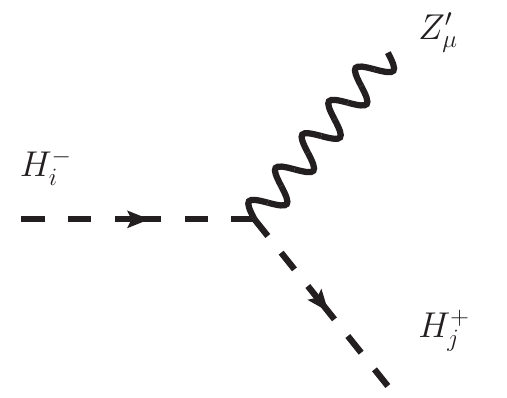}
\end{wrapfigure}~\\[-20pt]
\begin{flalign*}
    & \frac{i}{2} \left(\left(\left(g_1 s_{W} - g_2 c_{W}\right)s_{\theta'}+g_x'c_{\theta'}\right)Z_{C_{i1}}Z_{C_{j1}}+ \left(\left(2g_x+g_x'\right)c_{\theta'} \right.\right. && \\ & \left.\left. +\left(g_1s_{W} - g_2 c_{W}\right)s_{\theta'}\right)Z_{C_{i2}}Z_{C_{j2}}\right) \left(p^{H^-_{i}}_{\mu}-p^{H^+_{j}}_{\mu}\right)&&
\end{flalign*}
\\

Here, $Z_{h}$, $Z_{A}$ and $Z_{C}$ are mixing matrices for CP-even, CP-odd, and charged scalars, respectively.
\iffalse
\begin{table}[!h]
\begin{center}
\begin{tabular}{cc}
\multirow{1}{*}{
\includegraphics[width=0.16\textwidth,height=0.15\textwidth]{wwzp.pdf} }
&
\multirow{4}{*}{ $i g_2 c_{W}  s_{\theta'}  \Big(g_{\rho \mu} \Big(- p^{{Z'}_{{\mu}}}_{\sigma}  + p^{W^+_{{\rho}}}_{\sigma}\Big) + g_{\rho \sigma} \Big(- p^{W^+_{{\rho}}}_{\mu} + p^{W^-_{{\sigma}}}_{\mu}\Big)   + g_{\sigma \mu} \Big(- p^{W^-_{{\sigma}}}_{\rho}  + p^{{Z'}_{{\mu}}}_{\rho}\Big)\Big)  $ } 
\end{tabular}
\end{center}
\end{table}
\fi
\begin{wrapfigure}[4]{l}{0.18\textwidth}
	\includegraphics[width=0.18\textwidth]{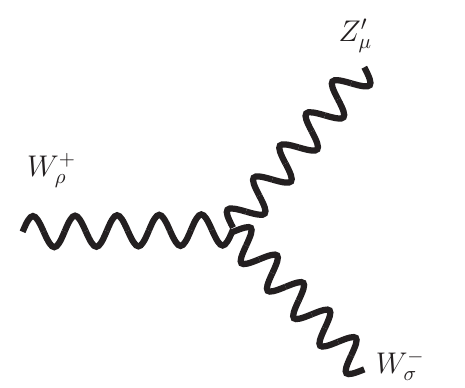}
\end{wrapfigure}~\\[-15pt]
\begin{flalign*}
    i g_2 c_{W}  s_{\theta'}  \Big(g_{\rho \mu} \Big(- p^{{Z'}_{{\mu}}}_{\sigma}  + p^{W^+_{{\rho}}}_{\sigma}\Big) + g_{\rho \sigma} \Big(- p^{W^+_{{\rho}}}_{\mu} + p^{W^-_{{\sigma}}}_{\mu}\Big)   + g_{\sigma \mu} \Big(- p^{W^-_{{\sigma}}}_{\rho}  + p^{{Z'}_{{\mu}}}_{\rho}\Big)\Big) 
\end{flalign*}
\\

\begin{wrapfigure}[4]{L}{0.15\textwidth}
	\includegraphics[width=0.2\textwidth]{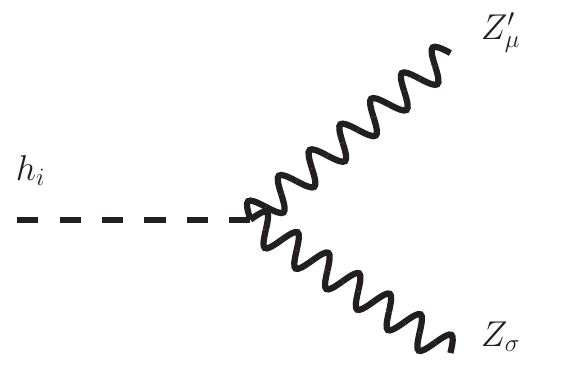}
\end{wrapfigure}~\\[-70pt]
\begin{flalign*}
  &\frac{i}{4} \Big(v_1 \Big(-2 {g_x'} g_2 {c_{W}}  c_{2 {\theta'}}   -2 g_1 {g_x'} c_{{\theta'}}^{2} {s_{W}}  +2 g_1 {g_x'} {s_{W}}  s_{\theta'}^{2} \nonumber \\ 
 &+{g_x'}^{2} s_{2 {\theta'}}   - g_{2}^{2} c_{\theta}^{2} s_{2 {\theta'}}   - g_{1}^{2} {s_{W}}^{2} s_{2 {\theta'}}  - g_1 g_2 s_{2 W}   s_{2 {\theta'}}   \Big)Z_{h_{i 1}} \nonumber \\ 
 &- v_2 \Big(g_2 {c_{W}}  \Big(2 \Big(2 {g_{x}}  + {g_x'}\Big)c_{2 {\theta'}}    - g_1 c_{2 {\theta'}+\theta_W}   + g_1 c_{2 {\theta'}-\theta_W}  \Big)\nonumber \\ 
 &+2 g_1 \Big(2 {g_{x}}  + {g_x'}\Big)  c_{{\theta'}}^{2} {s_{W}} + 2 {s_{W}}  s_{{\theta'}}  \Big( g_1\Big(- \Big(2 {g_{x}}  + {g_x'}\Big)s_{{\theta'}}   +    g_1c_{{\theta'}}  {s_{W}}  \Big)   \Big)\nonumber \\ 
 &- \Big(2 {g_{x}}  + {g_x'}\Big)^{2} s_{2 {\theta'}}   +g_{2}^{2} c_{{\theta'}}^{2} s_{2 {\theta'}}   \Big)Z_{h_{i 2}} + 16 v_S \Big(- {g_{x}}^{2} {s_{W}}^{2}    \Big)Z_{h_{i 3}} \Big)\Big(g_{\sigma \mu}\Big)
\end{flalign*}
\\
%\bibliographystyle{apsrev} 
%\bibliography{nuphilZ}

\begin{thebibliography}{70}%
	\makeatletter
	\providecommand \@ifxundefined [1]{%
		\@ifx{#1\undefined}
	}%
	\providecommand \@ifnum [1]{%
		\ifnum #1\expandafter \@firstoftwo
		\else \expandafter \@secondoftwo
		\fi
	}%
	\providecommand \@ifx [1]{%
		\ifx #1\expandafter \@firstoftwo
		\else \expandafter \@secondoftwo
		\fi
	}%
	\providecommand \natexlab [1]{#1}%
	\providecommand \enquote  [1]{``#1''}%
	\providecommand \bibnamefont  [1]{#1}%
	\providecommand \bibfnamefont [1]{#1}%
	\providecommand \citenamefont [1]{#1}%
	\providecommand \href@noop [0]{\@secondoftwo}%
	\providecommand \href [0]{\begingroup \@sanitize@url \@href}%
	\providecommand \@href[1]{\@@startlink{#1}\@@href}%
	\providecommand \@@href[1]{\endgroup#1\@@endlink}%
	\providecommand \@sanitize@url [0]{\catcode `\\12\catcode `\$12\catcode
		`\&12\catcode `\#12\catcode `\^12\catcode `\_12\catcode `\%12\relax}%
	\providecommand \@@startlink[1]{}%
	\providecommand \@@endlink[0]{}%
	\providecommand \url  [0]{\begingroup\@sanitize@url \@url }%
	\providecommand \@url [1]{\endgroup\@href {#1}{\urlprefix }}%
	\providecommand \urlprefix  [0]{URL }%
	\providecommand \Eprint [0]{\href }%
	\providecommand \doibase [0]{http://dx.doi.org/}%
	\providecommand \selectlanguage [0]{\@gobble}%
	\providecommand \bibinfo  [0]{\@secondoftwo}%
	\providecommand \bibfield  [0]{\@secondoftwo}%
	\providecommand \translation [1]{[#1]}%
	\providecommand \BibitemOpen [0]{}%
	\providecommand \bibitemStop [0]{}%
	\providecommand \bibitemNoStop [0]{.\EOS\space}%
	\providecommand \EOS [0]{\spacefactor3000\relax}%
	\providecommand \BibitemShut  [1]{\csname bibitem#1\endcsname}%
	\let\auto@bib@innerbib\@empty
	%</preamble>
	\bibitem [{\citenamefont {Zwicky}(1933)}]{Zwicky:1933gu}%
	\BibitemOpen
	\bibfield  {author} {\bibinfo {author} {\bibfnamefont {F.}~\bibnamefont
			{Zwicky}},\ }\href {\doibase 10.1007/s10714-008-0707-4} {\bibfield  {journal}
		{\bibinfo  {journal} {Helv. Phys. Acta}\ }\textbf {\bibinfo {volume} {6}},\
		\bibinfo {pages} {110} (\bibinfo {year} {1933})}\BibitemShut {NoStop}%
	\bibitem [{\citenamefont {Babcock}(1939)}]{1939LicOB..19...41B}%
	\BibitemOpen
	\bibfield  {author} {\bibinfo {author} {\bibfnamefont {H.~W.}\ \bibnamefont
			{Babcock}},\ }\href {\doibase 10.5479/ADS/bib/1939LicOB.19.41B} {\bibfield
		{journal} {\bibinfo  {journal} {Lick Observatory Bulletin}\ }\textbf
		{\bibinfo {volume} {498}},\ \bibinfo {pages} {41} (\bibinfo {year}
		{1939})}\BibitemShut {NoStop}%
	\bibitem [{\citenamefont {Rubin}\ \emph {et~al.}(1980)\citenamefont {Rubin},
		\citenamefont {Ford},\ and\ \citenamefont {Thonnard}}]{1980ApJ...238..471R}%
	\BibitemOpen
	\bibfield  {author} {\bibinfo {author} {\bibfnamefont {V.~C.}\ \bibnamefont
			{Rubin}}, \bibinfo {author} {\bibfnamefont {J.}~\bibnamefont {Ford},
			\bibfnamefont {W.~K.}}, \ and\ \bibinfo {author} {\bibfnamefont
			{N.}~\bibnamefont {Thonnard}},\ }\href {\doibase 10.1086/158003} {\bibfield
		{journal} {\bibinfo  {journal} {\apj}\ }\textbf {\bibinfo {volume} {238}},\
		\bibinfo {pages} {471} (\bibinfo {year} {1980})}\BibitemShut {NoStop}%
	\bibitem [{\citenamefont {Massey}\ \emph {et~al.}(2010)\citenamefont {Massey},
		\citenamefont {Kitching},\ and\ \citenamefont {Richard}}]{Massey:2010hh}%
	\BibitemOpen
	\bibfield  {author} {\bibinfo {author} {\bibfnamefont {R.}~\bibnamefont
			{Massey}}, \bibinfo {author} {\bibfnamefont {T.}~\bibnamefont {Kitching}}, \
		and\ \bibinfo {author} {\bibfnamefont {J.}~\bibnamefont {Richard}},\ }\href
	{\doibase 10.1088/0034-4885/73/8/086901} {\bibfield  {journal} {\bibinfo
			{journal} {Rept. Prog. Phys.}\ }\textbf {\bibinfo {volume} {73}},\ \bibinfo
		{pages} {086901} (\bibinfo {year} {2010})},\ \Eprint
	{http://arxiv.org/abs/1001.1739} {arXiv:1001.1739 [astro-ph.CO]} \BibitemShut
	{NoStop}%
	\bibitem [{\citenamefont {Clowe}\ \emph {et~al.}(2006)\citenamefont {Clowe},
		\citenamefont {Bradac}, \citenamefont {Gonzalez}, \citenamefont {Markevitch},
		\citenamefont {Randall}, \citenamefont {Jones},\ and\ \citenamefont
		{Zaritsky}}]{Clowe:2006eq}%
	\BibitemOpen
	\bibfield  {author} {\bibinfo {author} {\bibfnamefont {D.}~\bibnamefont
			{Clowe}}, \bibinfo {author} {\bibfnamefont {M.}~\bibnamefont {Bradac}},
		\bibinfo {author} {\bibfnamefont {A.~H.}\ \bibnamefont {Gonzalez}}, \bibinfo
		{author} {\bibfnamefont {M.}~\bibnamefont {Markevitch}}, \bibinfo {author}
		{\bibfnamefont {S.~W.}\ \bibnamefont {Randall}}, \bibinfo {author}
		{\bibfnamefont {C.}~\bibnamefont {Jones}}, \ and\ \bibinfo {author}
		{\bibfnamefont {D.}~\bibnamefont {Zaritsky}},\ }\href {\doibase
		10.1086/508162} {\bibfield  {journal} {\bibinfo  {journal} {Astrophys. J.
				Lett.}\ }\textbf {\bibinfo {volume} {648}},\ \bibinfo {pages} {L109}
		(\bibinfo {year} {2006})},\ \Eprint {http://arxiv.org/abs/astro-ph/0608407}
	{arXiv:astro-ph/0608407} \BibitemShut {NoStop}%
	\bibitem [{\citenamefont {Bennett}\ \emph {et~al.}(2013)\citenamefont {Bennett}
		\emph {et~al.}}]{WMAP:2012fli}%
	\BibitemOpen
	\bibfield  {author} {\bibinfo {author} {\bibfnamefont {C.~L.}\ \bibnamefont
			{Bennett}} \emph {et~al.} (\bibinfo {collaboration} {WMAP}),\ }\href
	{\doibase 10.1088/0067-0049/208/2/20} {\bibfield  {journal} {\bibinfo
			{journal} {Astrophys. J. Suppl.}\ }\textbf {\bibinfo {volume} {208}},\
		\bibinfo {pages} {20} (\bibinfo {year} {2013})},\ \Eprint
	{http://arxiv.org/abs/1212.5225} {arXiv:1212.5225 [astro-ph.CO]} \BibitemShut
	{NoStop}%
	\bibitem [{\citenamefont {Aghanim}\ \emph {et~al.}(2020)\citenamefont {Aghanim}
		\emph {et~al.}}]{Planck:2018vyg}%
	\BibitemOpen
	\bibfield  {author} {\bibinfo {author} {\bibfnamefont {N.}~\bibnamefont
			{Aghanim}} \emph {et~al.} (\bibinfo {collaboration} {Planck}),\ }\href
	{\doibase 10.1051/0004-6361/201833910} {\bibfield  {journal} {\bibinfo
			{journal} {Astron. Astrophys.}\ }\textbf {\bibinfo {volume} {641}},\ \bibinfo
		{pages} {A6} (\bibinfo {year} {2020})},\ \bibinfo {note} {[Erratum:
		Astron.Astrophys. 652, C4 (2021)]},\ \Eprint
	{http://arxiv.org/abs/1807.06209} {arXiv:1807.06209 [astro-ph.CO]}
	\BibitemShut {NoStop}%
	\bibitem [{\citenamefont {Aprile}\ \emph {et~al.}(2018)\citenamefont {Aprile}
		\emph {et~al.}}]{XENON:2018voc}%
	\BibitemOpen
	\bibfield  {author} {\bibinfo {author} {\bibfnamefont {E.}~\bibnamefont
			{Aprile}} \emph {et~al.} (\bibinfo {collaboration} {XENON}),\ }\href
	{\doibase 10.1103/PhysRevLett.121.111302} {\bibfield  {journal} {\bibinfo
			{journal} {Phys. Rev. Lett.}\ }\textbf {\bibinfo {volume} {121}},\ \bibinfo
		{pages} {111302} (\bibinfo {year} {2018})},\ \Eprint
	{http://arxiv.org/abs/1805.12562} {arXiv:1805.12562 [astro-ph.CO]}
	\BibitemShut {NoStop}%
	\bibitem [{\citenamefont {Aprile}\ \emph {et~al.}(2019)\citenamefont {Aprile}
		\emph {et~al.}}]{XENON:2019rxp}%
	\BibitemOpen
	\bibfield  {author} {\bibinfo {author} {\bibfnamefont {E.}~\bibnamefont
			{Aprile}} \emph {et~al.} (\bibinfo {collaboration} {XENON}),\ }\href
	{\doibase 10.1103/PhysRevLett.122.141301} {\bibfield  {journal} {\bibinfo
			{journal} {Phys. Rev. Lett.}\ }\textbf {\bibinfo {volume} {122}},\ \bibinfo
		{pages} {141301} (\bibinfo {year} {2019})},\ \Eprint
	{http://arxiv.org/abs/1902.03234} {arXiv:1902.03234 [astro-ph.CO]}
	\BibitemShut {NoStop}%
	\bibitem [{\citenamefont {Meng}\ \emph {et~al.}(2021)\citenamefont {Meng} \emph
		{et~al.}}]{PandaX-4T:2021bab}%
	\BibitemOpen
	\bibfield  {author} {\bibinfo {author} {\bibfnamefont {Y.}~\bibnamefont
			{Meng}} \emph {et~al.} (\bibinfo {collaboration} {PandaX-4T}),\ }\href@noop
	{} {\  (\bibinfo {year} {2021})},\ \Eprint {http://arxiv.org/abs/2107.13438}
	{arXiv:2107.13438 [hep-ex]} \BibitemShut {NoStop}%
	\bibitem [{\citenamefont {Kolb}\ and\ \citenamefont
		{Turner}(1990)}]{Kolb:1990vq}%
	\BibitemOpen
	\bibfield  {author} {\bibinfo {author} {\bibfnamefont {E.~W.}\ \bibnamefont
			{Kolb}}\ and\ \bibinfo {author} {\bibfnamefont {M.~S.}\ \bibnamefont
			{Turner}},\ }\href {\doibase 10.1201/9780429492860} {\emph {\bibinfo {title}
			{{The Early Universe}}}},\ Vol.~\bibinfo {volume} {69}\ (\bibinfo {year}
	{1990})\BibitemShut {NoStop}%
	\bibitem [{\citenamefont {Jungman}\ \emph {et~al.}(1996)\citenamefont
		{Jungman}, \citenamefont {Kamionkowski},\ and\ \citenamefont
		{Griest}}]{Jungman:1995df}%
	\BibitemOpen
	\bibfield  {author} {\bibinfo {author} {\bibfnamefont {G.}~\bibnamefont
			{Jungman}}, \bibinfo {author} {\bibfnamefont {M.}~\bibnamefont
			{Kamionkowski}}, \ and\ \bibinfo {author} {\bibfnamefont {K.}~\bibnamefont
			{Griest}},\ }\href {\doibase 10.1016/0370-1573(95)00058-5} {\bibfield
		{journal} {\bibinfo  {journal} {Phys. Rept.}\ }\textbf {\bibinfo {volume}
			{267}},\ \bibinfo {pages} {195} (\bibinfo {year} {1996})},\ \Eprint
	{http://arxiv.org/abs/hep-ph/9506380} {arXiv:hep-ph/9506380} \BibitemShut
	{NoStop}%
	\bibitem [{\citenamefont {Bertone}\ \emph {et~al.}(2005)\citenamefont
		{Bertone}, \citenamefont {Hooper},\ and\ \citenamefont
		{Silk}}]{Bertone:2004pz}%
	\BibitemOpen
	\bibfield  {author} {\bibinfo {author} {\bibfnamefont {G.}~\bibnamefont
			{Bertone}}, \bibinfo {author} {\bibfnamefont {D.}~\bibnamefont {Hooper}}, \
		and\ \bibinfo {author} {\bibfnamefont {J.}~\bibnamefont {Silk}},\ }\href
	{\doibase 10.1016/j.physrep.2004.08.031} {\bibfield  {journal} {\bibinfo
			{journal} {Phys. Rept.}\ }\textbf {\bibinfo {volume} {405}},\ \bibinfo
		{pages} {279} (\bibinfo {year} {2005})},\ \Eprint
	{http://arxiv.org/abs/hep-ph/0404175} {arXiv:hep-ph/0404175} \BibitemShut
	{NoStop}%
	\bibitem [{\citenamefont {Hall}\ \emph {et~al.}(2010)\citenamefont {Hall},
		\citenamefont {Jedamzik}, \citenamefont {March-Russell},\ and\ \citenamefont
		{West}}]{Hall:2009bx}%
	\BibitemOpen
	\bibfield  {author} {\bibinfo {author} {\bibfnamefont {L.~J.}\ \bibnamefont
			{Hall}}, \bibinfo {author} {\bibfnamefont {K.}~\bibnamefont {Jedamzik}},
		\bibinfo {author} {\bibfnamefont {J.}~\bibnamefont {March-Russell}}, \ and\
		\bibinfo {author} {\bibfnamefont {S.~M.}\ \bibnamefont {West}},\ }\href
	{\doibase 10.1007/JHEP03(2010)080} {\bibfield  {journal} {\bibinfo  {journal}
			{JHEP}\ }\textbf {\bibinfo {volume} {03}},\ \bibinfo {pages} {080} (\bibinfo
		{year} {2010})},\ \Eprint {http://arxiv.org/abs/0911.1120} {arXiv:0911.1120
		[hep-ph]} \BibitemShut {NoStop}%
	\bibitem [{\citenamefont {Edsj\"o}\ and\ \citenamefont
		{Gondolo}(1997)}]{PhysRevD.56.1879}%
	\BibitemOpen
	\bibfield  {author} {\bibinfo {author} {\bibfnamefont {J.}~\bibnamefont
			{Edsj\"o}}\ and\ \bibinfo {author} {\bibfnamefont {P.}~\bibnamefont
			{Gondolo}},\ }\href {\doibase 10.1103/PhysRevD.56.1879} {\bibfield  {journal}
		{\bibinfo  {journal} {Phys. Rev. D}\ }\textbf {\bibinfo {volume} {56}},\
		\bibinfo {pages} {1879} (\bibinfo {year} {1997})}\BibitemShut {NoStop}%
	\bibitem [{\citenamefont {Dutra}\ \emph {et~al.}(2015)\citenamefont {Dutra},
		\citenamefont {de~S.~Pires},\ and\ \citenamefont {Rodrigues~da
			Silva}}]{Dutra:2015vca}%
	\BibitemOpen
	\bibfield  {author} {\bibinfo {author} {\bibfnamefont {M.}~\bibnamefont
			{Dutra}}, \bibinfo {author} {\bibfnamefont {C.~A.}\ \bibnamefont
			{de~S.~Pires}}, \ and\ \bibinfo {author} {\bibfnamefont {P.~S.}\ \bibnamefont
			{Rodrigues~da Silva}},\ }\href {\doibase 10.1007/JHEP09(2015)147} {\bibfield
		{journal} {\bibinfo  {journal} {JHEP}\ }\textbf {\bibinfo {volume} {09}},\
		\bibinfo {pages} {147} (\bibinfo {year} {2015})},\ \Eprint
	{http://arxiv.org/abs/1504.07222} {arXiv:1504.07222 [hep-ph]} \BibitemShut
	{NoStop}%
	\bibitem [{\citenamefont {Mizukoshi}\ \emph {et~al.}(2011)\citenamefont
		{Mizukoshi}, \citenamefont {de~S.~Pires}, \citenamefont {Queiroz},\ and\
		\citenamefont {Rodrigues~da Silva}}]{Mizukoshi:2010ky}%
	\BibitemOpen
	\bibfield  {author} {\bibinfo {author} {\bibfnamefont {J.~K.}\ \bibnamefont
			{Mizukoshi}}, \bibinfo {author} {\bibfnamefont {C.~A.}\ \bibnamefont
			{de~S.~Pires}}, \bibinfo {author} {\bibfnamefont {F.~S.}\ \bibnamefont
			{Queiroz}}, \ and\ \bibinfo {author} {\bibfnamefont {P.~S.}\ \bibnamefont
			{Rodrigues~da Silva}},\ }\href {\doibase 10.1103/PhysRevD.83.065024}
	{\bibfield  {journal} {\bibinfo  {journal} {Phys. Rev. D}\ }\textbf {\bibinfo
			{volume} {83}},\ \bibinfo {pages} {065024} (\bibinfo {year} {2011})},\
	\Eprint {http://arxiv.org/abs/1010.4097} {arXiv:1010.4097 [hep-ph]}
	\BibitemShut {NoStop}%
	\bibitem [{\citenamefont {Fayet}(2006)}]{Fayet:2006xd}%
	\BibitemOpen
	\bibfield  {author} {\bibinfo {author} {\bibfnamefont {P.}~\bibnamefont
			{Fayet}},\ }\href@noop {} {\  (\bibinfo {year} {2006})},\ \Eprint
	{http://arxiv.org/abs/hep-ph/0607094} {arXiv:hep-ph/0607094} \BibitemShut
	{NoStop}%
	\bibitem [{\citenamefont {An}\ \emph {et~al.}(2015)\citenamefont {An},
		\citenamefont {Pospelov}, \citenamefont {Pradler},\ and\ \citenamefont
		{Ritz}}]{An:2014twa}%
	\BibitemOpen
	\bibfield  {author} {\bibinfo {author} {\bibfnamefont {H.}~\bibnamefont
			{An}}, \bibinfo {author} {\bibfnamefont {M.}~\bibnamefont {Pospelov}},
		\bibinfo {author} {\bibfnamefont {J.}~\bibnamefont {Pradler}}, \ and\
		\bibinfo {author} {\bibfnamefont {A.}~\bibnamefont {Ritz}},\ }\href {\doibase
		10.1016/j.physletb.2015.06.018} {\bibfield  {journal} {\bibinfo  {journal}
			{Phys. Lett. B}\ }\textbf {\bibinfo {volume} {747}},\ \bibinfo {pages} {331}
		(\bibinfo {year} {2015})},\ \Eprint {http://arxiv.org/abs/1412.8378}
	{arXiv:1412.8378 [hep-ph]} \BibitemShut {NoStop}%
	\bibitem [{\citenamefont {Pospelov}\ \emph {et~al.}(2008)\citenamefont
		{Pospelov}, \citenamefont {Ritz},\ and\ \citenamefont
		{Voloshin}}]{Pospelov:2007mp}%
	\BibitemOpen
	\bibfield  {author} {\bibinfo {author} {\bibfnamefont {M.}~\bibnamefont
			{Pospelov}}, \bibinfo {author} {\bibfnamefont {A.}~\bibnamefont {Ritz}}, \
		and\ \bibinfo {author} {\bibfnamefont {M.~B.}\ \bibnamefont {Voloshin}},\
	}\href {\doibase 10.1016/j.physletb.2008.02.052} {\bibfield  {journal}
		{\bibinfo  {journal} {Phys. Lett. B}\ }\textbf {\bibinfo {volume} {662}},\
		\bibinfo {pages} {53} (\bibinfo {year} {2008})},\ \Eprint
	{http://arxiv.org/abs/0711.4866} {arXiv:0711.4866 [hep-ph]} \BibitemShut
	{NoStop}%
	\bibitem [{\citenamefont {Chakraborty}\ \emph {et~al.}(2021)\citenamefont
		{Chakraborty}, \citenamefont {Ghosh}, \citenamefont {Ghosh},\ and\
		\citenamefont {Rai}}]{Chakraborty:2021tdo}%
	\BibitemOpen
	\bibfield  {author} {\bibinfo {author} {\bibfnamefont {I.}~\bibnamefont
			{Chakraborty}}, \bibinfo {author} {\bibfnamefont {D.~K.}\ \bibnamefont
			{Ghosh}}, \bibinfo {author} {\bibfnamefont {N.}~\bibnamefont {Ghosh}}, \ and\
		\bibinfo {author} {\bibfnamefont {S.~K.}\ \bibnamefont {Rai}},\ }\href
	{\doibase 10.1140/epjc/s10052-021-09446-5} {\bibfield  {journal} {\bibinfo
			{journal} {Eur. Phys. J. C}\ }\textbf {\bibinfo {volume} {81}},\ \bibinfo
		{pages} {679} (\bibinfo {year} {2021})},\ \Eprint
	{http://arxiv.org/abs/2104.03351} {arXiv:2104.03351 [hep-ph]} \BibitemShut
	{NoStop}%
	\bibitem [{\citenamefont {Dey}\ \emph {et~al.}(2022)\citenamefont {Dey},
		\citenamefont {Ghosh},\ and\ \citenamefont {Rai}}]{Dey:2022whc}%
	\BibitemOpen
	\bibfield  {author} {\bibinfo {author} {\bibfnamefont {S.}~\bibnamefont
			{Dey}}, \bibinfo {author} {\bibfnamefont {P.}~\bibnamefont {Ghosh}}, \ and\
		\bibinfo {author} {\bibfnamefont {S.~K.}\ \bibnamefont {Rai}},\ }\href
	{\doibase 10.1140/epjc/s10052-022-10778-z} {\bibfield  {journal} {\bibinfo
			{journal} {Eur. Phys. J. C}\ }\textbf {\bibinfo {volume} {82}},\ \bibinfo
		{pages} {876} (\bibinfo {year} {2022})},\ \Eprint
	{http://arxiv.org/abs/2202.11638} {arXiv:2202.11638 [hep-ph]} \BibitemShut
	{NoStop}%
	\bibitem [{\citenamefont {Gondolo}\ and\ \citenamefont
		{Gelmini}(1991)}]{Gondolo:1990dk}%
	\BibitemOpen
	\bibfield  {author} {\bibinfo {author} {\bibfnamefont {P.}~\bibnamefont
			{Gondolo}}\ and\ \bibinfo {author} {\bibfnamefont {G.}~\bibnamefont
			{Gelmini}},\ }\href {\doibase 10.1016/0550-3213(91)90438-4} {\bibfield
		{journal} {\bibinfo  {journal} {Nucl. Phys. B}\ }\textbf {\bibinfo {volume}
			{360}},\ \bibinfo {pages} {145} (\bibinfo {year} {1991})}\BibitemShut
	{NoStop}%
	\bibitem [{\citenamefont {Yaguna}(2011)}]{Yaguna:2011qn}%
	\BibitemOpen
	\bibfield  {author} {\bibinfo {author} {\bibfnamefont {C.~E.}\ \bibnamefont
			{Yaguna}},\ }\href {\doibase 10.1007/JHEP08(2011)060} {\bibfield  {journal}
		{\bibinfo  {journal} {JHEP}\ }\textbf {\bibinfo {volume} {08}},\ \bibinfo
		{pages} {060} (\bibinfo {year} {2011})},\ \Eprint
	{http://arxiv.org/abs/1105.1654} {arXiv:1105.1654 [hep-ph]} \BibitemShut
	{NoStop}%
	\bibitem [{\citenamefont {Chu}\ \emph {et~al.}(2012)\citenamefont {Chu},
		\citenamefont {Hambye},\ and\ \citenamefont {Tytgat}}]{Chu:2011be}%
	\BibitemOpen
	\bibfield  {author} {\bibinfo {author} {\bibfnamefont {X.}~\bibnamefont
			{Chu}}, \bibinfo {author} {\bibfnamefont {T.}~\bibnamefont {Hambye}}, \ and\
		\bibinfo {author} {\bibfnamefont {M.~H.~G.}\ \bibnamefont {Tytgat}},\ }\href
	{\doibase 10.1088/1475-7516/2012/05/034} {\bibfield  {journal} {\bibinfo
			{journal} {JCAP}\ }\textbf {\bibinfo {volume} {05}},\ \bibinfo {pages} {034}
		(\bibinfo {year} {2012})},\ \Eprint {http://arxiv.org/abs/1112.0493}
	{arXiv:1112.0493 [hep-ph]} \BibitemShut {NoStop}%
	\bibitem [{\citenamefont {Shakya}(2016)}]{Shakya:2015xnx}%
	\BibitemOpen
	\bibfield  {author} {\bibinfo {author} {\bibfnamefont {B.}~\bibnamefont
			{Shakya}},\ }\href {\doibase 10.1142/S0217732316300056} {\bibfield  {journal}
		{\bibinfo  {journal} {Mod. Phys. Lett. A}\ }\textbf {\bibinfo {volume}
			{31}},\ \bibinfo {pages} {1630005} (\bibinfo {year} {2016})},\ \Eprint
	{http://arxiv.org/abs/1512.02751} {arXiv:1512.02751 [hep-ph]} \BibitemShut
	{NoStop}%
	\bibitem [{\citenamefont {Pandey}\ \emph {et~al.}(2018)\citenamefont {Pandey},
		\citenamefont {Majumdar},\ and\ \citenamefont {Modak}}]{Pandey:2017quk}%
	\BibitemOpen
	\bibfield  {author} {\bibinfo {author} {\bibfnamefont {M.}~\bibnamefont
			{Pandey}}, \bibinfo {author} {\bibfnamefont {D.}~\bibnamefont {Majumdar}}, \
		and\ \bibinfo {author} {\bibfnamefont {K.~P.}\ \bibnamefont {Modak}},\ }\href
	{\doibase 10.1088/1475-7516/2018/06/023} {\bibfield  {journal} {\bibinfo
			{journal} {JCAP}\ }\textbf {\bibinfo {volume} {06}},\ \bibinfo {pages} {023}
		(\bibinfo {year} {2018})},\ \Eprint {http://arxiv.org/abs/1709.05955}
	{arXiv:1709.05955 [hep-ph]} \BibitemShut {NoStop}%
	\bibitem [{\citenamefont {Elahi}\ \emph {et~al.}(2015)\citenamefont {Elahi},
		\citenamefont {Kolda},\ and\ \citenamefont {Unwin}}]{Elahi:2014fsa}%
	\BibitemOpen
	\bibfield  {author} {\bibinfo {author} {\bibfnamefont {F.}~\bibnamefont
			{Elahi}}, \bibinfo {author} {\bibfnamefont {C.}~\bibnamefont {Kolda}}, \ and\
		\bibinfo {author} {\bibfnamefont {J.}~\bibnamefont {Unwin}},\ }\href
	{\doibase 10.1007/JHEP03(2015)048} {\bibfield  {journal} {\bibinfo  {journal}
			{JHEP}\ }\textbf {\bibinfo {volume} {03}},\ \bibinfo {pages} {048} (\bibinfo
		{year} {2015})},\ \Eprint {http://arxiv.org/abs/1410.6157} {arXiv:1410.6157
		[hep-ph]} \BibitemShut {NoStop}%
	\bibitem [{\citenamefont {Bernal}\ \emph {et~al.}(2017)\citenamefont {Bernal},
		\citenamefont {Heikinheimo}, \citenamefont {Tenkanen}, \citenamefont
		{Tuominen},\ and\ \citenamefont {Vaskonen}}]{Bernal:2017kxu}%
	\BibitemOpen
	\bibfield  {author} {\bibinfo {author} {\bibfnamefont {N.}~\bibnamefont
			{Bernal}}, \bibinfo {author} {\bibfnamefont {M.}~\bibnamefont {Heikinheimo}},
		\bibinfo {author} {\bibfnamefont {T.}~\bibnamefont {Tenkanen}}, \bibinfo
		{author} {\bibfnamefont {K.}~\bibnamefont {Tuominen}}, \ and\ \bibinfo
		{author} {\bibfnamefont {V.}~\bibnamefont {Vaskonen}},\ }\href {\doibase
		10.1142/S0217751X1730023X} {\bibfield  {journal} {\bibinfo  {journal} {Int.
				J. Mod. Phys. A}\ }\textbf {\bibinfo {volume} {32}},\ \bibinfo {pages}
		{1730023} (\bibinfo {year} {2017})},\ \Eprint
	{http://arxiv.org/abs/1706.07442} {arXiv:1706.07442 [hep-ph]} \BibitemShut
	{NoStop}%
	\bibitem [{\citenamefont {Merle}\ \emph {et~al.}(2014)\citenamefont {Merle},
		\citenamefont {Niro},\ and\ \citenamefont {Schmidt}}]{Merle:2013wta}%
	\BibitemOpen
	\bibfield  {author} {\bibinfo {author} {\bibfnamefont {A.}~\bibnamefont
			{Merle}}, \bibinfo {author} {\bibfnamefont {V.}~\bibnamefont {Niro}}, \ and\
		\bibinfo {author} {\bibfnamefont {D.}~\bibnamefont {Schmidt}},\ }\href
	{\doibase 10.1088/1475-7516/2014/03/028} {\bibfield  {journal} {\bibinfo
			{journal} {JCAP}\ }\textbf {\bibinfo {volume} {03}},\ \bibinfo {pages} {028}
		(\bibinfo {year} {2014})},\ \Eprint {http://arxiv.org/abs/1306.3996}
	{arXiv:1306.3996 [hep-ph]} \BibitemShut {NoStop}%
	\bibitem [{\citenamefont {Ghosh}\ \emph {et~al.}(2017)\citenamefont {Ghosh},
		\citenamefont {Mondal},\ and\ \citenamefont {Mukhopadhyaya}}]{Ghosh:2017vhe}%
	\BibitemOpen
	\bibfield  {author} {\bibinfo {author} {\bibfnamefont {A.}~\bibnamefont
			{Ghosh}}, \bibinfo {author} {\bibfnamefont {T.}~\bibnamefont {Mondal}}, \
		and\ \bibinfo {author} {\bibfnamefont {B.}~\bibnamefont {Mukhopadhyaya}},\
	}\href {\doibase 10.1007/JHEP12(2017)136} {\bibfield  {journal} {\bibinfo
			{journal} {JHEP}\ }\textbf {\bibinfo {volume} {12}},\ \bibinfo {pages} {136}
		(\bibinfo {year} {2017})},\ \Eprint {http://arxiv.org/abs/1706.06815}
	{arXiv:1706.06815 [hep-ph]} \BibitemShut {NoStop}%
	\bibitem [{\citenamefont {Cosme}\ \emph
		{et~al.}(2021{\natexlab{a}})\citenamefont {Cosme}, \citenamefont {Dutra},
		\citenamefont {Godfrey},\ and\ \citenamefont {Gray}}]{Cosme:2021baj}%
	\BibitemOpen
	\bibfield  {author} {\bibinfo {author} {\bibfnamefont {C.}~\bibnamefont
			{Cosme}}, \bibinfo {author} {\bibfnamefont {M.}~\bibnamefont {Dutra}},
		\bibinfo {author} {\bibfnamefont {S.}~\bibnamefont {Godfrey}}, \ and\
		\bibinfo {author} {\bibfnamefont {T.~R.}\ \bibnamefont {Gray}},\ }\href
	{\doibase 10.1007/JHEP09(2021)056} {\bibfield  {journal} {\bibinfo  {journal}
			{JHEP}\ }\textbf {\bibinfo {volume} {09}},\ \bibinfo {pages} {056} (\bibinfo
		{year} {2021}{\natexlab{a}})},\ \Eprint {http://arxiv.org/abs/2104.13937}
	{arXiv:2104.13937 [hep-ph]} \BibitemShut {NoStop}%
	\bibitem [{\citenamefont {Cosme}\ \emph
		{et~al.}(2021{\natexlab{b}})\citenamefont {Cosme}, \citenamefont {Dutra},
		\citenamefont {Ma}, \citenamefont {Wu},\ and\ \citenamefont
		{Yang}}]{Cosme:2020mck}%
	\BibitemOpen
	\bibfield  {author} {\bibinfo {author} {\bibfnamefont {C.}~\bibnamefont
			{Cosme}}, \bibinfo {author} {\bibfnamefont {M.}~\bibnamefont {Dutra}},
		\bibinfo {author} {\bibfnamefont {T.}~\bibnamefont {Ma}}, \bibinfo {author}
		{\bibfnamefont {Y.}~\bibnamefont {Wu}}, \ and\ \bibinfo {author}
		{\bibfnamefont {L.}~\bibnamefont {Yang}},\ }\href {\doibase
		10.1007/JHEP03(2021)026} {\bibfield  {journal} {\bibinfo  {journal} {JHEP}\
		}\textbf {\bibinfo {volume} {03}},\ \bibinfo {pages} {026} (\bibinfo {year}
		{2021}{\natexlab{b}})},\ \Eprint {http://arxiv.org/abs/2003.01723}
	{arXiv:2003.01723 [hep-ph]} \BibitemShut {NoStop}%
	\bibitem [{\citenamefont {Bhattacharyya}\ \emph {et~al.}(2018)\citenamefont
		{Bhattacharyya}, \citenamefont {Dutra}, \citenamefont {Mambrini},\ and\
		\citenamefont {Pierre}}]{Bhattacharyya:2018evo}%
	\BibitemOpen
	\bibfield  {author} {\bibinfo {author} {\bibfnamefont {G.}~\bibnamefont
			{Bhattacharyya}}, \bibinfo {author} {\bibfnamefont {M.}~\bibnamefont
			{Dutra}}, \bibinfo {author} {\bibfnamefont {Y.}~\bibnamefont {Mambrini}}, \
		and\ \bibinfo {author} {\bibfnamefont {M.}~\bibnamefont {Pierre}},\ }\href
	{\doibase 10.1103/PhysRevD.98.035038} {\bibfield  {journal} {\bibinfo
			{journal} {Phys. Rev. D}\ }\textbf {\bibinfo {volume} {98}},\ \bibinfo
		{pages} {035038} (\bibinfo {year} {2018})},\ \Eprint
	{http://arxiv.org/abs/1806.00016} {arXiv:1806.00016 [hep-ph]} \BibitemShut
	{NoStop}%
	\bibitem [{\citenamefont {Abdallah}\ \emph {et~al.}(2019)\citenamefont
		{Abdallah}, \citenamefont {Choubey},\ and\ \citenamefont
		{Khan}}]{Abdallah:2019svm}%
	\BibitemOpen
	\bibfield  {author} {\bibinfo {author} {\bibfnamefont {W.}~\bibnamefont
			{Abdallah}}, \bibinfo {author} {\bibfnamefont {S.}~\bibnamefont {Choubey}}, \
		and\ \bibinfo {author} {\bibfnamefont {S.}~\bibnamefont {Khan}},\ }\href
	{\doibase 10.1007/JHEP06(2019)095} {\bibfield  {journal} {\bibinfo  {journal}
			{JHEP}\ }\textbf {\bibinfo {volume} {06}},\ \bibinfo {pages} {095} (\bibinfo
		{year} {2019})},\ \Eprint {http://arxiv.org/abs/1904.10015} {arXiv:1904.10015
		[hep-ph]} \BibitemShut {NoStop}%
	\bibitem [{\citenamefont {Abdallah}\ \emph {et~al.}(2021)\citenamefont
		{Abdallah}, \citenamefont {Barik}, \citenamefont {Rai},\ and\ \citenamefont
		{Samui}}]{Abdallah:2021npg}%
	\BibitemOpen
	\bibfield  {author} {\bibinfo {author} {\bibfnamefont {W.}~\bibnamefont
			{Abdallah}}, \bibinfo {author} {\bibfnamefont {A.~K.}\ \bibnamefont {Barik}},
		\bibinfo {author} {\bibfnamefont {S.~K.}\ \bibnamefont {Rai}}, \ and\
		\bibinfo {author} {\bibfnamefont {T.}~\bibnamefont {Samui}},\ }\href@noop {}
	{\  (\bibinfo {year} {2021})},\ \Eprint {http://arxiv.org/abs/2106.01362}
	{arXiv:2106.01362 [hep-ph]} \BibitemShut {NoStop}%
	\bibitem [{\citenamefont {Robinett}\ and\ \citenamefont
		{Rosner}(1982{\natexlab{a}})}]{Robinett:1981yz}%
	\BibitemOpen
	\bibfield  {author} {\bibinfo {author} {\bibfnamefont {R.~W.}\ \bibnamefont
			{Robinett}}\ and\ \bibinfo {author} {\bibfnamefont {J.~L.}\ \bibnamefont
			{Rosner}},\ }\href {\doibase 10.1103/PhysRevD.27.679} {\bibfield  {journal}
		{\bibinfo  {journal} {Phys. Rev. D}\ }\textbf {\bibinfo {volume} {25}},\
		\bibinfo {pages} {3036} (\bibinfo {year} {1982}{\natexlab{a}})},\ \bibinfo
	{note} {[Erratum: Phys.Rev.D 27, 679 (1983)]}\BibitemShut {NoStop}%
	\bibitem [{\citenamefont {Robinett}\ and\ \citenamefont
		{Rosner}(1982{\natexlab{b}})}]{Robinett:1982tq}%
	\BibitemOpen
	\bibfield  {author} {\bibinfo {author} {\bibfnamefont {R.~W.}\ \bibnamefont
			{Robinett}}\ and\ \bibinfo {author} {\bibfnamefont {J.~L.}\ \bibnamefont
			{Rosner}},\ }\href {\doibase 10.1103/PhysRevD.26.2396} {\bibfield  {journal}
		{\bibinfo  {journal} {Phys. Rev. D}\ }\textbf {\bibinfo {volume} {26}},\
		\bibinfo {pages} {2396} (\bibinfo {year} {1982}{\natexlab{b}})}\BibitemShut
	{NoStop}%
	\bibitem [{\citenamefont {Langacker}\ \emph {et~al.}(1984)\citenamefont
		{Langacker}, \citenamefont {Robinett},\ and\ \citenamefont
		{Rosner}}]{Langacker:1984dc}%
	\BibitemOpen
	\bibfield  {author} {\bibinfo {author} {\bibfnamefont {P.}~\bibnamefont
			{Langacker}}, \bibinfo {author} {\bibfnamefont {R.~W.}\ \bibnamefont
			{Robinett}}, \ and\ \bibinfo {author} {\bibfnamefont {J.~L.}\ \bibnamefont
			{Rosner}},\ }\href {\doibase 10.1103/PhysRevD.30.1470} {\bibfield  {journal}
		{\bibinfo  {journal} {Phys. Rev. D}\ }\textbf {\bibinfo {volume} {30}},\
		\bibinfo {pages} {1470} (\bibinfo {year} {1984})}\BibitemShut {NoStop}%
	\bibitem [{\citenamefont {Hewett}\ and\ \citenamefont
		{Rizzo}(1989)}]{Hewett:1988xc}%
	\BibitemOpen
	\bibfield  {author} {\bibinfo {author} {\bibfnamefont {J.~L.}\ \bibnamefont
			{Hewett}}\ and\ \bibinfo {author} {\bibfnamefont {T.~G.}\ \bibnamefont
			{Rizzo}},\ }\href {\doibase 10.1016/0370-1573(89)90071-9} {\bibfield
		{journal} {\bibinfo  {journal} {Phys. Rept.}\ }\textbf {\bibinfo {volume}
			{183}},\ \bibinfo {pages} {193} (\bibinfo {year} {1989})}\BibitemShut
	{NoStop}%
	\bibitem [{\citenamefont {Abdallah}\ \emph {et~al.}(2023)\citenamefont
		{Abdallah}, \citenamefont {Barik}, \citenamefont {Rai},\ and\ \citenamefont
		{Samui}}]{Abdallah:2021dul}%
	\BibitemOpen
	\bibfield  {author} {\bibinfo {author} {\bibfnamefont {W.}~\bibnamefont
			{Abdallah}}, \bibinfo {author} {\bibfnamefont {A.~K.}\ \bibnamefont {Barik}},
		\bibinfo {author} {\bibfnamefont {S.~K.}\ \bibnamefont {Rai}}, \ and\
		\bibinfo {author} {\bibfnamefont {T.}~\bibnamefont {Samui}},\ }\href
	{\doibase 10.1103/PhysRevD.107.015026} {\bibfield  {journal} {\bibinfo
			{journal} {Phys. Rev. D}\ }\textbf {\bibinfo {volume} {107}},\ \bibinfo
		{pages} {015026} (\bibinfo {year} {2023})},\ \Eprint
	{http://arxiv.org/abs/2109.07980} {arXiv:2109.07980 [hep-ph]} \BibitemShut
	{NoStop}%
	\bibitem [{\citenamefont {Aad}\ \emph {et~al.}(2012)\citenamefont {Aad} \emph
		{et~al.}}]{ATLAS:2012yve}%
	\BibitemOpen
	\bibfield  {author} {\bibinfo {author} {\bibfnamefont {G.}~\bibnamefont
			{Aad}} \emph {et~al.} (\bibinfo {collaboration} {ATLAS}),\ }\href {\doibase
		10.1016/j.physletb.2012.08.020} {\bibfield  {journal} {\bibinfo  {journal}
			{Phys. Lett. B}\ }\textbf {\bibinfo {volume} {716}},\ \bibinfo {pages} {1}
		(\bibinfo {year} {2012})},\ \Eprint {http://arxiv.org/abs/1207.7214}
	{arXiv:1207.7214 [hep-ex]} \BibitemShut {NoStop}%
	\bibitem [{\citenamefont {Chatrchyan}\ \emph {et~al.}(2012)\citenamefont
		{Chatrchyan} \emph {et~al.}}]{CMS:2012qbp}%
	\BibitemOpen
	\bibfield  {author} {\bibinfo {author} {\bibfnamefont {S.}~\bibnamefont
			{Chatrchyan}} \emph {et~al.} (\bibinfo {collaboration} {CMS}),\ }\href
	{\doibase 10.1016/j.physletb.2012.08.021} {\bibfield  {journal} {\bibinfo
			{journal} {Phys. Lett. B}\ }\textbf {\bibinfo {volume} {716}},\ \bibinfo
		{pages} {30} (\bibinfo {year} {2012})},\ \Eprint
	{http://arxiv.org/abs/1207.7235} {arXiv:1207.7235 [hep-ex]} \BibitemShut
	{NoStop}%
	\bibitem [{\citenamefont {del Aguila}\ \emph {et~al.}(1995)\citenamefont {del
			Aguila}, \citenamefont {Masip},\ and\ \citenamefont
		{Perez-Victoria}}]{delAguila:1995rb}%
	\BibitemOpen
	\bibfield  {author} {\bibinfo {author} {\bibfnamefont {F.}~\bibnamefont {del
				Aguila}}, \bibinfo {author} {\bibfnamefont {M.}~\bibnamefont {Masip}}, \ and\
		\bibinfo {author} {\bibfnamefont {M.}~\bibnamefont {Perez-Victoria}},\ }\href
	{\doibase 10.1016/0550-3213(95)00511-6} {\bibfield  {journal} {\bibinfo
			{journal} {Nucl. Phys. B}\ }\textbf {\bibinfo {volume} {456}},\ \bibinfo
		{pages} {531} (\bibinfo {year} {1995})},\ \Eprint
	{http://arxiv.org/abs/hep-ph/9507455} {arXiv:hep-ph/9507455} \BibitemShut
	{NoStop}%
	\bibitem [{\citenamefont {Chankowski}\ \emph {et~al.}(2006)\citenamefont
		{Chankowski}, \citenamefont {Pokorski},\ and\ \citenamefont
		{Wagner}}]{Chankowski:2006jk}%
	\BibitemOpen
	\bibfield  {author} {\bibinfo {author} {\bibfnamefont {P.~H.}\ \bibnamefont
			{Chankowski}}, \bibinfo {author} {\bibfnamefont {S.}~\bibnamefont
			{Pokorski}}, \ and\ \bibinfo {author} {\bibfnamefont {J.}~\bibnamefont
			{Wagner}},\ }\href {\doibase 10.1140/epjc/s2006-02537-3} {\bibfield
		{journal} {\bibinfo  {journal} {Eur. Phys. J. C}\ }\textbf {\bibinfo {volume}
			{47}},\ \bibinfo {pages} {187} (\bibinfo {year} {2006})},\ \Eprint
	{http://arxiv.org/abs/hep-ph/0601097} {arXiv:hep-ph/0601097} \BibitemShut
	{NoStop}%
	\bibitem [{\citenamefont {Zyla}\ \emph {et~al.}(2020)\citenamefont {Zyla} \emph
		{et~al.}}]{ParticleDataGroup:2020ssz}%
	\BibitemOpen
	\bibfield  {author} {\bibinfo {author} {\bibfnamefont {P.~A.}\ \bibnamefont
			{Zyla}} \emph {et~al.} (\bibinfo {collaboration} {Particle Data Group}),\
	}\href {\doibase 10.1093/ptep/ptaa104} {\bibfield  {journal} {\bibinfo
			{journal} {PTEP}\ }\textbf {\bibinfo {volume} {2020}},\ \bibinfo {pages}
		{083C01} (\bibinfo {year} {2020})}\BibitemShut {NoStop}%
	\bibitem [{\citenamefont {Higaki}\ \emph {et~al.}(2014)\citenamefont {Higaki},
		\citenamefont {Kitano},\ and\ \citenamefont {Sato}}]{Higaki:2014dwa}%
	\BibitemOpen
	\bibfield  {author} {\bibinfo {author} {\bibfnamefont {T.}~\bibnamefont
			{Higaki}}, \bibinfo {author} {\bibfnamefont {R.}~\bibnamefont {Kitano}}, \
		and\ \bibinfo {author} {\bibfnamefont {R.}~\bibnamefont {Sato}},\ }\href
	{\doibase 10.1007/JHEP07(2014)044} {\bibfield  {journal} {\bibinfo  {journal}
			{JHEP}\ }\textbf {\bibinfo {volume} {07}},\ \bibinfo {pages} {044} (\bibinfo
		{year} {2014})},\ \Eprint {http://arxiv.org/abs/1405.0013} {arXiv:1405.0013
		[hep-ph]} \BibitemShut {NoStop}%
	\bibitem [{\citenamefont {Bhupal~Dev}\ \emph {et~al.}(2016)\citenamefont
		{Bhupal~Dev}, \citenamefont {Mohapatra},\ and\ \citenamefont
		{Zhang}}]{BhupalDev:2016gna}%
	\BibitemOpen
	\bibfield  {author} {\bibinfo {author} {\bibfnamefont {P.~S.}\ \bibnamefont
			{Bhupal~Dev}}, \bibinfo {author} {\bibfnamefont {R.~N.}\ \bibnamefont
			{Mohapatra}}, \ and\ \bibinfo {author} {\bibfnamefont {Y.}~\bibnamefont
			{Zhang}},\ }\href {\doibase 10.1007/JHEP11(2016)077} {\bibfield  {journal}
		{\bibinfo  {journal} {JHEP}\ }\textbf {\bibinfo {volume} {11}},\ \bibinfo
		{pages} {077} (\bibinfo {year} {2016})},\ \Eprint
	{http://arxiv.org/abs/1608.06266} {arXiv:1608.06266 [hep-ph]} \BibitemShut
	{NoStop}%
	\bibitem [{\citenamefont {Re~Fiorentin}\ \emph {et~al.}(2016)\citenamefont
		{Re~Fiorentin}, \citenamefont {Niro},\ and\ \citenamefont
		{Fornengo}}]{ReFiorentin:2016rzn}%
	\BibitemOpen
	\bibfield  {author} {\bibinfo {author} {\bibfnamefont {M.}~\bibnamefont
			{Re~Fiorentin}}, \bibinfo {author} {\bibfnamefont {V.}~\bibnamefont {Niro}},
		\ and\ \bibinfo {author} {\bibfnamefont {N.}~\bibnamefont {Fornengo}},\
	}\href {\doibase 10.1007/JHEP11(2016)022} {\bibfield  {journal} {\bibinfo
			{journal} {JHEP}\ }\textbf {\bibinfo {volume} {11}},\ \bibinfo {pages} {022}
		(\bibinfo {year} {2016})},\ \Eprint {http://arxiv.org/abs/1606.04445}
	{arXiv:1606.04445 [hep-ph]} \BibitemShut {NoStop}%
	\bibitem [{\citenamefont {Di~Bari}\ \emph {et~al.}(2016)\citenamefont
		{Di~Bari}, \citenamefont {Ludl},\ and\ \citenamefont
		{Palomares-Ruiz}}]{DiBari:2016guw}%
	\BibitemOpen
	\bibfield  {author} {\bibinfo {author} {\bibfnamefont {P.}~\bibnamefont
			{Di~Bari}}, \bibinfo {author} {\bibfnamefont {P.~O.}\ \bibnamefont {Ludl}}, \
		and\ \bibinfo {author} {\bibfnamefont {S.}~\bibnamefont {Palomares-Ruiz}},\
	}\href {\doibase 10.1088/1475-7516/2016/11/044} {\bibfield  {journal}
		{\bibinfo  {journal} {JCAP}\ }\textbf {\bibinfo {volume} {11}},\ \bibinfo
		{pages} {044} (\bibinfo {year} {2016})},\ \Eprint
	{http://arxiv.org/abs/1606.06238} {arXiv:1606.06238 [hep-ph]} \BibitemShut
	{NoStop}%
	\bibitem [{\citenamefont {Atre}\ \emph {et~al.}(2009)\citenamefont {Atre},
		\citenamefont {Han}, \citenamefont {Pascoli},\ and\ \citenamefont
		{Zhang}}]{Atre:2009rg}%
	\BibitemOpen
	\bibfield  {author} {\bibinfo {author} {\bibfnamefont {A.}~\bibnamefont
			{Atre}}, \bibinfo {author} {\bibfnamefont {T.}~\bibnamefont {Han}}, \bibinfo
		{author} {\bibfnamefont {S.}~\bibnamefont {Pascoli}}, \ and\ \bibinfo
		{author} {\bibfnamefont {B.}~\bibnamefont {Zhang}},\ }\href {\doibase
		10.1088/1126-6708/2009/05/030} {\bibfield  {journal} {\bibinfo  {journal}
			{JHEP}\ }\textbf {\bibinfo {volume} {05}},\ \bibinfo {pages} {030} (\bibinfo
		{year} {2009})},\ \Eprint {http://arxiv.org/abs/0901.3589} {arXiv:0901.3589
		[hep-ph]} \BibitemShut {NoStop}%
	\bibitem [{\citenamefont {Griest}\ and\ \citenamefont
		{Seckel}(1991)}]{Griest:1990kh}%
	\BibitemOpen
	\bibfield  {author} {\bibinfo {author} {\bibfnamefont {K.}~\bibnamefont
			{Griest}}\ and\ \bibinfo {author} {\bibfnamefont {D.}~\bibnamefont
			{Seckel}},\ }\href {\doibase 10.1103/PhysRevD.43.3191} {\bibfield  {journal}
		{\bibinfo  {journal} {Phys. Rev. D}\ }\textbf {\bibinfo {volume} {43}},\
		\bibinfo {pages} {3191} (\bibinfo {year} {1991})}\BibitemShut {NoStop}%
	\bibitem [{\citenamefont {B\'elanger}\ \emph {et~al.}(2018)\citenamefont
		{B\'elanger}, \citenamefont {Boudjema}, \citenamefont {Goudelis},
		\citenamefont {Pukhov},\ and\ \citenamefont {Zaldivar}}]{Belanger:2018ccd}%
	\BibitemOpen
	\bibfield  {author} {\bibinfo {author} {\bibfnamefont {G.}~\bibnamefont
			{B\'elanger}}, \bibinfo {author} {\bibfnamefont {F.}~\bibnamefont
			{Boudjema}}, \bibinfo {author} {\bibfnamefont {A.}~\bibnamefont {Goudelis}},
		\bibinfo {author} {\bibfnamefont {A.}~\bibnamefont {Pukhov}}, \ and\ \bibinfo
		{author} {\bibfnamefont {B.}~\bibnamefont {Zaldivar}},\ }\href {\doibase
		10.1016/j.cpc.2018.04.027} {\bibfield  {journal} {\bibinfo  {journal}
			{Comput. Phys. Commun.}\ }\textbf {\bibinfo {volume} {231}},\ \bibinfo
		{pages} {173} (\bibinfo {year} {2018})},\ \Eprint
	{http://arxiv.org/abs/1801.03509} {arXiv:1801.03509 [hep-ph]} \BibitemShut
	{NoStop}%
	\bibitem [{\citenamefont {Staub}(2014)}]{Staub:2013tta}%
	\BibitemOpen
	\bibfield  {author} {\bibinfo {author} {\bibfnamefont {F.}~\bibnamefont
			{Staub}},\ }\href {\doibase 10.1016/j.cpc.2014.02.018} {\bibfield  {journal}
		{\bibinfo  {journal} {Comput. Phys. Commun.}\ }\textbf {\bibinfo {volume}
			{185}},\ \bibinfo {pages} {1773} (\bibinfo {year} {2014})},\ \Eprint
	{http://arxiv.org/abs/1309.7223} {arXiv:1309.7223 [hep-ph]} \BibitemShut
	{NoStop}%
	\bibitem [{\citenamefont {Porod}\ and\ \citenamefont
		{Staub}(2012)}]{Porod:2011nf}%
	\BibitemOpen
	\bibfield  {author} {\bibinfo {author} {\bibfnamefont {W.}~\bibnamefont
			{Porod}}\ and\ \bibinfo {author} {\bibfnamefont {F.}~\bibnamefont {Staub}},\
	}\href {\doibase 10.1016/j.cpc.2012.05.021} {\bibfield  {journal} {\bibinfo
			{journal} {Comput. Phys. Commun.}\ }\textbf {\bibinfo {volume} {183}},\
		\bibinfo {pages} {2458} (\bibinfo {year} {2012})},\ \Eprint
	{http://arxiv.org/abs/1104.1573} {arXiv:1104.1573 [hep-ph]} \BibitemShut
	{NoStop}%
	\bibitem [{\citenamefont {Belanger}\ \emph {et~al.}(2010)\citenamefont
		{Belanger}, \citenamefont {Boudjema}, \citenamefont {Pukhov},\ and\
		\citenamefont {Semenov}}]{Belanger:2010pz}%
	\BibitemOpen
	\bibfield  {author} {\bibinfo {author} {\bibfnamefont {G.}~\bibnamefont
			{Belanger}}, \bibinfo {author} {\bibfnamefont {F.}~\bibnamefont {Boudjema}},
		\bibinfo {author} {\bibfnamefont {A.}~\bibnamefont {Pukhov}}, \ and\ \bibinfo
		{author} {\bibfnamefont {A.}~\bibnamefont {Semenov}},\ }\href {\doibase
		10.1393/ncc/i2010-10591-3} {\bibfield  {journal} {\bibinfo  {journal} {Nuovo
				Cim. C}\ }\textbf {\bibinfo {volume} {033N2}},\ \bibinfo {pages} {111}
		(\bibinfo {year} {2010})},\ \Eprint {http://arxiv.org/abs/1005.4133}
	{arXiv:1005.4133 [hep-ph]} \BibitemShut {NoStop}%
	\bibitem [{\citenamefont {Bechtle}\ \emph {et~al.}(2010)\citenamefont
		{Bechtle}, \citenamefont {Brein}, \citenamefont {Heinemeyer}, \citenamefont
		{Weiglein},\ and\ \citenamefont {Williams}}]{Bechtle:2008jh}%
	\BibitemOpen
	\bibfield  {author} {\bibinfo {author} {\bibfnamefont {P.}~\bibnamefont
			{Bechtle}}, \bibinfo {author} {\bibfnamefont {O.}~\bibnamefont {Brein}},
		\bibinfo {author} {\bibfnamefont {S.}~\bibnamefont {Heinemeyer}}, \bibinfo
		{author} {\bibfnamefont {G.}~\bibnamefont {Weiglein}}, \ and\ \bibinfo
		{author} {\bibfnamefont {K.~E.}\ \bibnamefont {Williams}},\ }\href {\doibase
		10.1016/j.cpc.2009.09.003} {\bibfield  {journal} {\bibinfo  {journal}
			{Comput. Phys. Commun.}\ }\textbf {\bibinfo {volume} {181}},\ \bibinfo
		{pages} {138} (\bibinfo {year} {2010})},\ \Eprint
	{http://arxiv.org/abs/0811.4169} {arXiv:0811.4169 [hep-ph]} \BibitemShut
	{NoStop}%
	\bibitem [{\citenamefont {Bechtle}\ \emph {et~al.}(2011)\citenamefont
		{Bechtle}, \citenamefont {Brein}, \citenamefont {Heinemeyer}, \citenamefont
		{Weiglein},\ and\ \citenamefont {Williams}}]{Bechtle:2011sb}%
	\BibitemOpen
	\bibfield  {author} {\bibinfo {author} {\bibfnamefont {P.}~\bibnamefont
			{Bechtle}}, \bibinfo {author} {\bibfnamefont {O.}~\bibnamefont {Brein}},
		\bibinfo {author} {\bibfnamefont {S.}~\bibnamefont {Heinemeyer}}, \bibinfo
		{author} {\bibfnamefont {G.}~\bibnamefont {Weiglein}}, \ and\ \bibinfo
		{author} {\bibfnamefont {K.~E.}\ \bibnamefont {Williams}},\ }\href {\doibase
		10.1016/j.cpc.2011.07.015} {\bibfield  {journal} {\bibinfo  {journal}
			{Comput. Phys. Commun.}\ }\textbf {\bibinfo {volume} {182}},\ \bibinfo
		{pages} {2605} (\bibinfo {year} {2011})},\ \Eprint
	{http://arxiv.org/abs/1102.1898} {arXiv:1102.1898 [hep-ph]} \BibitemShut
	{NoStop}%
	\bibitem [{\citenamefont {Bechtle}\ \emph {et~al.}(2014)\citenamefont
		{Bechtle}, \citenamefont {Heinemeyer}, \citenamefont {St\r{a}l},
		\citenamefont {Stefaniak},\ and\ \citenamefont {Weiglein}}]{Bechtle:2013xfa}%
	\BibitemOpen
	\bibfield  {author} {\bibinfo {author} {\bibfnamefont {P.}~\bibnamefont
			{Bechtle}}, \bibinfo {author} {\bibfnamefont {S.}~\bibnamefont {Heinemeyer}},
		\bibinfo {author} {\bibfnamefont {O.}~\bibnamefont {St\r{a}l}}, \bibinfo
		{author} {\bibfnamefont {T.}~\bibnamefont {Stefaniak}}, \ and\ \bibinfo
		{author} {\bibfnamefont {G.}~\bibnamefont {Weiglein}},\ }\href {\doibase
		10.1140/epjc/s10052-013-2711-4} {\bibfield  {journal} {\bibinfo  {journal}
			{Eur. Phys. J. C}\ }\textbf {\bibinfo {volume} {74}},\ \bibinfo {pages}
		{2711} (\bibinfo {year} {2014})},\ \Eprint {http://arxiv.org/abs/1305.1933}
	{arXiv:1305.1933 [hep-ph]} \BibitemShut {NoStop}%
	\bibitem [{\citenamefont {Aalbers}\ \emph {et~al.}(2016)\citenamefont {Aalbers}
		\emph {et~al.}}]{DARWIN:2016hyl}%
	\BibitemOpen
	\bibfield  {author} {\bibinfo {author} {\bibfnamefont {J.}~\bibnamefont
			{Aalbers}} \emph {et~al.} (\bibinfo {collaboration} {DARWIN}),\ }\href
	{\doibase 10.1088/1475-7516/2016/11/017} {\bibfield  {journal} {\bibinfo
			{journal} {JCAP}\ }\textbf {\bibinfo {volume} {11}},\ \bibinfo {pages} {017}
		(\bibinfo {year} {2016})},\ \Eprint {http://arxiv.org/abs/1606.07001}
	{arXiv:1606.07001 [astro-ph.IM]} \BibitemShut {NoStop}%
	\bibitem [{\citenamefont {Akerib}\ \emph {et~al.}(2020)\citenamefont {Akerib}
		\emph {et~al.}}]{LZ:2018qzl}%
	\BibitemOpen
	\bibfield  {author} {\bibinfo {author} {\bibfnamefont {D.~S.}\ \bibnamefont
			{Akerib}} \emph {et~al.} (\bibinfo {collaboration} {LZ}),\ }\href {\doibase
		10.1103/PhysRevD.101.052002} {\bibfield  {journal} {\bibinfo  {journal}
			{Phys. Rev. D}\ }\textbf {\bibinfo {volume} {101}},\ \bibinfo {pages}
		{052002} (\bibinfo {year} {2020})},\ \Eprint
	{http://arxiv.org/abs/1802.06039} {arXiv:1802.06039 [astro-ph.IM]}
	\BibitemShut {NoStop}%
	\bibitem [{\citenamefont {Aprile}\ \emph {et~al.}(2020)\citenamefont {Aprile}
		\emph {et~al.}}]{XENON:2020kmp}%
	\BibitemOpen
	\bibfield  {author} {\bibinfo {author} {\bibfnamefont {E.}~\bibnamefont
			{Aprile}} \emph {et~al.} (\bibinfo {collaboration} {XENON}),\ }\href
	{\doibase 10.1088/1475-7516/2020/11/031} {\bibfield  {journal} {\bibinfo
			{journal} {JCAP}\ }\textbf {\bibinfo {volume} {11}},\ \bibinfo {pages} {031}
		(\bibinfo {year} {2020})},\ \Eprint {http://arxiv.org/abs/2007.08796}
	{arXiv:2007.08796 [physics.ins-det]} \BibitemShut {NoStop}%
	\bibitem [{\citenamefont {Ahnen}\ \emph {et~al.}(2016)\citenamefont {Ahnen}
		\emph {et~al.}}]{MAGIC:2016xys}%
	\BibitemOpen
	\bibfield  {author} {\bibinfo {author} {\bibfnamefont {M.~L.}\ \bibnamefont
			{Ahnen}} \emph {et~al.} (\bibinfo {collaboration} {MAGIC, Fermi-LAT}),\
	}\href {\doibase 10.1088/1475-7516/2016/02/039} {\bibfield  {journal}
		{\bibinfo  {journal} {JCAP}\ }\textbf {\bibinfo {volume} {02}},\ \bibinfo
		{pages} {039} (\bibinfo {year} {2016})},\ \Eprint
	{http://arxiv.org/abs/1601.06590} {arXiv:1601.06590 [astro-ph.HE]}
	\BibitemShut {NoStop}%
	\bibitem [{\citenamefont {Daylan}\ \emph {et~al.}(2016)\citenamefont {Daylan},
		\citenamefont {Finkbeiner}, \citenamefont {Hooper}, \citenamefont {Linden},
		\citenamefont {Portillo}, \citenamefont {Rodd},\ and\ \citenamefont
		{Slatyer}}]{Daylan:2014rsa}%
	\BibitemOpen
	\bibfield  {author} {\bibinfo {author} {\bibfnamefont {T.}~\bibnamefont
			{Daylan}}, \bibinfo {author} {\bibfnamefont {D.~P.}\ \bibnamefont
			{Finkbeiner}}, \bibinfo {author} {\bibfnamefont {D.}~\bibnamefont {Hooper}},
		\bibinfo {author} {\bibfnamefont {T.}~\bibnamefont {Linden}}, \bibinfo
		{author} {\bibfnamefont {S.~K.~N.}\ \bibnamefont {Portillo}}, \bibinfo
		{author} {\bibfnamefont {N.~L.}\ \bibnamefont {Rodd}}, \ and\ \bibinfo
		{author} {\bibfnamefont {T.~R.}\ \bibnamefont {Slatyer}},\ }\href {\doibase
		10.1016/j.dark.2015.12.005} {\bibfield  {journal} {\bibinfo  {journal} {Phys.
				Dark Univ.}\ }\textbf {\bibinfo {volume} {12}},\ \bibinfo {pages} {1}
		(\bibinfo {year} {2016})},\ \Eprint {http://arxiv.org/abs/1402.6703}
	{arXiv:1402.6703 [astro-ph.HE]} \BibitemShut {NoStop}%
	\bibitem [{\citenamefont {Barman}\ \emph {et~al.}(2021)\citenamefont {Barman},
		\citenamefont {B\'elanger}, \citenamefont {Bhattacherjee}, \citenamefont
		{Godbole}, \citenamefont {Sengupta},\ and\ \citenamefont
		{Tata}}]{Barman:2020vzm}%
	\BibitemOpen
	\bibfield  {author} {\bibinfo {author} {\bibfnamefont {R.~K.}\ \bibnamefont
			{Barman}}, \bibinfo {author} {\bibfnamefont {G.}~\bibnamefont {B\'elanger}},
		\bibinfo {author} {\bibfnamefont {B.}~\bibnamefont {Bhattacherjee}}, \bibinfo
		{author} {\bibfnamefont {R.}~\bibnamefont {Godbole}}, \bibinfo {author}
		{\bibfnamefont {D.}~\bibnamefont {Sengupta}}, \ and\ \bibinfo {author}
		{\bibfnamefont {X.}~\bibnamefont {Tata}},\ }\href {\doibase
		10.1103/PhysRevD.103.015029} {\bibfield  {journal} {\bibinfo  {journal}
			{Phys. Rev. D}\ }\textbf {\bibinfo {volume} {103}},\ \bibinfo {pages}
		{015029} (\bibinfo {year} {2021})},\ \Eprint
	{http://arxiv.org/abs/2006.07854} {arXiv:2006.07854 [hep-ph]} \BibitemShut
	{NoStop}%
	\bibitem [{\citenamefont {Blennow}\ \emph {et~al.}(2014)\citenamefont
		{Blennow}, \citenamefont {Fernandez-Martinez},\ and\ \citenamefont
		{Zaldivar}}]{Blennow:2013jba}%
	\BibitemOpen
	\bibfield  {author} {\bibinfo {author} {\bibfnamefont {M.}~\bibnamefont
			{Blennow}}, \bibinfo {author} {\bibfnamefont {E.}~\bibnamefont
			{Fernandez-Martinez}}, \ and\ \bibinfo {author} {\bibfnamefont
			{B.}~\bibnamefont {Zaldivar}},\ }\href {\doibase
		10.1088/1475-7516/2014/01/003} {\bibfield  {journal} {\bibinfo  {journal}
			{JCAP}\ }\textbf {\bibinfo {volume} {01}},\ \bibinfo {pages} {003} (\bibinfo
		{year} {2014})},\ \Eprint {http://arxiv.org/abs/1309.7348} {arXiv:1309.7348
		[hep-ph]} \BibitemShut {NoStop}%
	\bibitem [{\citenamefont {Aaboud}\ \emph {et~al.}(2017)\citenamefont {Aaboud}
		\emph {et~al.}}]{ATLAS:2017fih}%
	\BibitemOpen
	\bibfield  {author} {\bibinfo {author} {\bibfnamefont {M.}~\bibnamefont
			{Aaboud}} \emph {et~al.} (\bibinfo {collaboration} {ATLAS}),\ }\href
	{\doibase 10.1007/JHEP10(2017)182} {\bibfield  {journal} {\bibinfo  {journal}
			{JHEP}\ }\textbf {\bibinfo {volume} {10}},\ \bibinfo {pages} {182} (\bibinfo
		{year} {2017})},\ \Eprint {http://arxiv.org/abs/1707.02424} {arXiv:1707.02424
		[hep-ex]} \BibitemShut {NoStop}%
	\bibitem [{\citenamefont {Sirunyan}\ \emph {et~al.}(2020)\citenamefont
		{Sirunyan} \emph {et~al.}}]{CMS:2019buh}%
	\BibitemOpen
	\bibfield  {author} {\bibinfo {author} {\bibfnamefont {A.~M.}\ \bibnamefont
			{Sirunyan}} \emph {et~al.} (\bibinfo {collaboration} {CMS}),\ }\href
	{\doibase 10.1103/PhysRevLett.124.131802} {\bibfield  {journal} {\bibinfo
			{journal} {Phys. Rev. Lett.}\ }\textbf {\bibinfo {volume} {124}},\ \bibinfo
		{pages} {131802} (\bibinfo {year} {2020})},\ \Eprint
	{http://arxiv.org/abs/1912.04776} {arXiv:1912.04776 [hep-ex]} \BibitemShut
	{NoStop}%
	\bibitem [{\citenamefont {ACCOMANDO}\ \emph {et~al.}(2013)\citenamefont
		{ACCOMANDO}, \citenamefont {Becciolini}, \citenamefont {Belyaev},
		\citenamefont {De~Curtis}, \citenamefont {Dominici}, \citenamefont {King},
		\citenamefont {Moretti},\ and\ \citenamefont
		{Shepherd-Themistocleous}}]{ACCOMANDO:2013zz}%
	\BibitemOpen
	\bibfield  {author} {\bibinfo {author} {\bibfnamefont {E.}~\bibnamefont
			{ACCOMANDO}}, \bibinfo {author} {\bibfnamefont {D.}~\bibnamefont
			{Becciolini}}, \bibinfo {author} {\bibfnamefont {A.}~\bibnamefont {Belyaev}},
		\bibinfo {author} {\bibfnamefont {S.}~\bibnamefont {De~Curtis}}, \bibinfo
		{author} {\bibfnamefont {D.}~\bibnamefont {Dominici}}, \bibinfo {author}
		{\bibfnamefont {S.~F.}\ \bibnamefont {King}}, \bibinfo {author}
		{\bibfnamefont {S.}~\bibnamefont {Moretti}}, \ and\ \bibinfo {author}
		{\bibfnamefont {C.~H.}\ \bibnamefont {Shepherd-Themistocleous}},\ }\href
	{\doibase 10.22323/1.191.0125} {\bibfield  {journal} {\bibinfo  {journal}
			{PoS}\ }\textbf {\bibinfo {volume} {DIS 2013}},\ \bibinfo {pages} {125}
		(\bibinfo {year} {2013})}\BibitemShut {NoStop}%
	\bibitem [{\citenamefont {Aad}\ \emph {et~al.}(2016)\citenamefont {Aad} \emph
		{et~al.}}]{ATLAS:2016neq}%
	\BibitemOpen
	\bibfield  {author} {\bibinfo {author} {\bibfnamefont {G.}~\bibnamefont
			{Aad}} \emph {et~al.} (\bibinfo {collaboration} {ATLAS, CMS}),\ }\href
	{\doibase 10.1007/JHEP08(2016)045} {\bibfield  {journal} {\bibinfo  {journal}
			{JHEP}\ }\textbf {\bibinfo {volume} {08}},\ \bibinfo {pages} {045} (\bibinfo
		{year} {2016})},\ \Eprint {http://arxiv.org/abs/1606.02266} {arXiv:1606.02266
		[hep-ex]} \BibitemShut {NoStop}%
\end{thebibliography}
\end{document}